    [2016/10/16 v3.2016%
     Template file for NDdiss2e class]
\documentclass[final,sort&compress,noinfo]{nddiss2e}

\usepackage{tikz}
\usepackage{amsmath}
\usepackage{cancel}
\usepackage{algorithm2e,algorithmic}
\usepackage{hyperref}
\usepackage{color,colortbl}
\usepackage{amssymb}
\usepackage{booktabs}
\usepackage{graphicx}
\usepackage{doi}
\usetikzlibrary{calc}
\newcommand{\diff}{\text{d}}
\newcommand{\nuc}[2]{$^{#1}$#2}
\newcommand{\expect}[1]{\left<#1\right>}
\newcommand{\asymmerror}[3]{$#1^{+#2}_{-#3}$}
\renewcommand{\vec}{\mathbf}
\newcommand{\ket}[1]{\left|#1\right>}
\newcommand{\bra}[1]{\left<#1\right|}
\newcommand{\braketop}[3]{\left<#1\right|#2\left|#3\right>}

\definecolor{Gray}{gray}{0.9}

\begin{document}

\frontmatter             

\title{Structure effects on the giant monopole resonance and determinations of the nuclear incompressibility}  

\author{Kevin B. Howard}      
\work{Dissertation}    
\degaward{Doctor of Philosophy}  
\advisor{Umesh Garg}  
\department{Physics}           

\maketitle               

 \makecopyright        


  \begin{abstract}
    Giant resonances are archetypal forms of collective nuclear motion which provide a unique laboratory setting to probe the bulk properties of the nuclear force. One of the isoscalar compressional modes -- namely, the isoscalar giant monopole resonance (ISGMR) -- has been studied extensively with the goal of constraining the density dependence of the equation of state (EoS) for infinite nuclear matter. For example, the nuclear incompressibility, $K_\infty$, is a fundamental quantity in the EoS and is directly correlated with the energies of the ISGMR in finite nuclei.

    Previous work has shown that interactions with $K_\infty$ which reproduce the centroid energies of the ISGMR in \nuc{208}{Pb} and \nuc{90}{Zr} well, overestimate the ISGMR response of the tin and cadmium nuclei. To further elucidate this question as also to examine \emph{when} this ``softness'' appears in moving away from the doubly-closed nucleus \nuc{90}{Zr}, and \emph{how} this effect develops, the first portion of this thesis consists of measurements and analyses of the ISGMR within the molybdenum isotopes. The experiments were performed for $^{94,96,97,98,100}$Mo, using inelastic scattering of 100 MeV/u $\alpha$-particles at the Research Center for Nuclear Physics at Osaka University. The strength distributions for the giant resonances were extracted using multipole decomposition analyses within a Markov-Chain Monte Carlo framework to quantify the uncertainties in the strength distributions and the ISGMR energies. Comparison of the measured ISGMR strengths with Random Phase Approximation calculations demonstrates that the molybdenum nuclei have ISGMR energies which are overestimated to a similar degree as seen in the tin and cadmium nuclei, while the strength of \nuc{208}{Pb} is precisely reproduced. This suggests clearly that the molybdenum nuclei exhibit the same open-shell softness which has been documented previously.

    Studies of the ISGMR in isotopic chains encompassing a broad range of proton-neutron asymmetries allow for extraction of the dependence of the finite nuclear incompressibility on the isospin asymmetry, as quantified by the asymmetry term of the nuclear incompressibility, $K_\tau$.  Recent data on the ISGMR in \nuc{40,44,48}{Ca} have contradicted prior results for $K_\tau$. To reconcile the otherwise highly concerning conclusion that $K_\tau = + 582$ MeV, the second portion of this thesis is focused upon independently studying this claim. A simultaneous measurement of the ISGMR in \nuc{40,42,44,48}{Ca} was completed and has resulted in a high-confidence exclusion of the possibility of a positive value for the asymmetry term, and indeed has found consistency with previous data, placing $K_\tau$ at $-510 \pm 115$ MeV.
  \end{abstract}


\tableofcontents
\listoffigures
\listoftables



 \begin{acknowledge}

I would be remiss to submit this dissertation and complete my PhD without appropriately acknowledging the underlying support system that has helped me throughout my training as a physicist. I first thought to put off writing this section until the very end, so that I could collect my thoughts and present a carefully-constructed, thoughtful narrative about precisely how I have benefited from all of my closest family, friends, and collaborators. In so doing, I've found that it would be futile to try and delineate each and every person and the exact roles which they have played in helping me along this journey. So, this section is not meant to be exhaustive in that it will not list everyone who has influenced me along the way, but instead to credit those --- in no particular order --- without whom I feel the pathway to this degree would simply not have existed.

I owe my gratitude to Professor Umesh Garg, who has served as my supervisor and the director of this dissertation research for many years. His flexible supervisory style is the reason that I credit for having developed into a successful researcher. Furthermore, it allowed me to pursue research tangents and teaching opportunities when I found them interesting. Altogether, I'm grateful to him for doing his part in making the arduous path towards a PhD on average, to the extent possible,  a pleasant one. 

Along those lines, I'd like to thank Professor Mark Caprio for trusting me to take on substantial teaching opportunities over the last few years. The courses I have taught have been a welcome break from writing my dissertation, and I have had positively wonderful experiences with the students and in the curriculum development. I also feel that our work with the University Writing Center has invariably made the actual dissertation-writing process a more straightforward one.

I thank each of my dissertation committee members: Professors Tan Ahn, Mark Caprio, Umesh Garg, and Grant Mathews. Throughout my PhD candidacy, they have provided me with candid advice and constructive feedback on my progress, both at our regularly-scheduled annual meetings, as well as whenever I cornered them in the hallways or in their offices to discuss matters of research. For the latter, I also apologize; but I nonetheless think highly of each of them that they were all generally willing to engage me under such circumstances.

It would have been impossible to complete any of this research without the logistical support --- and in many cases, sage life-advice --- of the office staff. Shari Herman, Lori Fuson, and Janet Weikel have each been invaluable in these respects, and so I thank them for always going above and beyond in helping students in any way possible. Each of them made working in the department extremely enjoyable for me, and for that I owe each of them a deep debt of gratitude.

My prior mentors deserve special credit. The enthusiasm and teaching efforts of Mike Wood, Rob Selkowitz, Ken Scherkoske, and Ben DeJonge were hugely impactful in my decision to continue my physics education and pursue graduate study. In many ways, I have emulated the best features of their teaching styles with the hope that I might have a comparable impact on my own students. I'd like to additionally thank Richard Escobales, Tony Weston, and Christine Kinsey for their advice and support throughout my undergraduate mathematics education. It was a happy accident that I found a second home in the Mathematics Department as an undergraduate, and I think that my experiences there were quite instrumental in my subsequent academic pursuits.

I would like to credit all of the individuals --- whether they are named here, or not --- who have helped me with taking care of Clover during the last few years. She is a wonderfully sweet dog, and her company during my PhD studies has been a joy. My research has, however, given me great opportunity to travel the globe, sometimes for months at a time. It would have simply been impossible to succeed if not for the constant reassurance that she was in good hands during my absence. Thank you.

The friends and colleagues that I have made here, both within and outside of the department, have formed a bedrock of support without which I would have burned out years ago. For the members of my research group who have withstood the tests of time: Nirupama, Joe, and Joseph; each of your personalities contributed to a research-group culture that combined levity with discipline, and have collectively made even the most stressful periods of time seem manageable. Patrick, Orlando, Jake O., and Erika: you are individuals whom I hold in extremely high esteem, as I cannot begin to count the number of enjoyable hours I have spent discussing interesting science, curious problems, bad code, or myriad other topics with each of you. Better yet, you all seemed to have a sixth sense for knowing exactly when I needed a break from my work, and so these conversations were both therapeutic and engaging. Bryce, Laura, and Jacob: each of you have been wonderful friends over the years; I've reliably found the company of you all to be among the very best catalysts for forgetting the stressors associated with the completion of this degree.

The friends who have been around for a while longer have been equally as instrumental. I expected to grow apart from some of my friends from my undergraduate years when I moved away from Western New York. Ashleigh, and Jake D: you both have, despite your own hectic schedules, always made time to meaningfully stay in touch. Despite it having been nearly five years since we've spent significant time together, I have never thought twice about reaching out to share news --- good or bad --- with, or to blow off some steam by venting to, either of you. I know each of you feel the same, and so our friendships have been priceless.

Theron: words cannot really completely describe how thankful I am that you have been as steadfastly positive and unwavering in your support as you have been during this journey, but I hope that these ones begin to. I am further grateful to Ken, Anita, Evan, Emily, Vivian, and Brenda: you have all welcomed me into your family. In the Midwest, as far from home as I am, it has been a constant comfort to know that you all are only a short drive away.

Finally, I absolutely must credit my family back home for forming the foundation on which this accomplishment is built. Janie, Kevin, Owen and Elle: watching your family grow up as I have moved through my undergraduate and graduate school years has been heartwarming, and a constant reminder that there is so much more to life beyond my studies and research. My parents, Kevin and Vicki: you have never failed to pick up the phone. You have taught me to be independent, but have at the same time been willing to drop everything to help me if I have ever needed it. I would not be where I am today if not for you both.

This work was supported in part by National Science Foundation grant numbers PHY-1419765 and PHY-1713857, the Arthur J. Schmitt Foundation, as well as the Liu Institute for Asia and Asian Studies and the College of Science, each of the University of Notre Dame.



 \end{acknowledge}
 \include{acknowledgement}

\mainmatter

%
%
%
%
%
%
%
%
%
%

%
%

\chapter{Background of the physical problem}\label{intro}

\section{Properties of bulk nuclear matter} \label{intro:bulk properties}
Since its infancy, the field of nuclear physics has endeavored to characterize the strong nuclear force and to predict the nuclear ground- and excited-state properties which arise from its features; those emergent phenomena of the nucleon-nucleon interaction have resulted in great interplay between efforts to constrain both collective properties of finite nuclei as well as observed astrophysical features of bulk nuclear matter. \footnote{The concept of \emph{nuclear matter}, as referred to in this thesis work, is the theoretical limit in which the nucleon number goes to infinity with a fixed proton-neutron asymmetry.} Along these lines, there has been substantial effort by the nuclear physics community to constrain properties of the nucleon-nucleon interaction using limits from both nuclear and astrophysical measurements, while simultaneously constraining the microscopic observables which depend on the interactions and are used as input for benchmarking the theories; these topics have evolved into independent fields of research in their own rights, with laudable progress over the recent years \cite{dutra_stone_skyrme,steiner_mass_radius_EoS_neutron_star,schneider_EoS_core_collapse}.

One of the ultimate goals for these fields is the calculation of the nuclear \emph{equation of state} (EoS), denoted $\epsilon(\rho,\eta)$, which relates the energy per nucleon, $\epsilon$, to the nucleon density, $\rho$, and the proton-neutron asymmetry, $\eta=(N-Z)/A$. This EoS is a constitutive relationship which is uniquely characterized by the nucleon-nucleon interaction, and yields a fundamental relation between the particle density and energy per nucleon. In systems which are well-modeled in the limit of nuclear matter, the EoS provides the full thermodynamic description of the variation of extensive and intensive properties of the system. Further applications of the EoS and symmetry energy lie within the fields of heavy-ion collisions \cite{symmetry_energy_heavy_ion_collisions}, modeling of core-collapse stellar events \cite{schneider_EoS_core_collapse}, recent gravitational-wave observations corresponding to GS170817 \cite{gravitational_waves_nuclear_eos}, and modeling the structure of neutron stars \cite{lattimer_neutron_stars,piekarewicz_neutron_stars}.

As one example: these equations of state are of especial importance as astrophysical input for calculating the dynamical properties of neutron stars. Indeed, they serve as the sole input in solving the Tolman Oppenheimer Volkoff (TOV) equations that describe the hydrostatic equilibrium between gravitational collapse and the pressure arising from the nucleon-nucleon interaction within a general-relativistic framework \cite{steiner_mass_radius_EoS_neutron_star,schneider_EoS_core_collapse,fattoyev_neutron_stars_EoS,piekarewicz_neutron_stars,lattimer_neutron_stars,todd-rutel_piekarewicz_relativistic_interactions_and_neutron_stars,zhang_neutron_stars_symmetry_energy}.

The TOV equations are a set of first order, non-linear, coupled differential equations which arise from the condition of hydrodynamical stability between the internal pressure and gravitational collapse, and which collectively model the pressure and mass profile of a stellar body of general-relativistic mass scale. For spherical, non-rotating neutron stars, the TOV equations are:

\begin{align}
  \frac{\diff P}{\diff r} &= - G \frac{{\mathcal{E}}(r) m(r)}{r^2} \left(1 + \frac{P(r)}{{\mathcal{E}(r)}} \right) \left(1 + \frac{4\pi r^3 P(r)}{m(r)} \right) \left [1-2 G \frac{m(r)}{r} \right]^{-1}, \notag \\
  \frac{\diff m}{\diff r} &= 4 \pi r^2 \mathcal{E}(r).
  \label{TOV_equations}
\end{align}

Here, $G$ is the gravitational constant, and $P$, $\mathcal{E}$, and $m$ are, respectively, the pressure, relativistic mass density, and enclosed mass as functions of the radius, $r$. The radial dependence of each of these three quantities is unknown, and the third equation which allows for simultaneously solving for the functional forms of $P$, $\mathcal{E}$, and $m$ is precisely the EoS of asymmetric nuclear matter.

Given an EoS $\epsilon(\rho,\eta)$, in which $\rho$ is the nucleon number density, the pressure $P(\rho)$ is given by
\begin{align}
  P(\rho) & = \rho^2 \frac{\diff \epsilon }{\diff \rho}.
\end{align}
Moreover, the relationship between the nucleon number density and the relativistic mass density is:
\begin{align}
  \mathcal{E}(\rho,\eta) = \rho \left [m + \epsilon(\rho,\eta ) \right]
\end{align}
with $m$ being the nucleon mass. The acquisition of an EoS that relates the density to the pressure allows one to iteratively solve Eqs. \eqref{TOV_equations} for the neutron star structure. A highly-important relation that can be extracted from these solutions is the relationship between the mass and radius of a neutron star: the gravitational compression of the neutron star is directly balanced by the pressure which, in this case, arises directly from the nucleon-nucleon interaction.

\begin{figure}[t!]
  \centering
  \includegraphics[width=\linewidth]{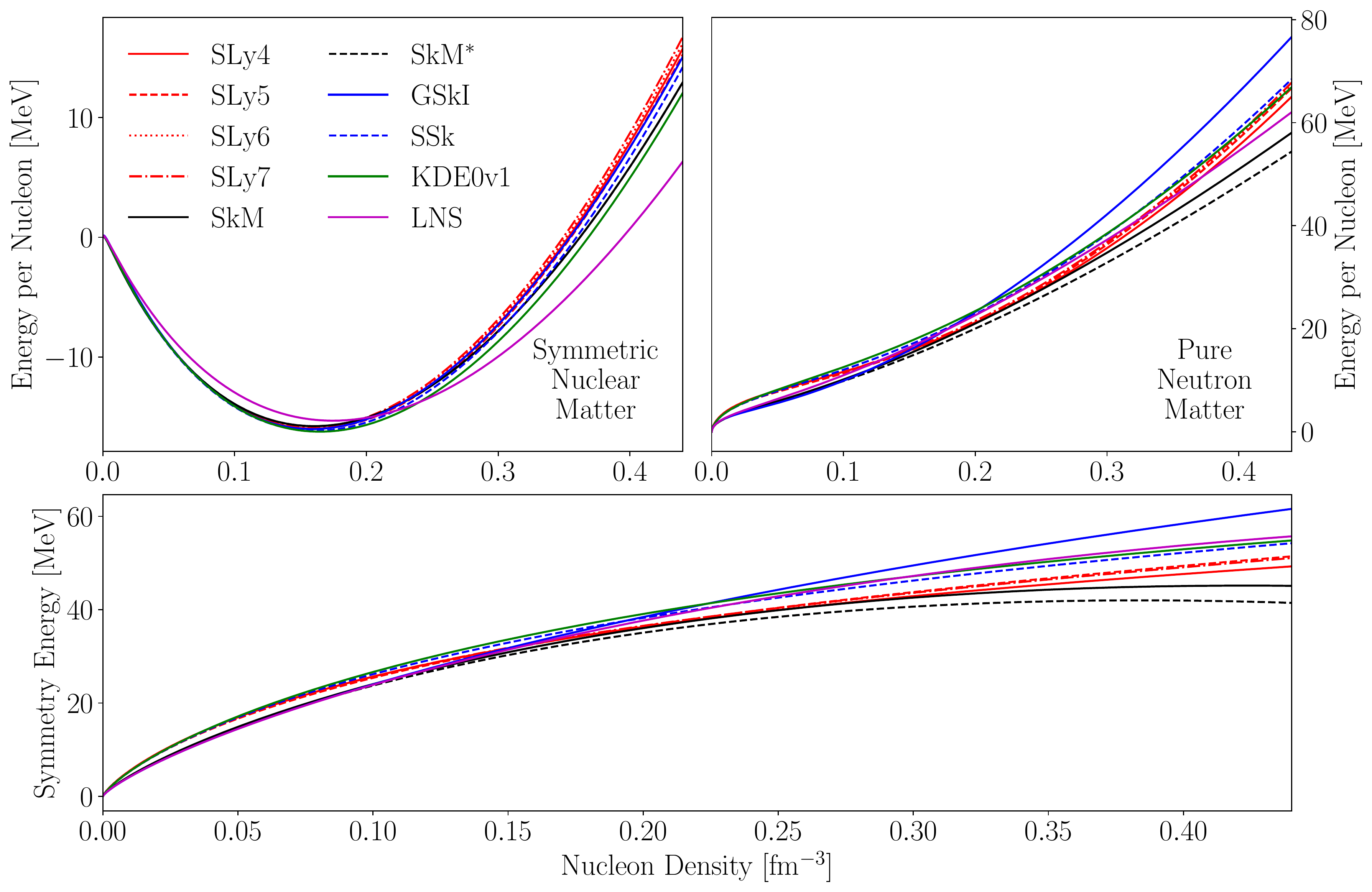}
  \caption[Equations of state and symmetry energy for symmetric nuclear and pure neutron matter as calculated with Skyrme interactions.]{Top left: sample equations of state depicting the density dependence of the energy per nucleon in symmetric nuclear matter as calculated with various nonrelativistic Skyrme interactions \cite{SLYx_commissioned,SkM_commissioned,SkM_star_commissioned,GSKI_and_SSk_commissioned,KDE0v1_commissioned,LNS_commissioned}. Top right: same, but for pure neutron matter. Bottom: density dependence of the symmetry energy for each of the Skyrme interactions.}
  \label{intro:eos_symmetric_figure}
\end{figure}

Figure \ref{intro:eos_symmetric_figure} shows sample EoS, calculated with some modern nonrelativistic parametrizations of the nuclear force \cite{SLYx_commissioned,SkM_commissioned,SkM_star_commissioned,GSKI_and_SSk_commissioned,KDE0v1_commissioned,LNS_commissioned}, for the cases of \emph{symmetric} nuclear matter ($\eta=0$) as well as for \emph{pure neutron} matter ($\eta =1$).

\begin{figure}[t!]
  \centering
  \includegraphics[width=\linewidth]{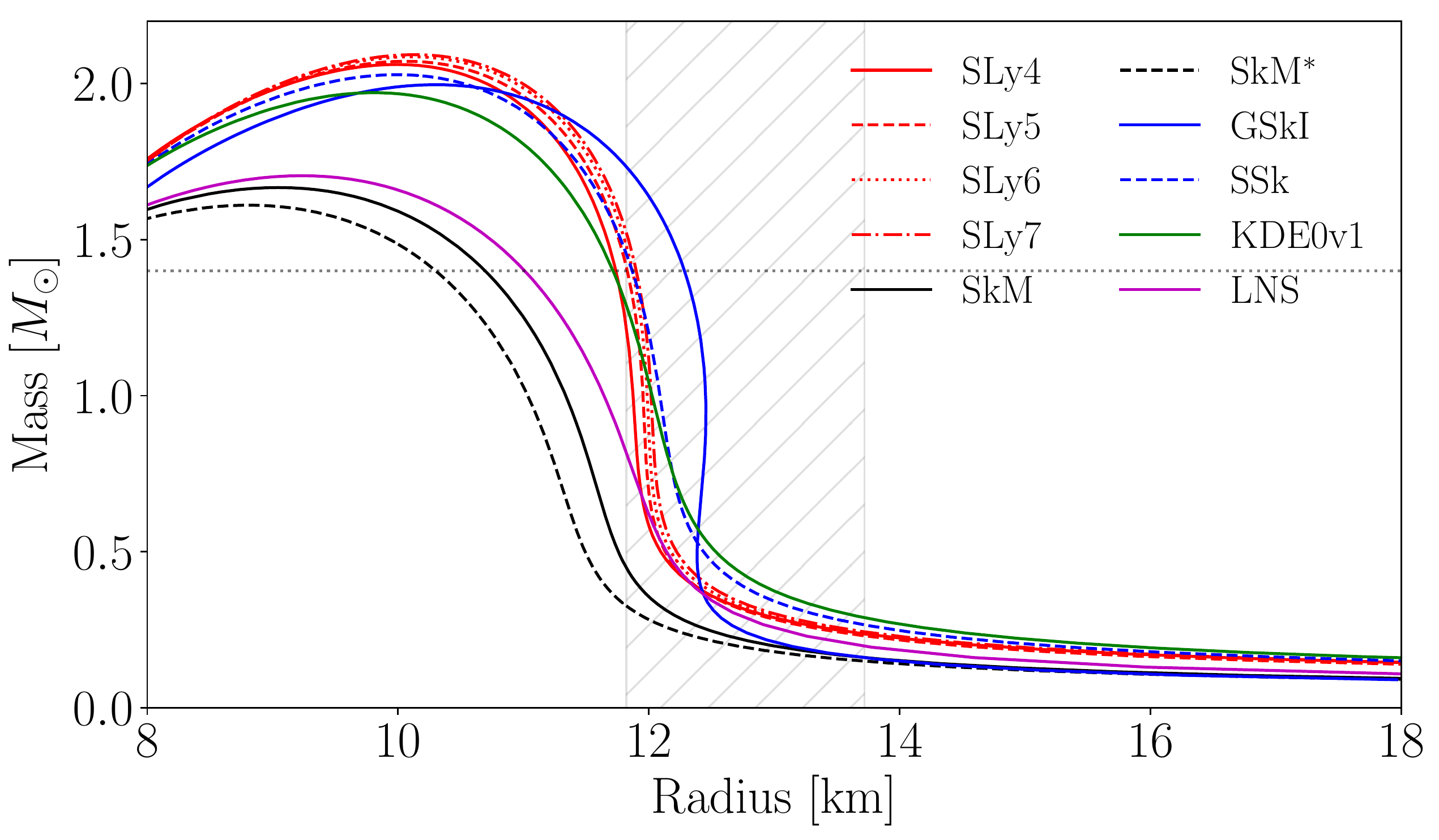}
  \caption[Neutron star mass-radius relation calculated using various Skyrme interactions.]{Neutron star mass-radius relations calculated using various Skyrme interactions \cite{SLYx_commissioned,SkM_commissioned,SkM_star_commissioned,GSKI_and_SSk_commissioned,KDE0v1_commissioned,LNS_commissioned}. The horizontal line corresponds to $M=1.4 M_{{\odot}}$, the Chandrasekhar limit. The shaded region corresponds to the constraints placed on the radius of a $1.4 M_{{\odot}}$ neutron star by the recent gravitational waves observation GS170817 \cite{gravitational_waves_nuclear_eos}. }
  \label{mass_radius_plot}
\end{figure}

Figure \ref{intro:eos_symmetric_figure} also shows a quantity called the \emph{symmetry energy}, $S(\rho)$:
\begin{align}
  \epsilon(\rho,\eta) = \epsilon(\rho,0) + \eta^2 S(\rho), \label{eos_asymmetric}
\end{align}
which determines the energy cost associated with having a neutron excess within a bulk collection of nucleons (\emph{i.e.} the symmetry energy is the difference between the EoS for pure neutron matter and that of symmetric nuclear matter). Inspection of the various curves in Fig. \ref{intro:eos_symmetric_figure} suggests a general agreement among the various interactions in reproducing the static saturation properties of nuclear matter --- namely, $\epsilon(\rho_0,0) \approx -16$ MeV and the saturation density $\rho_0 \approx$ 0.16 fm$^{-3}$ --- whereas the density dependence of each curve (symmetric matter, pure neutron matter, and the symmetry energy) is heavily interaction-dependent. The implications of this density dependence are significant: considering again the case of the dynamics of neutron stars, one finds that each EoS yields, when used in the solution of the TOV equations, a \emph{unique} relationship between the mass and radius of a given neutron star \cite{lindblom_unique_mass_radius}.

Simple calculations which relate the mass and radii of neutron stars using these sample equations of state as input to the TOV equations for pure neutron matter\interfootnotelinepenalty=10000\footnote{One should note that a more accurate calculation would impose the constraint of $\beta$-equilibrium and thus include the effects of a nonzero proton fraction in the stellar composition. However, such considerations are outside the scope of this thesis and the presentation of the existing calculations conveys the salient points.} are presented in Fig. \ref{mass_radius_plot}. One should note that these calculations do not include the effects due to phase transitions in the high-density stellar interior \cite{todd-rutel_piekarewicz_relativistic_interactions_and_neutron_stars} and have been completed with an assumption of pure neutron matter for the stellar composition. Nonetheless, even with a fairly narrow selection of interactions, the figure shows clearly the marked variation of the astrophysical observables which arise due to a lack of constraint on the density dependence of the EoS. It is thus the charge of terrestrial nuclear physicists to endeavor to place limits on the possible nuclear equations of state which then, in turn, better reproduce astrophysical properties.

To these ends, one can isolate properties of the EoS that can be most directly measured within a laboratory setting. Considering first the case of symmetric nuclear matter, one can expand in a Taylor series about its minimum at saturation density:

\begin{align}
  \epsilon(\rho,0) &= \epsilon(\rho_0,0) + \cancelto{0}{\frac{\diff \epsilon}{{\diff}{\rho}} \bigg|_{\rho=\rho_0}} \left(\rho-\rho_0\right) +  \frac{1}{2} \frac{\diff^2 \epsilon}{\diff \rho^2} \bigg|_{\rho=\rho_0} \left(\rho-\rho_0 \right)^2 +  \frac{1}{6} \frac{\diff^3 \epsilon}{\diff \rho^3} \bigg|_{\rho=\rho_0} \left(\rho-\rho_0 \right)^3 + \ldots \notag \\
  &= \epsilon(\rho_0,0) + \frac{1}{2} \frac{1}{9 \rho_0^2} K_\infty (\rho-\rho_0)^2 + \frac{1}{6} \frac{1}{27\rho_0^3} Q_\infty (\rho-\rho_0)^3 +  \ldots  \label{intro:symmetric eos}
\end{align}
wherein we have defined the quantities
\begin{align}
  K_\infty &= 9 \rho_0^2 \frac{\diff^2 \epsilon}{\diff \rho^2} \bigg|_{\rho=\rho_0}, \label{intro:K_defined} \\
  Q_\infty &= 27 \rho_0^3 \frac{\diff^3 \epsilon}{\diff \rho^3} \bigg|_{\rho=\rho_0}. \label{intro:Q_defined}
\end{align}

\begin{table}[b!]
\centering
\caption{Nuclear-matter properties extracted from selected interactions}
\label{interactions_table}
\resizebox{\textwidth}{!}{%
\begin{tabular}{@{}ccccccccccc@{}}
\toprule
            &  & \multicolumn{4}{c}{Symmetric Nuclear Matter}                    &  & \multicolumn{4}{c}{Symmetry Energy}              \\ \cmidrule(lr){3-6} \cmidrule(l){8-11}
Interaction &  & $\rho_0$    & $\varepsilon(\rho_0,0)$ & $K_\infty$ & $Q_\infty$ &  & $J$   & $L$   & $K_\text{sym}$ & $K_\tau^\infty$ \\
            &  & [fm$^{-3}$] & [MeV]                   & [MeV]      & [MeV]      &  & [MeV] & [MeV] & [MeV]          & [MeV]           \\ \cmidrule(r){1-1}  \cmidrule(lr){3-6} \cmidrule(l){8-11}
SLy4 \cite{SLYx_commissioned}       &  & 0.160       & -15.97                  & 229.91     & 363.11     &  & 32.00 & 45.94 & -119.73        & -322.83         \\
SLy5 \cite{SLYx_commissioned}       &  & 0.161       & -15.99                  & 229.92     & 364.16     &  & 32.01 & 48.15 & -112.76        & -325.38         \\
SLy6 \cite{SLYx_commissioned}       &  & 0.159       & -15.92                  & 229.86     & 360.24     &  & 31.96 & 47.45 & -112.71        & -323.03         \\
SLy7 \cite{SLYx_commissioned}       &  & 0.158       & -15.90                  & 229.75     & 359.22     &  & 31.99 & 46.94 & -114.34        & -322.60         \\
SkM  \cite{SkM_commissioned}       &  & 0.160       & -15.77                  & 216.61     & 386.09     &  & 30.75 & 49.34 & -148.81        & -356.91         \\
SkM$^*$  \cite{SkM_star_commissioned}   &  & 0.160       & -15.77                  & 216.61     & 386.09     &  & 30.03 & 45.78 & -155.94        & -349.00         \\
GSkI  \cite{GSKI_and_SSk_commissioned}      &  & 0.159       & -16.02                  & 230.21     & 405.58     &  & 32.03 & 63.45 & -95.29         & -364.19         \\
SSk   \cite{GSKI_and_SSk_commissioned}      &  & 0.161       & -16.16                  & 229.31     & 375.38     &  & 33.50 & 52.78 & -119.15        & -349.42         \\
KDE0v1 \cite{KDE0v1_commissioned}     &  & 0.165       & -16.23                  & 227.54     & 384.86     &  & 34.58 & 54.69 & -127.12        & -362.78         \\
LNS   \cite{LNS_commissioned}      &  & 0.175       & -15.32                  & 210.78     & 382.55     &  & 33.43 & 61.45 & -127.36        & -384.55         \\
FSUGarnet \cite{fsugarnet_commissioned}  &  & 0.153       & -16.23                  & 229.50     & 4.50       &  & 30.92 & 51.00 & 59.50          & -247.3          \\ \bottomrule
\end{tabular}%
}
\end{table}

Here, $K_\infty$ is the \emph{nuclear incompressibility}, which is essentially a measure of the curvature of the EoS of symmetric nuclear matter at saturation density. This quantity, thus, is a characterization of the leading-order energy cost associated with increasing or decreasing the nucleon number density. Similarly, $Q_\infty$ is a measure of the higher-order skewness of the EoS.

Previous work has shown that the combination of Eqs. \eqref{eos_asymmetric} and \eqref{intro:symmetric eos} with a Taylor expansion of $S(\rho)$ around $\rho_0$ is sufficient for modeling the EoS for isospin-asymmetric nuclear matter \cite{isospin_dependence_EOS}. In so doing, one acquires the following:
\begin{align}
S(\rho) = J + \frac{1}{3 \rho_0} L \left(\rho-\rho_0 \right) + \frac{1}{2} \frac{1}{9\rho_0^2} K_\text{sym} \left(\rho-\rho_0 \right)^2 +\ldots
\end{align}

Here, $J$, $L$, and $K_\text{sym}$ are respectively the symmetry energy at saturation density, the symmetry pressure at saturation density, and the curvature of the symmetry energy at saturation density. These saturation properties, in combination with the corresponding quantities for symmetric nuclear matter, constitute a set of easily-calculable features of the EoS which can be experimentally probed in carefully-executed measurements of finite nuclei \cite{garg_colo_review,cao_sagawa_colo,piek_physG,Colo2014a}. Within this context, experimental nuclear physics has the capabilities to simultaneously restrict the classes of proposed nuclear interactions, in addition to the stellar equations of state, by using the dynamical properties of the EoS as limiting constraints.

For comparison purposes, Table \ref{interactions_table} shows the static and dynamic nuclear-matter properties discussed here --- as well as the asymmetry term $K_\tau^\infty$, which will be introduced shortly --- as calculated with the nonrelativistic Skyrme interactions characterized in Fig. \ref{intro:eos_symmetric_figure} and for which the neutron-star mass-radius relations are shown in Fig. \ref{mass_radius_plot}; a comprehensive listing of modern-day Skyrme interactions and their associated properties is given in Ref. \cite{dutra_stone_skyrme}. This table also shows the corresponding properties as calculated with the FSUGarnet relativistic interaction \cite{fsugarnet_commissioned}, as results for this interaction will also be compared against experimental data in subsequent chapters.

\section{Giant resonances and collectivity as lenses for bulk nuclear properties}

The epitome of nuclear collectivity is manifest within the \emph{giant resonances}, which are high-frequency excitations that typically involve the participation of a majority of the nucleons which constitute the atomic nucleus \cite{harakeh_book}. Owing to the bulk nature of these modes of nuclear oscillation which are generally independent of microscopic effects, they generate an ideal environment in which the bulk properties of the nucleus can be probed. The resonances are damped, harmonic oscillations about the equilibrium constitution of the system in which the properties of the strength distribution of the resonance are directly related to the ground-state properties of the system and the weak external field, $\mathcal{O}$, that initializes the oscillation \cite{harakeh_book,bortignon_book}.

Myriad giant resonances have been observed, and here only a brief presentation will be made on the general features of the various modes. A polychotomy of the resonances can be constructed using the quantum numbers of the external field that induces the oscillation and by characterizing the manner in which the nucleons participate in the vibration:
\begin{itemize}
  \item the \emph{electric, isoscalar} resonances are oscillations in which the external field couples neither to the isospin nor the spin projection of the nucleons;
  \item the \emph{electric, isovector} resonances are oscillations in which the external field couples to the isospin projection, but not to the spin projection;
  \item the \emph{magnetic, isoscalar} resonances are oscillations in which the external field does not couple to the isospin projection, but does couple to the spin projection; and
  \item the \emph{magnetic, isovector} resonances are oscillations in which the external field couples both to the isospin projection as well as to the spin projection.
\end{itemize}
The effects of the couplings of these fields to the nucleons is that different nucleons will oscillate in or out of phase with one another; for example, the electric isoscalar giant resonances consist of modes in which all nucleons oscillate in phase with one another. In contrast, the electric isovector resonances consist of oscillations in which the protons and neutrons oscillate directly out of phase with one another, irrespective of their spin projections. This thesis work will not dwell further upon features of the magnetic resonances, as they are not at all a focus of the work which was undertaken in this study.

\begin{figure}[t!]
  \centering
  \includegraphics[width=0.8\linewidth]{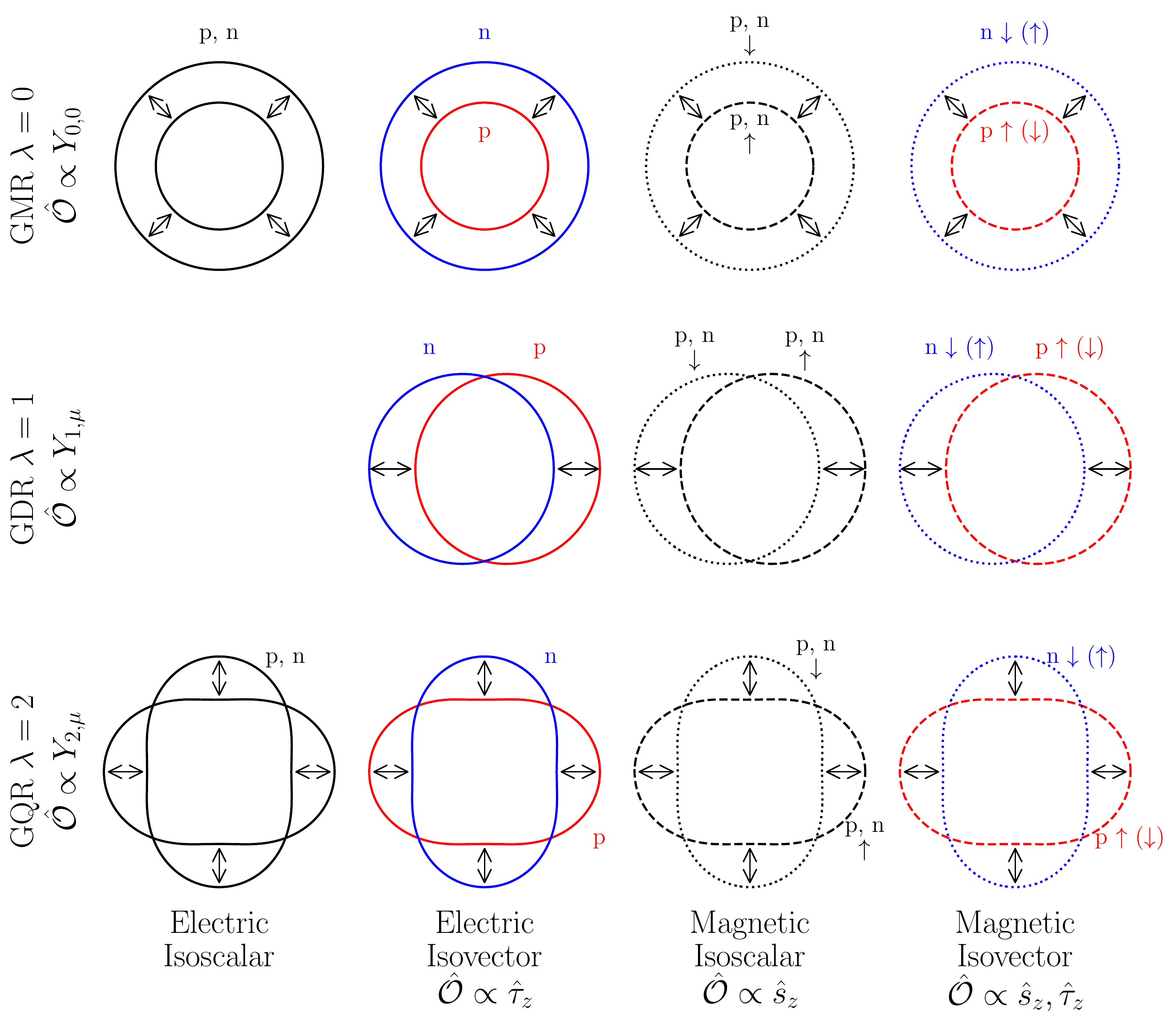}
  \caption{Schematic showing various geometries of collective motion.}
  \label{multipole figure}
\end{figure}

Schematics of these different giant resonances are shown in Fig. \ref{multipole figure}. In addition to the aforementioned organizational schema for the giant resonances, the multipolarity of the resonance geometry can vary according to the angular momentum transfer from the external field to the nucleus itself. This gives rise to the classifications of, for example, the \textbf{i}so\textbf{s}calar \textbf{g}iant \textbf{m}onopole \textbf{r}esonance (ISGMR), \textbf{i}so\textbf{v}ector \textbf{g}iant \textbf{d}ipole \textbf{r}esonance (IVGDR), and so on.

In this thesis work, the primary focus is on the electric giant resonances, and more specifically, the compressional behavior which manifests within the ISGMR.\footnote{The isoscalar giant dipole resonance, which is not shown in Fig. \ref{multipole figure}, is also a compressional mode. However, this mode was not a significant focus of this thesis work.} As shown in Fig. \ref{multipole figure}, in spherical nuclei, the ISGMR is characterized by a radially symmetric vibration in which the protons and neutrons rapidly expand and contract in the nuclear volume. With the particle number remaining constant in such a process, the nucleon density rapidly oscillates while the incompressibility modulus, or finite incompressibility, $K_A$ of the nucleus gives rise to the restoring force. The value of $K_A$ can be directly related to the driving frequency of the oscillation and in turn, the excitation energy of the resonance $E_\text{ISGMR}$:
\begin{align}
  E_\text{ISGMR} = \hbar \sqrt{\frac{K_A}{m \expect{r_0^2}}} . \label{KA_to_ISGMR}
\end{align}
Here, $m$ is the free-nucleon mass, while $\expect{r_0^2}$ is the ground-state mean-square radius. Among the different typical macroscopic models for the density vibration, the ISGMR energies would be associated with one of the moment ratios $\sqrt{m_3/m_1}$, $m_1/m_0$, or $\sqrt{m_1/m_{-1}}$, where the moments $m_k$ of the strength function $S_\lambda (E_x)$ are defined as
\begin{align}
  m_k = \int S_\lambda(E_{x}) E_{x}^k \,  \diff E_{x}, \label{moments_defined}
\end{align}
with $\lambda$ being the multipolarity of the resonance in question and $S_\lambda(E_x)$ its associated strength distribution \cite{stringari_sum_rules,stringari_lipparini_sum_rules,harakeh_book}. The value of $m_1$ is constrained by the \emph{energy-weighted sum rule} (EWSR) \cite{harakeh_book,stringari_sum_rules,stringari_lipparini_sum_rules}. The EWSR will be discussed in greater detail in Chapter \ref{theory}; for the moment, it is necessary only to assert that the EWSR depends solely on the ground-state properties of the nucleus in question and the features of the external field $\mathcal{O}$, and is model-independent. Furthermore, the percentage of the EWSR exhausted within a strength distribution is a typical metric of collectivity in characterizing the giant resonances. The percentage of the EWSR that is found to be exhausted within a given state is a quantitative measure of the collectivity of that state, and giant resonances are typically understood to exhaust a large percentage of the EWSR \cite{harakeh_book}.

\section{From properties of finite nuclei to those of nuclear matter}

Being that $K_A$ is itself a measure of the incompressibility modulus of a finite nucleus, one might expect that its values can be related to the values of the incompressibility coefficients $K_\infty$ and $K_\text{sym}$, owing to the commonalities in the nuclear force which gives rise to the phenomena on both finite and bulk scales. Under such a presumption, one would expect that as the nuclear incompressibility is the measure of the curvature of the nuclear equation of state, $K_\infty$ and $K_\text{sym}$ are bulk properties of the nuclear force and thus should be invariant to the choice of the finite nucleus one uses to constrain their values. Indeed, this is the case, provided that approximately 100\% of the energy-weighted sum rule (EWSR) is exhausted within the peak of the ISGMR response \cite{harakeh_book}.

\begin{figure}[t!]
  \includegraphics[width=\linewidth]{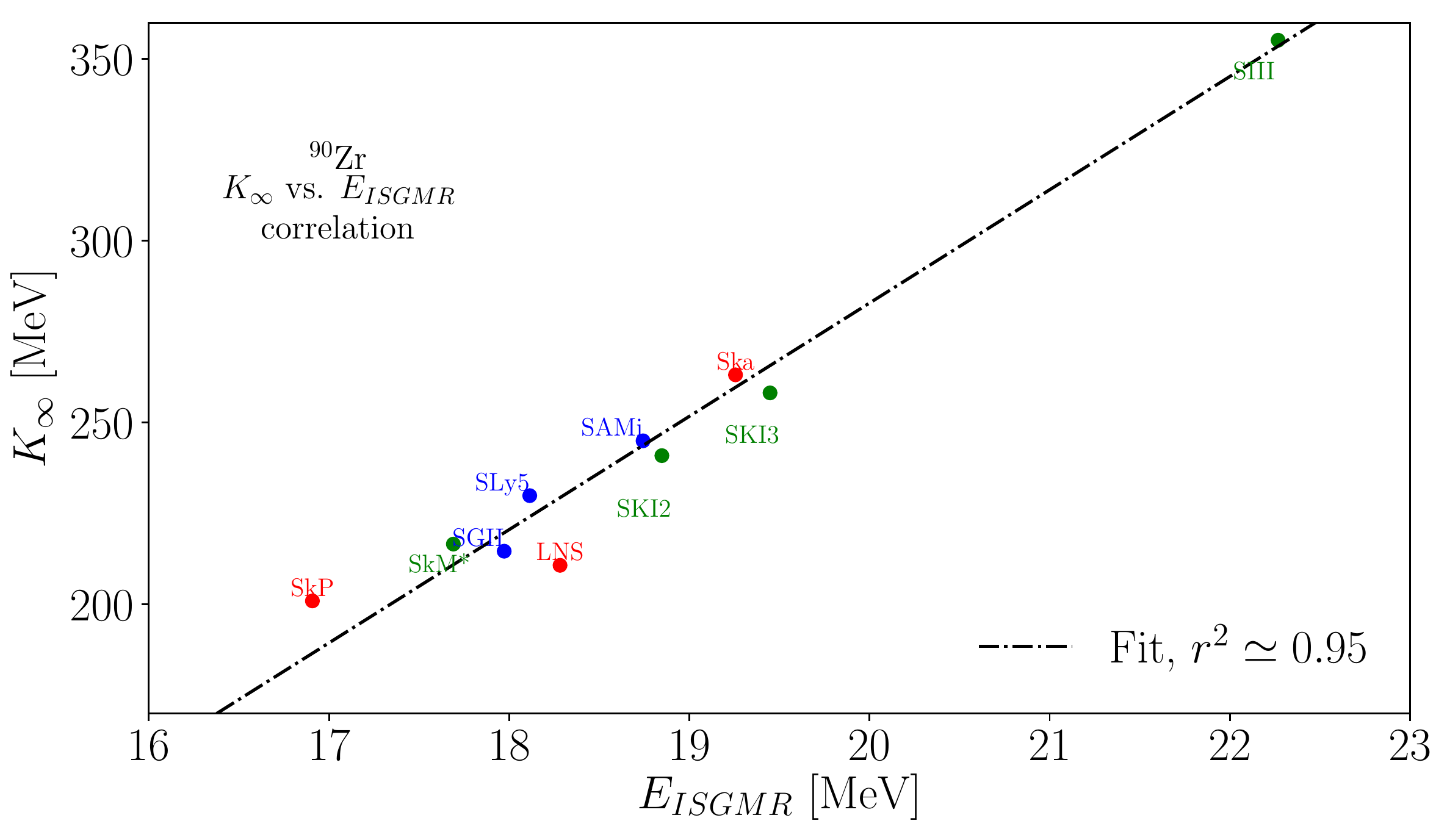}
  \caption{Typical correlation between the calculated $K_\infty$ values for nonrelativistic Skyrme interactions and the corresponding predicted $E_\text{ISGMR}$ in \nuc{90}{Zr} \cite{colo_simone_private}.}
  \label{kinf_correlation_plot}
\end{figure}

For a detailed discussion about how one obtains values of $K_\infty$ from finite nuclei, we refer the reader to Refs. \cite{blaizot,colo_2004a}; for further exposition on the ISGMR and for the models for extracting $K_A$ from experimental ISGMR strength distributions, Refs. \cite{stringari_sum_rules,stringari_lipparini_sum_rules,harakeh_book,bortignon_book,garg_colo_review} are most comprehensive. To summarize the procedure: one first takes any number of microscopic theories and interactions which are capable of describing, in a self-consistent way, the ISGMR responses of both finite nuclei and the bulk properties of nuclear matter (as an example, RPA calculations with a given effective interaction) \cite{Colo2014a,garg_colo_review}. Within each model framework, one then compares the calculated ISGMR response for a given nucleus, as well as its corresponding moments and moment ratios, to those which are experimentally available. The prescription is to then isolate the corresponding interactions which are capable of reproducing the experimental data and to use their corresponding values of $K_\infty$ as ``true'' values for the quantity. Extractions of this nature, as well as comparable analyses for other key saturation parameters in the EoS discussed in Section \ref{intro:bulk properties}, yield direct constraints on properties of the EoS and in turn, the interactions whence they arise \cite{Colo2014a}. Figure \ref{kinf_correlation_plot} shows some typical calculations for this procedure in the case of \nuc{90}{Zr} which were completed using a class of Skyrme interactions \cite{colo_simone_private}.

For slightly more than a decade, the accepted range of $K_\infty$ which has been extracted using this methodology has stood as $K_\infty = 240 \pm 20$ MeV \cite{shlomo_knm_240,garg_colo_review}; this value was obtained from analyses of compressional-mode resonance data on \nuc{90}{Zr} and \nuc{208}{Pb} from Refs. \cite{youngblood_90Zr_knm_240,youngblood_208Pb_knm_240} which included the effects of variations between relativistic and nonrelativistic interactions.

In any event, it is the general consensus of the field that microscopic calculations of $K_\infty$ are strongly correlated with the ISGMR response of finite nuclei \cite{garg_colo_review,blaizot,van_giai_correlations}. Under this assumption, \emph{any} local structure effects which are shown to influence the distribution of ISGMR strength --- and consequently, the corresponding value of $K_A$ --- within a finite system could in turn have implications on the extracted values for $K_\infty$ and by extension, the density dependence of the EoS.

\section{Open problems and the status of the field} \label{open problems}

Along these lines, to date, the only nuclear structure effect which has been adequately described and modeled within the existing collective-model framework is that of axial deformation on the giant monopole and quadrupole (and to a lesser extent, the dipole and octupole) resonances \cite{garg_sm_PRL,itoh_sm_PRC,gupta_24Mg_plb,gupta_24Mg_prc,peach_28Si_prc,kvasil_deformation_analysis,garg_colo_review}. A number of open problems have existed within the field which are, at present, unexplainable within existing theory. These problems are the general focus of this thesis work and are described in greater detail in the following subsections.

\subsection{Anomalous structure of the ISGMR in the $A=90$ region} \label{intro:A92_anomalous}

\begin{figure}[t!]
  \centering
\includegraphics[width=1\linewidth]{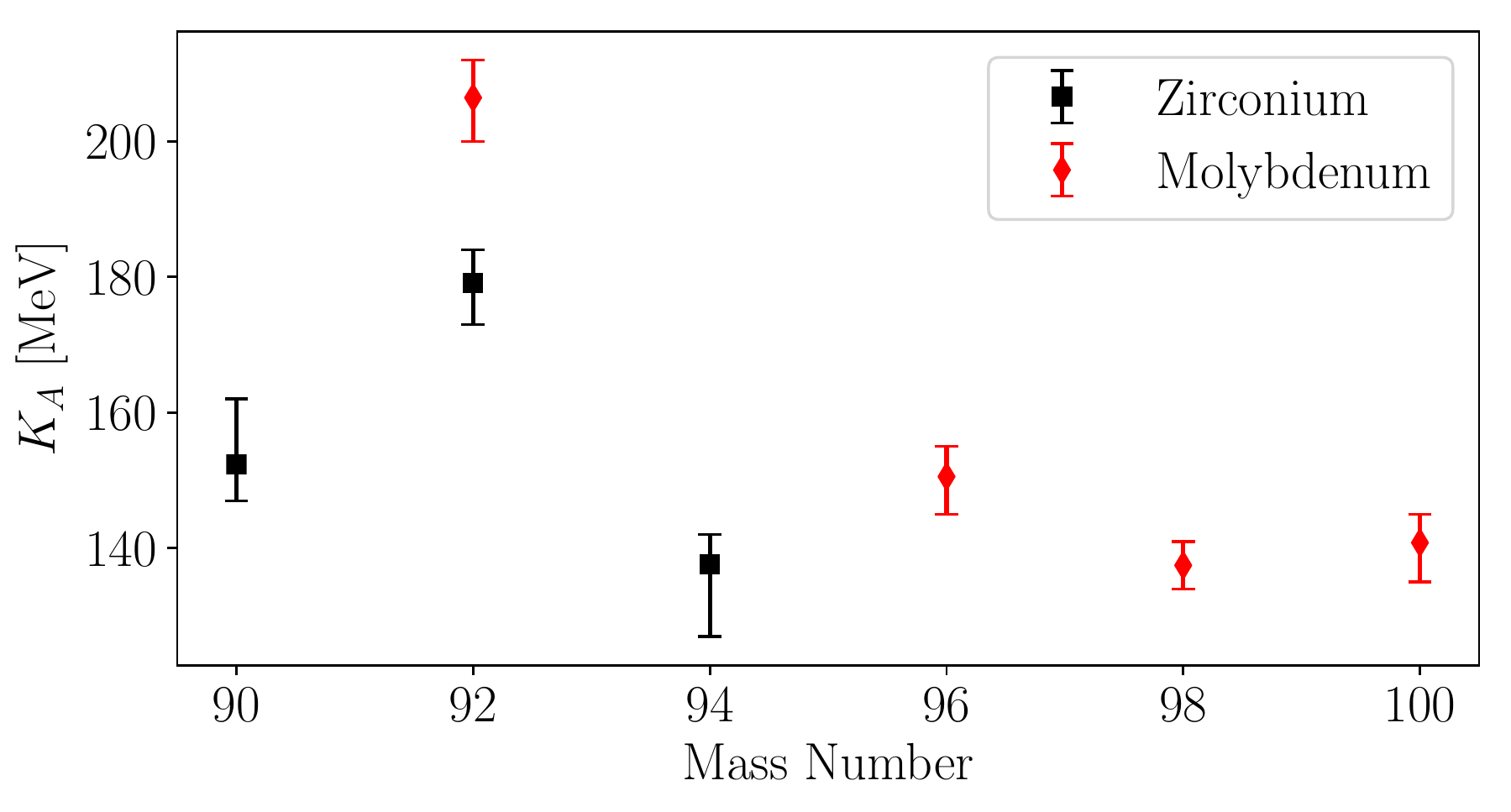}
\caption[Experimental $K_A$ values extracted within the scaling model using the methodology of Refs.  \cite{youngblood_A90_unexpected,krishichayan_Zr,youngblood_92_100Mo,button_94Mo} for \nuc{90,92,94}{Zr} and \nuc{92,96,98,100}{Mo}.]{Experimental $K_A$ values extracted within the scaling model using the methodology of Refs.  \cite{youngblood_A90_unexpected,krishichayan_Zr,youngblood_92_100Mo,button_94Mo} for \nuc{90,92,94}{Zr} and \nuc{92,96,98,100}{Mo}. Shown is the reportedly stark disparity between extracted values of $K_A$ for the $A=92$ isobars relative to the other nuclei in this mass region. Data adapted from Ref. \cite{youngblood_A90_unexpected}.}
\label{AM_KA}
\end{figure}

As argued by the Texas A\&M (TAMU) group in Refs. \cite{youngblood_A90_unexpected,krishichayan_Zr,youngblood_92_100Mo,button_94Mo}, the independence of the bulk nuclear properties to the choice of reference nucleus has been challenged on the basis of experimental observations of the ISGMR strength in even-even isotopes of zirconium and molybdenum, namely, \nuc{90-94}{Zr} and \nuc{92-100}{Mo}. Figure \ref{AM_KA} illustrates these results. In particular, the results indicated that for \nuc{92}{Zr} and \nuc{92}{Mo}, a large portion of the $E0$ strength lies above the main ISGMR peak, resulting in $K_A$ values which are commensurately large for $A=92$ isobars. While the structure of the ISGMR in these nuclei is indeed important to the understanding of collective excitations, it should be kept in mind that the association of $K_A$ with the ISGMR energies demands care, and can become untenable within the framework of Eq. \eqref{KA_to_ISGMR} for multiply-peaked distributions of ISGMR strength.

This question has been resolved in a previous experimental campaign \cite{gupta_A90_PLB,gupta_A90_PRC} into determining the nature of ISGMR strength for nuclei within this mass region seem. The results of Refs. \cite{gupta_A90_PLB,gupta_A90_PRC} conclusively disagree with the aforementioned conclusions posed by Texas A\&M. Nonetheless, we mention the motivation for the experimental efforts of Refs. \cite{gupta_A90_PLB,gupta_A90_PRC}, as the reported experimental data provides a foundation not only for answering the question as to the anomalous structure of the ISGMR in the $A=92$ isobars, but also for the question posed in the following subsection regarding the softness of open-shell nuclei.

\subsection{Softness of open-shell nuclei} \label{intro:softness_question}

The aforementioned accepted value of $K_\infty = 240 \pm 20$ MeV was produced using the methodology described by Blaizot \emph{et al.} \cite{blaizot}, wherein a self-consistent Random Phase Approximation (RPA) calculation is completed using an interaction with the goal of first modeling the response of the ISGMR in a given finite nucleus \cite{garg_colo_review,colo_2004a,cao_sagawa_colo}. With that same interaction, one then calculates the EoS for an infinite nuclear system using the same self-consistent framework and extracts the properties which are correlated with the finite nuclear response for comparison with the experimental data. Experiments on \nuc{208}{Pb} and \nuc{90}{Zr} \cite{youngblood_90Zr_knm_240,youngblood_208Pb_knm_240} are typically utilized as benchmark cases owing to their doubly-closed shell structure and commensurate computational ease; in both cases, the response of the ISGMR is well-developed and in the case of \nuc{90}{Zr}, contributions by the proton-neutron asymmetry to the ISGMR response are small in relation to the case of \nuc{208}{Pb} \cite{todd-rutel_piekarewicz_relativistic_interactions_and_neutron_stars}. From these procedures, $K_\infty = 240 \pm 20$ MeV was obtained using myriad interactions which adequately reproduced the position and structure of the ISGMR strength of these nuclei \cite{shlomo_knm_240}.

\begin{figure}[t!]
  \centering
  \includegraphics[width=\linewidth]{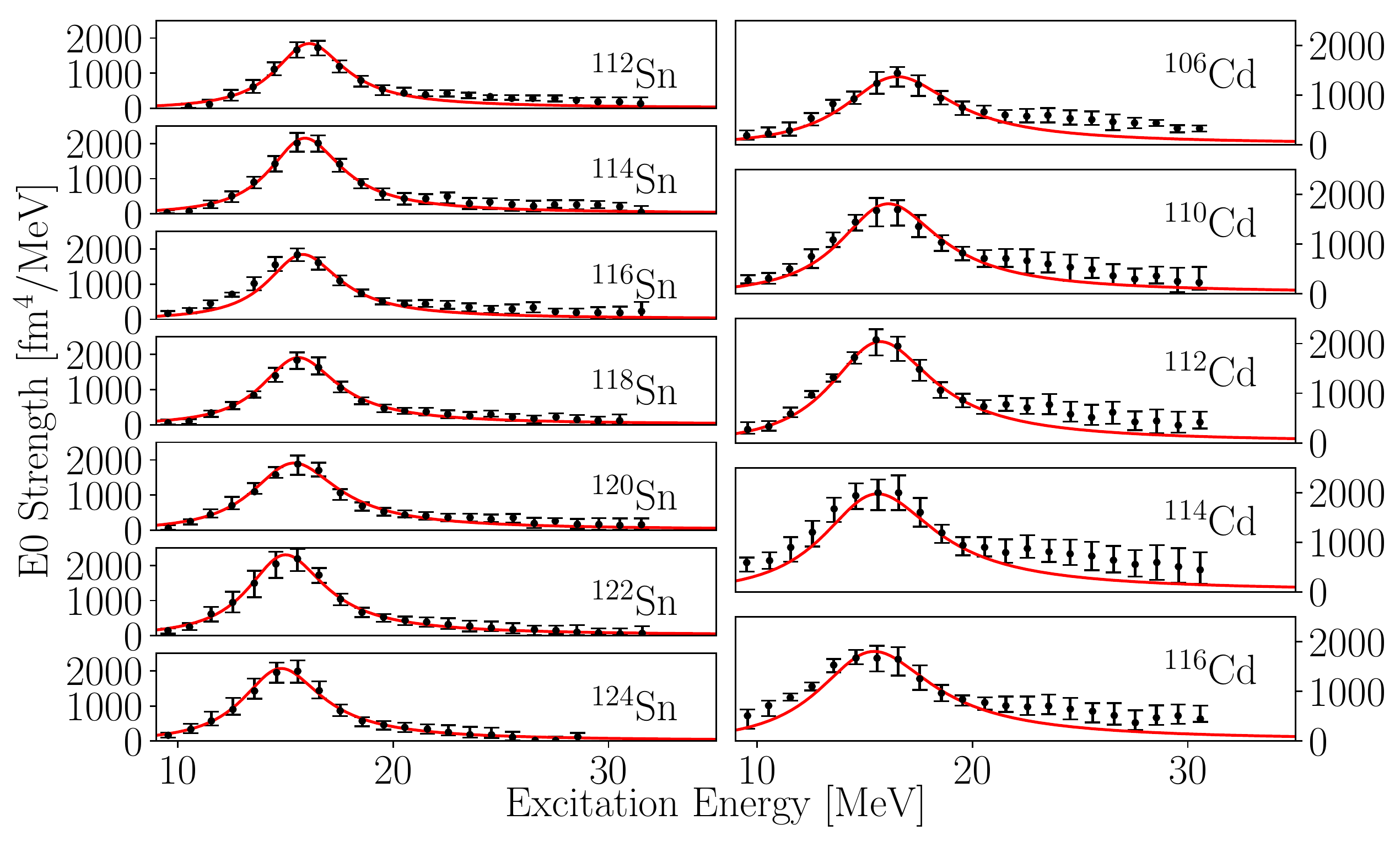}
  \caption[ISGMR strengths for \nuc{112,114,116,118,120,122,124}{Sn} \cite{Li_PRL,Li_PRC} and \nuc{106,110,112,114,116}{Cd} \cite{patel_cd}.]{Adaptation of reported ISGMR strengths for \nuc{112,114,116,118,120,122,124}{Sn} \cite{Li_PRL,Li_PRC} and \nuc{106,110,112,114,116}{Cd} \cite{patel_cd}, along with Breit-Wigner distributions fit to the experimental data points.}
  \label{intro:tin_cad_ISGMR}
\end{figure}

\begin{figure}[t!]
  \centering
  \includegraphics[width=\linewidth]{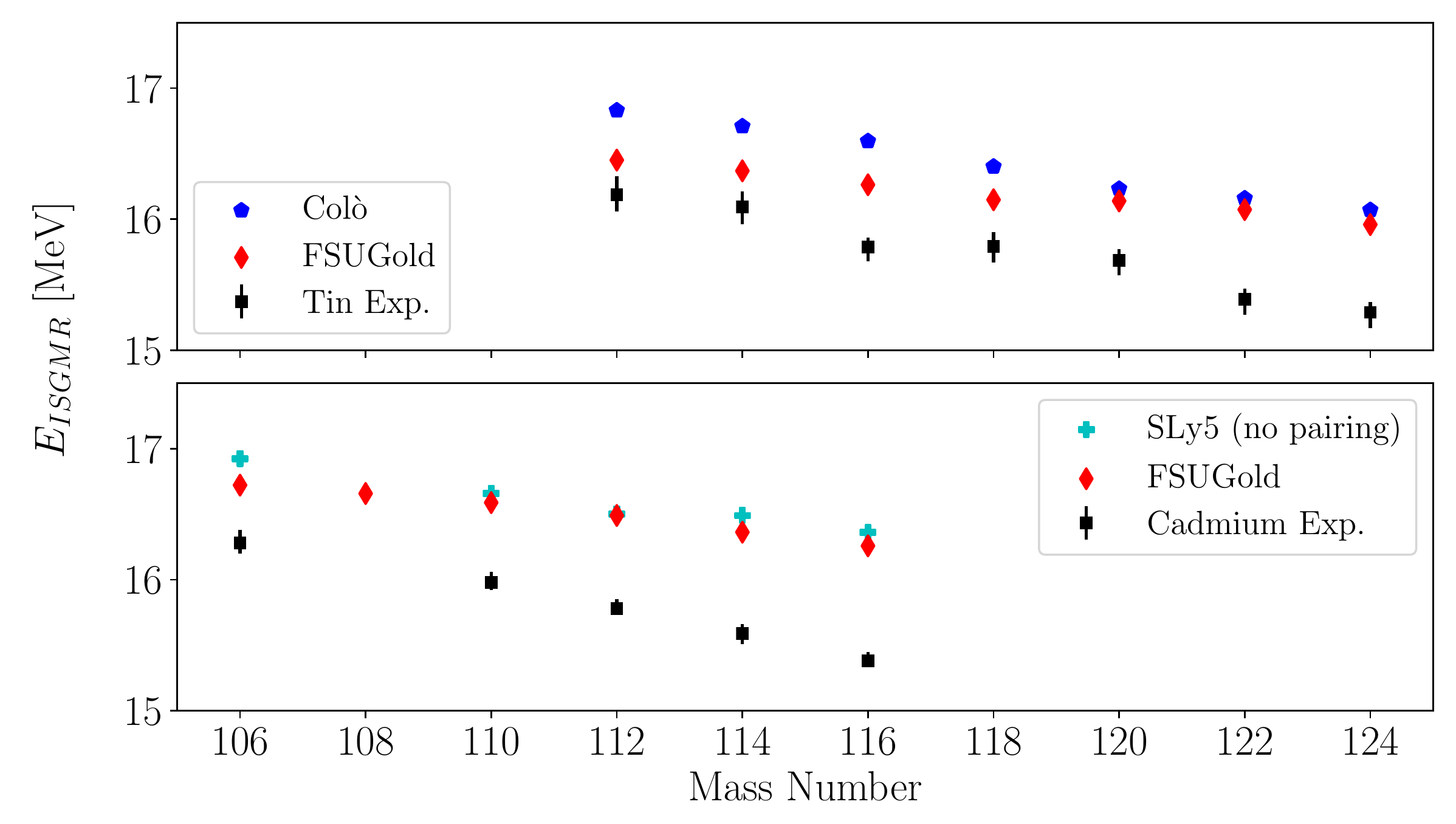}
  \caption{Comparison of the excitation energy of the ISGMR extracted from Refs. \cite{Li_PRL,Li_PRC,patel_cd} with those resulting from nonrelativistic (Col\'o Calc. and SLy5) and relativistic (FSUGold) RPA calculations.}
  \label{intro:tin_cad_interactions_softness}
\end{figure}

Figure \ref{intro:tin_cad_ISGMR} shows the ISGMR strength distributions which were extracted for the tin isotopes and cadmium isotopes. Inspection of the extracted ISGMR strengths for each isotopic chain absent a comparison with theoretical results or the results for other nuclei would fail to indicate any disagreement with the then-current understanding and descriptions of the giant resonances. However, examination of the ISGMR energies presented in Fig. \ref{intro:tin_cad_interactions_softness} paints a different picture: it is evident therein that both nonrelativistic and relativistic models which are benchmarked in the aforementioned manner against both ground- and excited-state observables, including the ISGMR of \nuc{90}{Zr} and \nuc{208}{Pb}, \emph{overestimate} the ISGMR energies of \nuc{108-116}{Cd} and \nuc{112-124}{Sn}. The effect is on the order of $\sim 500$ keV, and is clearly present for all nuclei in each of the isotopic chains.

As a result, in stark contrast to the previously-mentioned assertion that determinations of $K_\infty$ ought to be independent of the choice of nucleus, it was clearly observed that the extracted $K_\infty$ would be substantially lower than the presently accepted value of $240 \pm 20$ MeV \cite{piek_fluffy_physRev}. Thus, the tin and cadmium isotopes, were deemed to be ``soft'' in comparison to \nuc{90}{Zr} and \nuc{208}{Pb} \cite{garg_fluffy_nucA,piek_fluffy_physRev}. To these ends, a number of solutions were posed to explain this observation, such as the notion of mutually-enhanced magicity (MEM) in doubly-closed shell nuclei \cite{khan_MEM}, as well as contributions due to superfluid pairing interactions \cite{junli_pairing,khan_pairing,tselyaev_Sn_quasiparticle}. The MEM effect was refuted by experimental observations by Patel \emph{et al.} \cite{patel_MEM}, and the exact effects of pairing on the ISGMR are still somewhat uncertain \cite{khan_pairing} but nonetheless have been determined to be insufficient for accounting for the softness of open-shell nuclei. This open question, aptly posed as: ``why are the tin [and cadmium] isotopes so fluffy?'' \cite{piek_fluffy_physRev,garg_fluffy_nucA} has been deemed a fundamental open problem in nuclear structure physics and to this day, remains unanswered \cite{garg_fluffy_nucA,Li_PRL,Li_PRC,patel_cd,piek_fluffy_physRev,tselyaev_Sn_quasiparticle,piek_physG,cao_sagawa_colo,vesely_phys_rev_c}.

This question can naturally be extended to the molybdenum isotopic chain as well: said simply, if the tin and cadmium isotopes (respectively $Z=50$ and $Z=48$) are soft in relation to \nuc{90}{Zr} ($Z=40$) as measured by their ISGMR responses, and the latter are consistently reproduced by interactions with the same bulk properties and nuclear incompressibilities as those which well-model the ISGMR response of \nuc{208}{Pb}, then what changes in between zirconium, cadmium, and tin in the nuclear chart, and where does that change manifest?

To investigate this question, this thesis reports on an experiment on \nuc{94,96,97,98,100}{Mo}; the goal of this endeavor is to determine when, and how, this softness might appear as one moves away from \nuc{90}{Zr}. In combination with previous experimental data on \nuc{90,92}{Zr} and \nuc{92}{Mo}, the extraction of these ISGMR responses in these nuclei have been postulated to have the potential to provide substantial insight as to the origin of the softness open shell nuclei, and are therefore critical for accurately describing features of collective motion. Even further, these measurements and the resultant theoretical efforts are critical for maintaining well-founded extrapolations from finite nuclei to bulk nuclear systems, as the presently-available frameworks are predicated on the insensitivity of the resulting bulk properties to the choice of benchmark nucleus. 

\subsection{Increasing ISGMR energies within the calcium isotopes}

We will now turn our attention to the macroscopic leptodermous expansion of $K_A$ in terms of properties of infinite nuclear matter:

\begin{align}
  K_A \approx K_\infty + K_\text{surf} A^{-1/3} + K_\tau \eta^2 + K_\text{Coul} \frac{Z^2}{A^{4/3}}. \label{lepto}
\end{align}

\begin{figure}[t!]
  \centering
  \includegraphics[width=\linewidth]{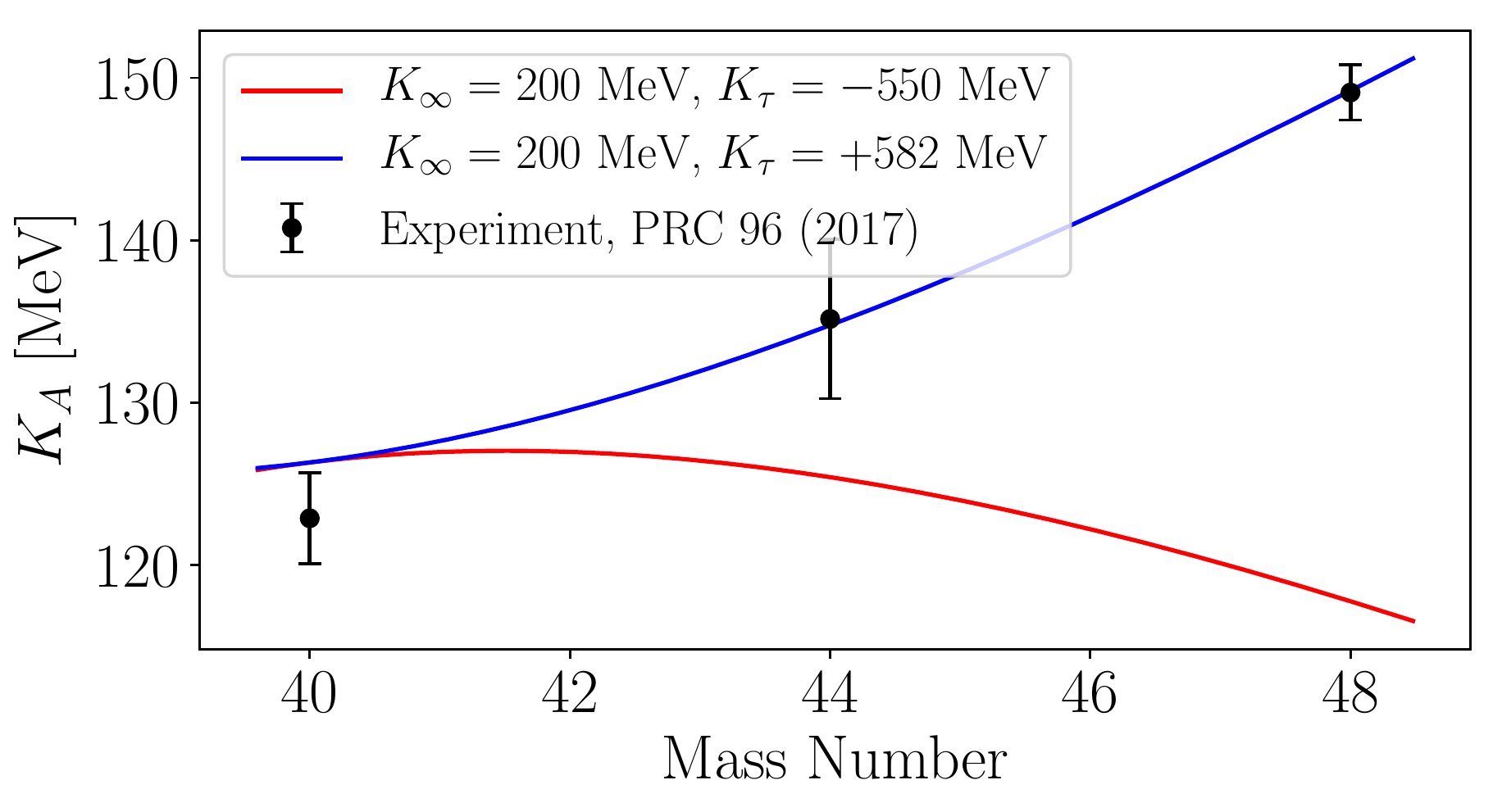}
  \caption[Extractions of $K_A$ from the calcium isotopes adapted from Ref \cite{TAMU_44Ca}, with data originating from Refs. \cite{TAMU_40Ca, TAMU_44Ca, TAMU_48Ca}.]{Extractions of $K_A$ from the calcium isotopes adapted from Ref \cite{TAMU_44Ca}, with data originating from Refs. \cite{TAMU_40Ca, TAMU_44Ca, TAMU_48Ca}. Shown are two macroscopic models in the style of Eq. \eqref{lepto}: for the same value of the nuclear incompressibility $K_\infty$, the previously accepted value of $K_\tau = -550 $ MeV wildly diverges from the reported behavior of $K_A$ for \nuc{44,48}{Ca}, while $K_\tau \gg 0 $ seems to well-model the reported results.}
  \label{AM_KA_calcium}
\end{figure}

Equation \eqref{lepto} can be useful in determining the value of the asymmetry term, $K_\tau$, for finite nuclei, owing in part to the isolated dependence on $\eta$ within the expression as well as the fairly minimal changes in the surface term, $K_\text{surf}$, within an isotopic chain. The general prescription for doing so is detailed in Refs. \cite{Li_PRL,Li_PRC}, and involves quadratically fitting the dependence of $K_A - K_\text{Coul} Z^2/A^{4/3}$ on $\eta$ with a model function of the form $K_\tau \eta^2 + c$, with $c$ being a constant. The values of $K_\tau$ which have been extracted utilizing this method are consistent with one another and have been found to be, for the even-$A$ \nuc{112-124}{Sn} and \nuc{106-116}{Cd} isotopes respectively, $K_\tau = - 550 \pm 100$ MeV and $K_\tau = - 555 \pm 75$ MeV \cite{Li_PRL,Li_PRC,patel_cd}. Even further, an independent reanalysis of the combined tin and cadmium ISGMR data was completed by Stone \emph{et al.}, which eventually came to the conclusion that $K_\tau = -595 \pm 154$ MeV \cite{dutra_stone_skyrme}.

The corresponding definition of $K_\tau^\infty$ in terms of properties of the EoS for infinite nuclear matter is \cite{Piekarewicz_centelles}:

\begin{align}
  K_\tau^\infty = K_\text{sym} - 6 L - \frac{Q_\infty}{K_\infty}L
  \label{KT_defined}
\end{align}
within which $Q_\infty/K_\infty$ is the skewness parameter for the EoS of symmetric nuclear matter.\footnote{One should take care to note that $K_\tau^\infty$ is \emph{not} equal to the value of $K_\tau$ extracted from finite nuclei utilizing the methodology of Eq. \eqref{lepto}, just as $K_\infty \neq K_A$. However, through the same self-consistent mechanisms by which measurements of $K_A$ serve to constrain $K_\infty$ as described by Blaizot \cite{blaizot}, determining values of $K_\tau$ from finite nuclei is critical for constraining the EoS for asymmetric infinite nuclear matter \cite{patel_cd}.} The implications of this are that experimental constraints on $K_\tau$ arising from measurements of $K_A$ on finite nuclei are critical on determining the density dependence of the symmetry energy; this argument is predicated on the smoothness with which the values of $K_A$ vary across the nuclear chart.  Indeed, as has been argued in Ref. \cite{KBH_EPJA}, \emph{any} structure effects which arise within a locus of the chart of nuclides would materially alter our understanding of the collective model upon which decades of understanding of these resonances is built.

In light of all this, recently-reported results for \nuc{40,44,48}{Ca} \cite{TAMU_40Ca,TAMU_44Ca,TAMU_48Ca} were very surprising: the moment ratios  for the ISGMR and, therefore, the $K_A$ values for \nuc{40,44,48}{Ca} \emph{increased} with increasing mass number. The most immediate consequence of this, considering Eq. \eqref{lepto}, is that $K_\tau$ is a \emph{positive} quantity, and it was shown in Ref. \cite{TAMU_44Ca} that a large positive value of $K_\tau$ models the data well. In a test of hundreds of energy-density functionals currently in use in the literature, the values of $K_\tau$ extracted were consistently between $-800$ MeV $ \leq K_\tau \leq -100$ MeV \cite{sagawa_private}. Table \ref{interactions_table}, as well as the more comprehensive presentations within Ref. \cite{dutra_stone_skyrme}, each show clearly consistently negative values for the asymmetry term. Examination of Eq. \eqref{KT_defined} also directly suggests that the symmetry energy would nonetheless need to be extremely soft in order to accommodate $K_\tau > 0$ \cite{piekarewicz_private}. Finally, the hydrodynamical model predicts $E_\text{ISGMR} \sim A^{-1/3}$, while the results of Refs. \cite{TAMU_40Ca,TAMU_44Ca,TAMU_48Ca} indicated exactly the opposite: the ISGMR energies increasing with mass number over the isotopic chain.

%

In light of such concerns, these results clearly demanded an independent verification before significant theoretical efforts were expended in understanding, and explaining, this unusual and unexplained phenomenon. For example, macroscopic models which have attempted to find consistency with these results have met with little success in reproducing the other saturation properties of nuclear matter \cite{jun_su_ISGMR_bohr_mottelson_model}. Even more gravely, inspection of Figs. \ref{intro:eos_symmetric_figure} and \ref{mass_radius_plot} shows how even slight deviations in the density dependence of the EoS and symmetry energy can result in significant deviations in predicted astrophysical properties, as illustrated by the given example of using the EoS to decouple Eqs. \eqref{TOV_equations} to extract the mass profile of a neutron star. The second part of this dissertation deals with experimentally extracting $K_\tau$ from the ISGMR responses within the calcium isotopic chain \nuc{40,42,44,48}{Ca} as a means of independently verifying an otherwise highly-surprising result.

%
%
%
%
%
%
%
%
%
%

%
%
\chapter{Theoretical tools for the study of giant resonances}\label{theory}

\section{Giant resonances as responses to external fields}
As alluded to in Fig. \ref{multipole figure} and its surrounding discussion, the giant resonances can be regarded macroscopically as nuclear vibrations of varying multipolarity. The microscopic basis for this description is rooted in the notion that the external field, $\mathcal{O}$, which induces the transitions, can be likewise expanded in terms of its multipole moments. The multipole moments $\mathcal{M}(E\lambda;\mu)$ of isoscalar and electric nuclear transitions of multipolarity $\lambda$ and projection $\mu$ were given by Bohr and Mottelson \cite{bohr_mottelson_vI}:
\begin{align}
  \mathcal{M}(E\lambda;\mu) &= \frac{(2\lambda+1)!!} {q^\lambda (\lambda+1)} \int \diff^3 \vec{r} \, \rho(\vec{r}) \frac{\partial}{\partial r}  \left[r j_\lambda(q r) \right] Y_\lambda^\mu(\Omega) \notag \\ &+ i \frac{(2\lambda+1)!!}{c q^\lambda (\lambda+1)} \int \diff^3 \vec{r} \, \left(q \vec{r}\right) \cdot \vec{J}(\vec{r}) j_\lambda(q r) Y_\lambda^\mu (\Omega).
\end{align}
Here, $q$ is the momentum transfer, whereas $j_\lambda$ is the regular spherical Bessel function and $\rho$ and $\vec{J}$ are, respectively, the mass and current density. The convention is to employ the long-wavelength approximation, which is to say that $qr \ll 1$ (rendering the first term in the above equation dominant). With this, the expansion of $j_\lambda$ is
\begin{align}
  j_\lambda(qr) & = \frac{ (qr)^\lambda}{(2\lambda + 1)!!} \left[ 1 - \frac{1}{2} \frac{(qr)^2}{2\lambda+3} +\ldots \right]. \label{bessel_expansion}
\end{align}
In the long-wavelength approximation, the leading-order term in this expansion is dominant and therefore the corresponding approximation for the multipole moment takes the form, for $\lambda \geq 2$:
\begin{align}
  \mathcal{M}(E\lambda;\mu) &= \int \diff^3  \vec{r}\, \rho(\vec{r}) r^\lambda Y_\lambda^\mu(\Omega). \label{general_multipole}
\end{align}
These expressions become trivial for $\lambda = 0$ and $\lambda = 1$; the monopole case results in a constant monopole moment (the static nuclear mass) and cannot induce any transitions, whereas the dipole case corresponds to a center-of-mass translation (\emph{e.g.} $r Y_1^0 \sim z$). The next-to-leading-order contribution from Eq. \eqref{bessel_expansion} is required in deriving the expressions for the electric multipole moments for monopole and dipole transitions \cite{sitenko_book}:
\begin{align}
  \mathcal{M}(E0;0) & \approx \int \diff^3  \vec{r}\, \rho(\vec{r}) \left[ 1 - \frac{q^2}{2} r^2 \right] \notag \\
  & = A - \frac{q^2}{2} \int \diff^3  \vec{r}\, \rho(\vec{r}) r^2 \notag \\
  \mathcal{M}(E1;\mu) & \approx  \int \diff^3  \vec{r}\, \rho(\vec{r}) \left[ r - \frac{q^2}{5} r^3 \right]Y_{1}^\mu (\Omega) \notag \\
  & = \int \diff^3  \vec{r}\, \rho(\vec{r}) r Y_{1}^\mu (\Omega) - \frac{q^2}{5} \int \diff^3  \vec{r}\, \rho(\vec{r}) r^3 Y_{1}^\mu (\Omega).\label{monopole and dipole}
\end{align}
In these expressions, the second terms are those which are responsible for inducing the isoscalar giant monopole and dipole resonances.

If the nucleons are considered to be pointlike, the corresponding nucleon density distribution for the $A$-nucleon system is of the form:
\begin{align}
  \rho(\vec{r}) & = \sum_{k=1}^A \delta^3(\vec{r}-\vec{r}_k),
\end{align}
and the Eqs. \eqref{general_multipole} and transition terms of \eqref{monopole and dipole} are (in the latter cases, up to a momentum-transfer dependent prefactor):

\begin{align}
  \mathcal{M}(E\lambda;\mu) &= \sum_{k=1}^A r_k^\lambda Y_\lambda^\mu(\Omega_k), & (\lambda \geq 2) \notag \\
  \mathcal{M}(E1;\mu) &= \sum_{k=1}^A r_k^{3} Y_1^\mu(\Omega_k), & (\lambda = 1) \notag \\
  \mathcal{M}(E0;0) &= \sum_{k=1}^A r_k^{2}. & (\lambda = 0) \label{multipole_moments_general}
\end{align}

\begin{table}[t!]
  \centering
  \caption{Excitation energies for giant resonances \cite{harakeh_book}}
  \label{selection_rules}
  \begin{tabular}{l c c c c c c }
    \hline
    monopole & $\lambda = 0$ & &  &$ 2 \hbar \omega$  &  & \\
    dipole & $\lambda = 1$ & & $1 \hbar \omega$  &  &$ 3 \hbar \omega$  & 		\\
    quadrupole & $\lambda = 2$ &$ (0 \hbar \omega)$ &  &$ 2 \hbar \omega$  &  & \\
    octupole & $\lambda = 3$ & & $1 \hbar \omega$  &  &$ 3 \hbar \omega$  & 	\\
    hexadecapole & $\lambda = 4$ &$ (0 \hbar \omega)$ &  &$ 2 \hbar \omega$  &  & $4 \hbar \omega$ \\						 \hline
  \end{tabular}
\end{table}

\begin{figure}[t!]
  \centering
  \includegraphics[width=0.85\linewidth]{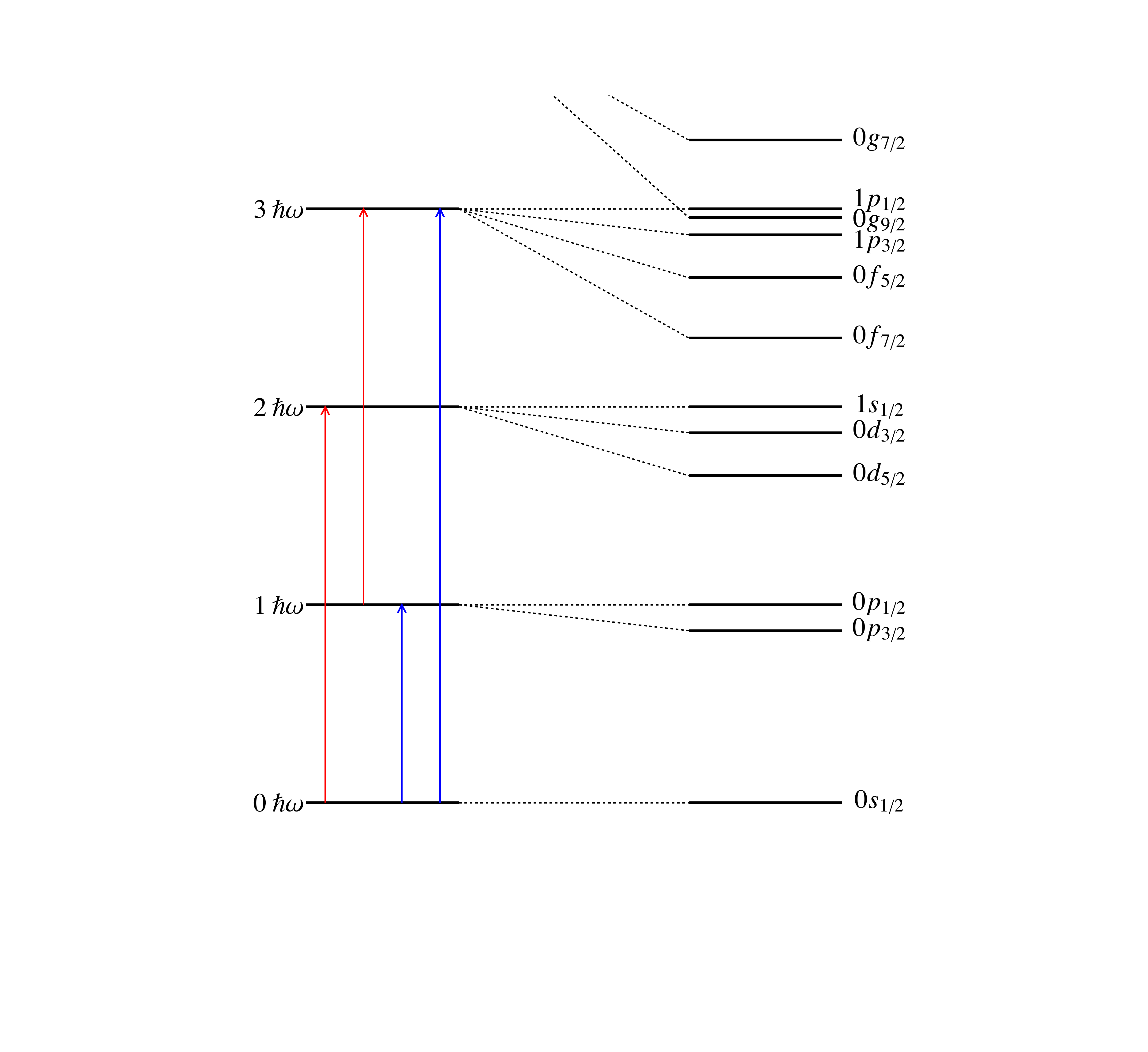}
  \caption[Shell-model schematic of possible giant resonance excitations.]{Shell-model schematic of possible giant resonance excitations. Shown in red are possible ISGMR excitations corresponding to $2\hbar \omega$, whereas blue correspond to possible ISGDR excitations at $1\hbar \omega$ and $3\hbar \omega$.}
  \label{theory:level_diagram}
\end{figure}

In these cases, one can write the total multipole moment as the sum of the individual responses of the constituent nucleons to an external one-body operator\footnote{We will make this clear shortly, but the case of the electric $\lambda = 1$ response demands special care due to spurious center-of-mass corrections. This has been done by Harakeh \emph{et al.} \cite{harakeh_ISGDR}, and subsequently in this chapter we will impose the corrections prescribed in the aforementioned reference.}, $\mathcal{O}^{\lambda,\mu}_{k}$:
\begin{align}
  \mathcal{M}(E\lambda;\mu) & =  \sum_{k=1}^A \mathcal{O}^{\lambda,\mu}_k(\vec{r}_k) \notag \\
  &=\sum_{k=1}^A f(\vec{r}_k) Y_\lambda^\mu,
\end{align}
where the external field $\hat{F}$ is defined as
\begin{align}
  \hat{F}_{\lambda,\mu} (\vec{r}) &= f(r) Y_{\lambda}^\mu (\Omega). \label{external fields}
\end{align}

This formalism has the additional benefit that, within the harmonic oscillator description of the nucleus, the effect of the operators in Eq. \eqref{multipole_moments_general} on the nuclear ground-state is to coherently excite particles (and holes) across the major oscillator shells. In such a case, the possible excitation energies of each multipole operator can be interpreted within the context of creation and annihilation operators generating energy quanta in multiples of $\hbar \omega$, thereby yielding the excitation energies shown in Table \ref{selection_rules}; typically, $\hbar \omega \approx 40 A^{-1/3}$ MeV is used as a coarse estimate for the giant resonance excitation energies. Furthermore, some possible transitions of nucleons within this picture are depicted in Fig.  \ref{theory:level_diagram}.

\subsection{Derivations of the energy-weighted sum rules}\label{theory:ewsr_derived}

The impact of writing the multipole moments in this way is somewhat nuanced, but in no way insignificant. The external field which induces these transitions can be used to great effect in deriving the \emph{energy-weighted sum rules} (EWSR) and associated observables for each of the above-mentioned multipolarities. Owing to a result that is known as Thouless' Theorem \cite{thouless_theorem,bohigas_sum_rules}, the total linear energy-weighted strength (with the strength being a measure of the reduced transition probability) of an operator acting on the ground\footnote{This theorem is actually general insofar that the ground state can directly be replaced with any excited state upon which one wishes to study a resonance structure.} state $\ket{0}$ is related to the nested commutator of the external field, applied to each constituent nucleon, with the ground-state nuclear Hamiltonian:
\begin{align}
  \sum_{n} \left| \bra{n} \hat{F}_{\lambda,\mu}  \ket{0} \right|^2 (E_n-E_0)= \frac{1}{2} \bigg < 0 \bigg| \left [ \sum_{k=1}^A \hat{F}_{\lambda,\mu}(\vec{r}_k) , \left[\hat{H}, \hat{F}_{\lambda,\mu}  \right]  \right] \bigg| 0 \bigg>. \label{EWSR_thouless}
\end{align}

This result is significant. The left-hand side of Eq. \eqref{EWSR_thouless} is the transition strength $|\bra{n} \hat{F}_{\lambda,\mu} \ket{0}|^2$ (later in the text, this will frequently be referred to as $S_\lambda(E_x)$ between the ground and $n^\text{th}$ excited state (beyond the particle threshold, this summation passes to an integral) weighted according to the energy of the transition itself.  This energy-weighted strength is the first energy-weighted moment of the distribution, denoted $m_1$. The right-hand side is a nested double-commutator in which the external field $\hat{F}$ and the nuclear Hamiltonian act \emph{only} on the ground state of the system. For a large class of nuclear potentials, the one-body external field commutes with the interaction potential energy components of the nuclear Hamiltonian (providing that the nuclear potential energy is translationally-invariant and does not contain velocity-dependent forces\footnote{While the addition of velocity-dependent forces technically breaks translational invariance, this can be corrected by imposing an \emph{effective mass}, as is commonly done in Skyrme models \cite{dutra_stone_skyrme,harakeh_book,bortignon_book}.}), as well as each of the other one-body kinetic energies \cite{harakeh_book,bortignon_book}:

\begin{align}
  \sum_{n} \left| \bra{n} \hat{F}_{\lambda,\mu}  \ket{0}  \right|^2 (E_n-E_0) &= \frac{1}{2} \bigg<0\bigg| \left [ \sum_{k=1}^A \hat{F}_{\lambda,\mu}(\vec{r}_k) , \left[\hat{T}_k, \hat{F}_{\lambda,\mu}(\vec{r}_k)  \right]  \right]  \bigg| 0 \bigg> \notag \\
  &= \frac{1}{2}  \sum_{k=1}^A  \expect { \left [ \hat{F}_{\lambda,\mu}(\vec{r}_k) , \left[ \frac{\hat{p}_k^2}{2m}, \hat{F}_{\lambda,\mu}(\vec{r}_k)  \right]  \right] }  \notag \\
  &= -\frac{1}{2} \frac{\hbar^2}{2m} \sum_{k=1}^A   \expect{\left [ \hat{F}_{\lambda,\mu} , \left[ \nabla_k^2, \hat{F}_{\lambda,\mu}  \right]  \right] }\notag \\
   &= -\frac{1}{2} \frac{\hbar^2}{2m} \sum_{k=1}^A   \expect{ \left [ \hat{F}_{\lambda,\mu} , \left[ \nabla_k \cdot [\nabla_k, \hat{F}_{\lambda,\mu}] + [\nabla_k,\hat{F}_{\lambda,\mu}] \cdot \nabla_k  \right]  \right]  }\notag \\
   &= -\frac{1}{2} \frac{\hbar^2}{2m} \sum_{k=1}^A   \expect{ \left [ \hat{F}_{\lambda,\mu} , \left[ \nabla_k \cdot \left( \nabla_k \hat{F}_{\lambda,\mu}  \right) + \left( \nabla_k \hat{F}_{\lambda,\mu}  \right) \cdot \nabla_k  \right]  \right]  }. \label{EWSR_pre_almost}
 \end{align}

Since the gradients of the external fields $\left( \nabla_k \hat{F}_{\lambda,\mu}  \right)$ are themselves only functions of position coordinates, they commute with the external field itself. Equation \eqref{EWSR_pre_almost} then becomes:
\begin{align}
  \sum_{n} \left| \bra{n} \hat{F}_{\lambda,\mu}  \ket{0}  \right|^2 (E_n-E_0) &=   -\frac{1}{2} \frac{\hbar^2}{2m} \sum_{k=1}^A   \bigg< \bigg[ [\hat{F}_{\lambda,\mu},\nabla_k] \cdot \left( \nabla_k \hat{F}_{\lambda,\mu}  \right) + \nabla_k \cdot \left[\hat{F}_{\lambda,\mu},\nabla_k \hat{F}_{\lambda,\mu}\right] \notag \\ &+ \left[\hat{F}_{\lambda,\mu},\nabla_k \hat{F}_{\lambda,\mu}\right] \cdot \nabla_k + \left(\nabla_k \hat{F}_{\lambda,\mu}\right) \cdot [\hat{F}_{\lambda,\mu},\nabla_k ]  \bigg] \bigg> \notag \intertext{as the middle two terms commute, the outer two terms $[\hat{F}_{\lambda,\mu},\nabla_k] = - [\nabla_k,\hat{F}_{\lambda,\mu}] = - \nabla_k \hat{F}_{\lambda,\mu}$. Thus, }
  & = \frac{\hbar^2}{2 m} \sum_{k=1}^A  \bigg<0\bigg|  \left| \nabla_k \hat{F}_{\lambda,\mu} \right|^2  \bigg| 0 \bigg>. \label{EWSR_almost}
\end{align}

 The EWSR is a measure of transition strength which depends essentially only upon properties of the ground state of the nucleus in question as well as those of the field which is inducing the transition; the interpretation of this is that the total energy-weighted strength of the transition in question is limited by the momentum transfer from the external field to the nucleus in its initial state. The gradient of the external field $\hat{F}$\footnote{We caution the reader: there is little consistency in the literature as to the exact definitions of the external field $\hat{F}$, and therefore there are myriad equivalent expressions for the EWSRs that simply have different prefactors. Which conventions are used are generally of little significance as these prefactors only influence the magnitudes of the strength distribution and  cancel out in subsequent derived expressions for the transition amplitudes \emph{etc.}, but it is nonetheless always prudent to pay close attention to this dynamic feature of the literature.} is calculable as
 \begin{align}
   \nabla \hat{F} & = \nabla \left ( f(r) Y_{\lambda,\mu} \right) \notag \\
   & = \frac{df}{dr} \vec{Y}_{\lambda,\mu} + \frac{f(r)}{r} \text{\boldmath$\Psi$}_{\lambda,\mu}.
 \end{align}
The $\vec{Y}_{\lambda,\mu}$ and $\text{\boldmath$\Psi$}_{\lambda,\mu}$ are the vector spherical harmonics, which are orthogonal and obey normalization conditions such that
\begin{align}
  \int d\Omega\, \vec{Y}_{\lambda,\mu} \cdot \vec{Y}_{\lambda^\prime,\mu^\prime} & = \delta_{\lambda,\lambda^\prime} \delta_{\mu,\mu^\prime} \intertext{and} \notag \\
  \int d\Omega\, \text{\boldmath$\Psi$}_{\lambda,\mu} \cdot \text{\boldmath$\Psi$}_{\lambda^\prime,\mu^\prime} & = \lambda (\lambda+1) \delta_{\lambda,\lambda^\prime} \delta_{\mu,\mu^\prime}.
\end{align}

Furthermore, each of the vector harmonics satisfy the addition theorem that:
\begin{align}
  \frac{2\lambda+1}{4\pi} & = \sum_{\mu=-\lambda}^\lambda {\vec{Y}_\lambda^\mu}^\dagger(\Omega)\cdot \vec{Y}_{\lambda}^\mu(\Omega).
\end{align}

With this, the summation over magnetic substates can be completed and the right hand side of Eq. \eqref{EWSR_almost} is calculable in generality:
\begin{align}
  \sum_{\mu=-\lambda}^\lambda \sum_{n} \left| \bra{n} \hat{F}_{\lambda,\mu}  \ket{0}   \right|^2 (E_n-E_0) &=  \sum_{\mu=-\lambda}^\lambda \frac{\hbar^2}{2m} \sum_{k=1}^A \bigg<0\bigg| \left(\frac{\diff f}{\diff r} \right)^2 (\vec{Y}_{\lambda}^\mu)^\dagger \cdot \vec{Y}_{\lambda}^\mu \notag \\  &+ \lambda (\lambda+1 ) \left( \frac{f(r)}{r}\right)^2  (\text{\boldmath$\Psi$}_{\lambda,\mu})^\dagger \cdot \text{\boldmath$\Psi$}_{\lambda,\mu}  \bigg|0 \bigg> \notag \\
  \sum_{n} \left| \bra{n} \hat{F}_{\lambda}  \ket{0}   \right|^2 (E_n-E_0)&= \frac{2\lambda+1}{4\pi} \frac{\hbar^2}{2m} \sum_{k=1}^A \bigg<0\bigg| \left(\frac{\diff f}{\diff r} \right)^2  + \lambda (\lambda+1 ) \left( \frac{f(r)}{r}\right)^2  \bigg|0 \bigg> \label{ewsr_almost_almost}
\end{align}

After summing over nucleons, one achieves the final EWSR which is proportional directly to a combination of expectation values of radial moments, calculated relative only to the ground state of the nucleus in question:
\begin{align}
    \sum_{n} \left| \bra{n} \hat{F}_{\lambda}  \ket{0}   \right|^2 (E_n-E_0)     &= \frac{2\lambda+1}{4\pi} \frac{\hbar^2 A}{2 m} \expect{ \left(\frac{\diff f}{\diff r} \right)^2 + \lambda (\lambda+1 ) \left( \frac{f(r)}{r}\right)^2}. \label{final_EWSR}
\end{align}

The conventions for $\hat{F}_{\lambda, \mu}$ which are used in this thesis work within its formalism are given below\footnote{N.B. Since the operator for the monopole is defined without the factor $Y_0^0$ in this work, the corresponding prefactor $(2\lambda+1) / 4\pi$, which manifests in the final line of Eq. \eqref{ewsr_almost_almost}, is absent from the corresponding EWSR for the monopole transition.}:
\begin{align}
  \hat{F}_{\lambda,\mu}(\vec{r}) &= f(r) Y_{\lambda}^\mu(\Omega) \notag \\
  &=  r^2,  & (\lambda = 0) \notag \\
  & = \frac{1}{2} r^3 Y_1^\mu(\Omega),  &(\lambda = 1) \notag \\
  & = r^\lambda Y_\lambda^\mu (\Omega),  &(\lambda \geq 2) \label{one_body_operators}
\end{align}
and correspondingly, the EWSRs from Eq. \eqref{final_EWSR} are:
\begin{align}
  \sum_{n} \left| \bra{n} \hat{F}_{\lambda}  \ket{0}   \right|^2 (E_n-E_0) &= m_1^\lambda \notag \\
  &= \frac{2 \hbar^2 \, A}{m} \expect{r^2}, & (\lambda = 0) \notag \\
  &= \frac{\hbar^2}{2m} \frac{3}{16\pi} A \left( 11 \expect{r^4} - \frac{25}{3} \expect{r^2}^2 - 10 \epsilon \expect{r^2} \right), & (\lambda = 1) \notag \\
  &=  \frac{\hbar^2}{8 \pi m} \lambda (2\lambda+1)^2 A \expect{r^{2\lambda - 2}}. & (\lambda \geq 2) \label{EWSR_very_final}
\end{align}
In the case of $\lambda = 1$, the center-of-mass contributions have been accounted for as described in Ref. \cite{harakeh_ISGDR} and \footnote{The shell-model description for $E_\text{ISGMR} = 80 A^{-1/3}$ MeV and $E_\text{ISGQR} = 65 A^{-1/3}$ are typically used in this expression.}
\begin{align}
\epsilon = (4/E_\text{ISGMR} + 5/E_\text{ISGQR}) \hbar^2/3mA.
\end{align}
For the case of the IVGDR, the EWSR is the well-known Thomas-Reiche-Kuhn (TRK) sum rule \cite{harakeh_book,satchler_isospin}:
\begin{align}
  m_1^\text{IVGDR} & = \frac{9}{4 \pi} \frac{\hbar^2}{2m} \frac{NZ}{A}e^2. \label{TRK_sum_rule}
\end{align}

\subsection{Transition densities}\label{theory:transition_densities}

A deformed nuclear surface can be parametrized by a multipole expansion, with a set of deformation parameters, which are the dynamical variables $\left\{\alpha_{\lambda,\mu}\right\}$ that describe the amplitudes of each multipolarity in the deformed system \cite{bohr_mottelson_vII,Satchler_direct_nuclear_reactions}. The expressions derived and provided in this section will be in terms of these $\alpha_{\lambda,\mu}$; the means by which one calculates their values for each given multipolarity will be presented in the subsequent section. Within such a description, the nuclear radius $R(\theta,\phi)$ deviates from a constant $R_0$ to
\begin{align}
  R(\theta,\phi) & = R_0 + \overbrace{R_0 \sum_{\lambda=0}^\infty \sum_{\mu = -\lambda}^\lambda \alpha_{\lambda,\mu} Y_{\lambda}^\mu(\theta,\phi)}^\text{$\delta R(\theta,\phi)$}.
\end{align}
Furthermore, the density distribution $\rho(\vec{r})$ likewise changes:
\begin{align}
  \rho(\vec{r}) &= \rho(r + \delta R(\theta,\phi)  ) \notag \\
  &\approx \rho(r) + \delta \rho(\delta R(\theta,\phi)).
\end{align}
The quantity $\delta \rho$ is the \emph{transition density}, and is necessary for the penultimate calculation of transition potentials and subsequently, for modeling angular distributions within the DWBA framework.

In the macroscopic scaling model for the ISGMR in a spherical nucleus, for example, the Tassie-model \cite{Tassie_transitions} transition density $\delta \rho$ can be calculated assuming a radially symmetric and uniform scaling of the nuclear surface by a vibrational amplitude $\beta_0$:
\begin{align}
  r^\prime &= r (1-\beta_0) \notag \\
  \rho (r^\prime) & = \mathcal{N} (\rho(r) + \delta \rho)
\end{align}
wherein $\mathcal{N}$ is a renormalization factor for the transition density. Expanding to first order:
\begin{align}
   \rho (r^\prime) \approx \mathcal{N} \rho(r) + \mathcal{N} \beta_0 r \frac{\diff \rho}{\diff r}.
\end{align}


The transition density $\delta \rho$ can be written in terms of the ground-state density and the renormalization factor:
\begin{align}
  \delta \rho = \left(\mathcal{N} - 1 \right) \rho(r) + \mathcal{N} \beta_0 r \frac{\diff \rho}{\diff r}. \label{monopole_density_unnormed}
\end{align}
As the integral over all space of $\rho(r)$ does not change --- the number of constituent nucleons is a constant --- the following condition on $\delta \rho$ should hold:
\begin{align}
  \int \diff^3 \vec{r} \, \delta \rho & = 0. \label{particle_conserve}
\end{align}
Equations \eqref{monopole_density_unnormed} and \eqref{particle_conserve} allow for the solution of the renormalization factor $\mathcal{N}$, and consequently the expression of the transition density in terms of the vibrational amplitude. Imposing the latter condition on particle conservation and integrating the rightmost term by parts:
\begin{align}
  0 & = \int \diff^3 \vec{r} \, \left[ \left(\mathcal{N} - 1 \right) \rho(r) + \mathcal{N} \beta_0 r \frac{\diff \rho}{\diff r}\right] \notag \\
  & = \left(\mathcal{N} - 1 \right) \int \diff \Omega \, \int_0^\infty \diff r \, r^2 \rho(r) + \mathcal{N} \beta_0 \int \diff \Omega \, \int_0^\infty \diff r \, r^2 \left(r \frac{\diff \rho}{\diff r} \right) \notag \\
  &= \left(\mathcal{N} - 1 \right) \int \diff \Omega \, \int_0^\infty \diff r \, r^2 \rho(r) + \mathcal{N} \beta_0 \int \diff \Omega \bigg [ \cancelto{0}{r^3\, \rho(r) \big |_{0}^{\infty}} - 3 \int_{0}^\infty \diff r \, r^2 \rho  \bigg].
\end{align}
Equating the remaining integrands yields
\begin{align}
\left(\mathcal{N} - 1 \right) \int \diff r \, r^2 \rho(r) & =  3 \mathcal{N} \beta_0 \int \diff r \, r^2 \rho(r) \notag \\
\mathcal{N} = \frac{1}{1+3\beta_0}.
\end{align}
Insertion of this into Eq. \eqref{monopole_density_unnormed} yields the desired transition density for the monopole transitions:
\begin{align}
\delta \rho_0 &= -\beta_0 \left ( \frac{3}{1+3\beta} + \frac{r}{1+3\beta_0}  \frac{\diff}{\diff r} \right) \rho(r) \notag \\
& \approx  - \beta_0 \left(3 + r \frac{\diff}{\diff r} \right)\rho(r) \notag \\
& = - \frac{\beta_0}{r^2} \frac{d}{dr} \left(r^3 \rho(r) \right) \label{monopole transition density}
\end{align}
wherein the last expression, the binomial expansion of the denominator was used in combination with the harmonic assumption that $\beta_0 \ll 1$ --- that is to say, terms of order $\mathcal{O}(\beta_0^2) \rightarrow 0$.

An analogous derivation for the Tassie-type transition density for the center-of-mass-corrected ISGDR was derived, partially by Ref. \cite{deal_ISGDR} and later, in its correct and final form, by Ref. \cite{harakeh_ISGDR}, with the result given below in terms of the Fermi half-mass radius $c$ and the deformation parameter\footnote{Henceforth, $\beta_\lambda = \expect{\sum_{\mu} \alpha_{\lambda,\mu}}.$} $\beta_1$:
\begin{align}
  \delta \rho_1 = - \frac{\beta_1}{ c } \left[3 r^2 \frac{\diff}{\diff r} + 10 r - \frac{5}{3} \expect{r^2} \frac{\diff}{\diff r} + \epsilon \left (r \frac{\diff^2}{\diff r^2} + 4 \frac{\diff}{\diff r} \right)\right] \rho(r). \label{ISGDR transition density}
\end{align}

These Tassie-type transition densities are most appropriate for compressional states which exhibit high degrees of collectivity as measured by the percentage of the EWSR exhausted by the transition \cite{harakeh_book,Satchler_direct_nuclear_reactions}, and are the standard transition densities in use for experimental studies of the ISGMR and ISGDR.

For higher-multipolarity isoscalar transitions which are shape vibrations rather than compressional-mode oscillations, the transition densities which are used most commonly in giant resonance studies are given by the form derived by Bohr and Mottelson for surface vibrations \cite{bohr_mottelson_vII,harakeh_book}:
\begin{align}
\delta \rho_\lambda = - \beta_\lambda c \frac{\diff \rho}{\diff r}. \label{transition_density_leq2}
\end{align}

Finally, for the IVGDR, the transition density is given by the Goldhaber-Teller model \cite{satchler_isospin,harakeh_book} as
\begin{align}
  \delta \rho_\text{IVGDR} & = -\beta_\text{IVGDR} \gamma \left(\frac{N-Z}{A} \right) \left[ \frac{\diff }{\diff r} + \frac{1}{3} c \frac{\diff^2 }{\diff r^2} \right] \rho(r). \label{IVGDR transition density}
\end{align}
This implementation of the Goldhaber-Teller model presumes the same shape between the proton and neutron densities, but allows for different radial extensions of the distributions. Here, $\gamma = 3 (c_n-c_p)/(c_n+c_p)/\eta$ is a measure of the difference in  ground-state proton and neutron radii within the isospin-asymmetric ($\eta = (N-Z)/A$) nucleus.

\subsection{Transition amplitudes and deformation parameters} \label{transition_amplitudes_section}

To briefly recapitulate the theory developed so far: as discussed in Section \ref{theory:ewsr_derived}, the EWSR provides a model-independent metric by which one can characterize the collectivity of a given excitation; by describing the strength of a particular multipole transition in terms of multiples of ``single-particle'' strength, for example, one can crudely characterize the number of nucleons which participate in that transition. In Section \ref{theory:ewsr_derived}, the EWSR was shown to put a direct constraint on the amount of strength, or reduced transition probability, that can be exhausted over a set of transitions. What will be developed in the following section is a description of how one calculates the nuclear physics observables which arise when a given fraction of the EWSR is exhausted within a collective excitation.

A generalization of the EWSR developed in Section \ref{theory:ewsr_derived} exists as a constraint on the magnitude of the transition density itself \cite{fallieros_amplitude_proceedings,Satchler_direct_nuclear_reactions,suzuki_amplitude_derivation,blaizot_nuclear_compressibilities}. In examining the development of, for example, the macroscopic transition density of Eq. \eqref{monopole transition density}, one should take note that there is an unspoken-for transition amplitude, $\beta_0$, which in the case of the ISGMR can be macroscopically understood to be the percentage fluctuation in the nuclear radius. As we will see in this subsection, the value of $\beta_0$ is itself limited by the EWSR.

A reference value for $\beta_0$ can be derived under the presumption that the transition in question exhausts the full EWSR; one can then determine the amount of the reference value of $\beta_0$ that is realized in an experimentally-observed transition, and in so doing, determine the fraction of the EWSR that is exhausted in that transition. The derivation of the $\beta_0$ which exhausts the EWSR is the case on which the following discussion will be focused; similar derivations for the ISGDR, higher-order isoscalar multipoles, and the IVGDR can be found elsewhere \cite{harakeh_ISGDR,Satchler_direct_nuclear_reactions,satchler_isospin}.

Let us assume that there is a single state, $\ket{k}$, which exhausts the entirety of the $m_1$ EWSR. If this is the case, and taking the ground-state energy $E_0$ as a reference value, then the sum rule simplifies:
\begin{align}
  m_1 & = \sum_{n=1}^A E_n \bigg | \big < n \big| \hat{F} \big| 0 \big> \bigg |^2 \notag \\
  & = E_k  \bigg | \big < k \big| \hat{F} \big| 0 \big> \bigg |^2. \label{ewsr_one_state}
\end{align}
Passing to a position-space representation,  this can be expressed in terms of the transition density $\delta \rho$ for the $\ket{0} \rightarrow \ket{n}$ transition:
\begin{align}
  \big < k \big| \hat{F} \big| 0 \big> & = \int \diff^3 \vec{r} \, \delta \rho(\vec{r}) F(\vec{r}). \label{matrix_element}
\end{align}
Here, we use the previously defined monopole operator of Eq. \eqref{one_body_operators}. The transition density $\delta \rho(\vec{r})$ is that which was derived in the previous section and is given by Eq. \eqref{monopole transition density}. For the sake of analytical tractability, we will assume a uniform ground-state nuclear mass-density distribution without loss of generality\footnote{The result generalizes to arbitrary density distributions; see, for example, the treatments of \cite{satchler_isospin,schlomo_overtone_amplitude,blaizot_nuclear_compressibilities,Satchler_direct_nuclear_reactions,suzuki_amplitude_derivation} for details.}:
\begin{align}
  \rho(r) & = \rho_0 \left[ \Theta(r) - \Theta(r-R) \right],
\end{align}
in which $R$ is the nuclear radius\footnote{For the uniform distribution, one should recall $\expect{r^2} = 3/5 R^2$.}, and $\Theta$ is the Heaviside step function. With this, the transition density takes the form
\begin{align}
  \delta \rho_0 (r) & = -\frac{\beta_0^\text{100\% EWSR}}{r^2} \left[3 r^2 \rho(r) + r^3 \left(\delta(r) - \delta(r-R) \right) \right] ; \label{transition_density_uniform}
\end{align}
upon insertion of Eqs. \eqref{transition_density_uniform} into \eqref{matrix_element} and again into Eq. \eqref{ewsr_one_state}, one finds that
\begin{align}
  m_1 & = 4 \left[\beta_0^\text{100\% EWSR}\right]^2 A^2 \expect{r^2}^2 E_k
\end{align}

One should note, for practical purposes, that there is a presumption by coupled-channels and DWBA codes that the internal and external transition potentials are normalized by $1/\sqrt{4\pi}$ (see, \emph{e.g.}, Ref. \cite{satchler_isospin} for comments along these lines). The prescription by Ref. \cite{satchler_isospin} in handling this is to pragmatically scale the $\beta_\lambda^2 \mapsto 4\pi \beta_\lambda^2$ in order to preserve the magnitude of the coupling.\footnote{Alternatively, one could --- perhaps more neatly --- omit the factors of $1/\sqrt{4\pi}$ in the transition potential calculation if the entire optical potential is externally calculated. As some optical models --- discussed in greater detail in Chapter \ref{Data Analysis} --- are not amenable to this (\emph{e.g.} a potential that uses externally-calculated volume potentials but internally-calculated surface or spin-orbit potentials), we will instead use the more general solution described by Ref. \cite{satchler_isospin} henceforth.} With this, the value of $\beta_0$ which exhausts the monopole EWSR for a transition of excitation energy $E_k$, and which is further directly compatible with most modern DWBA codes is:
\begin{align}
  \left[\beta_0^\text{100\% EWSR}\right]^2 & = 4 \pi \frac{\hbar^2}{2 m A \expect{r^2} E_k} \notag \\
  & = \frac{2 \pi \hbar^2}{A m \expect{r^2} E_k}. \; \; \; (\lambda = 0) \label{amplitudes_full_sum_rule}
\end{align}

Similar derivations can be done with the transition densities for the ISGDR and higher-order isoscalar giant resonances to acquire, respectively, the amplitude ($\lambda = 1$) and deformation parameters ($\lambda \geq 2$) which exhaust their corresponding sum rules:

\begin{align}
  \left[\beta^\text{100\% EWSR}_1\right]^2 & = \frac{6 \pi \hbar^2}{A m E_k} \frac{c^2}{11 \expect{r^4} - \frac{25}{3} \expect{r^2}^2 - 10 \epsilon \expect{r^2}}, \; \; \; &(\lambda = 1)  \notag \\
  \left[\beta^\text{100\% EWSR}_\lambda\right]^2 & = \frac{2\pi \hbar^2 }{A m c^2 E_k} \frac{\lambda(2\lambda+1)^2}{(\lambda+2)^2} \frac{\expect{r^{2\lambda-2}}}{\expect{r^{\lambda-1}}^2}. \; \; \; &(\lambda \geq 2)
\end{align}

Finally, for the IVGDR, the Goldhaber-Teller model yields a transition amplitude for the exhaustion of the TRK sum rule \cite{satchler_isospin,harakeh_book}:
\begin{align}
  \left[\beta^\text{100\% EWSR}_\text{IVGDR}\right]^2 & = \frac{\pi \hbar^2}{2m E_k} \frac{N Z}{A}. \label{TRK_amplitude}
\end{align}

%

\section{Direct reaction theory, distorted waves, and the distorted-wave Born approximation} \label{direct reaction theory}

\subsection{Development of the distorted-wave Born approximation}

As will be discussed in Chapter \ref{experimental}, the technique utilized in this work to isolate the features of the ISGMR in stable nuclei is based upon the analysis of experimental angular distribution data. In this section, we will briefly outline the general theory of direct nuclear reactions relevant to our methodology and data analysis. This material is sourced primarily from Refs. \cite{harakeh_book,Satchler_direct_nuclear_reactions}; further exposition into the general theories of direct reactions relevant for giant resonance studies may be found therein.

For a reaction of the form $a(A,B)b$, or $a+A \rightarrow b+B$, we write the single particle wavefunctions as $\psi_a(r_a)$, $\psi_A(r_A)$, and the total wavefunction for the incoming channel in the partition $a+A$ as $\psi_\alpha = \psi_a \psi_A$. The quantity $\vec{r}_\alpha$ is the relative distance coordinate between $a$ and $A$; the quantity $\vec{x}_\alpha$ denotes the combination of position coordinates, $\vec{x}_a$ and $ \vec{x}_A$, in channel $\alpha$. The incoming reaction channel for $a+A$, specified by a set of relevant quantum numbers and denoted by a collective index $\alpha$, is asymptotically related to an outgoing channel for $b+B$, with quantum numbers specified by $\beta$ within a spherical basis via:
\begin{align}
  \xi_\beta(\vec{r}_\beta) \sim \exp\left(i \vec{k}_\alpha \cdot \vec{r}_\alpha\right) \delta_{\alpha,\beta} +  f_{\beta,\alpha}(\vec{k}_\beta,\vec{k}_\alpha) \frac{\exp\left(i k_\beta r_\beta\right)}{r_\beta}. \label{measured_wave}
\end{align}
Here, $k$ is the wavenumber or momentum in the associated channel, and $\delta$ is the Kronecker delta. Equation \eqref{measured_wave} lends itself to the interpretation that the measured wavefunction is itself a superposition of the incoming plane wave (if $\beta=\alpha$) with a spherical outgoing wave. The quantity $f_{\beta,\alpha}$ is the complex \emph{scattering amplitude}\footnote{The definition for the transition matrix ($T$ matrix) $T_{\beta,\alpha}$ used here is $T_{\beta,\alpha} = - 2\pi \hbar^2 f_{\beta,\alpha}/\mu_\beta$, wherein $\mu_\beta$ is the reduced mass of the outgoing channel.} which connects the incoming and outgoing channels and is directly proportional to the transition matrix element $T_{\beta,\alpha}$; this quantity also is related to the measured differential cross section $d\sigma/d\Omega$ by 

\begin{align}
  \frac{d\sigma_{\alpha,\beta}}{d\Omega} = \frac{v_\beta}{v_\alpha} \left | f_{\beta,\alpha}(\vec{k}_\beta,\vec{k}_\alpha) \right|^2.\label{cs_defined}
\end{align}

It is upon this basis that we can  introduce the concept of distorted waves. In general, the total wavefunction for channel $\beta$ is the product of wavefunctions for the ejectile and recoil nuclei. A similar expansion can be done for the incoming channel. In any event, one can thus represent the total incident wavefunction $\Psi_\alpha^+(\vec{k}_\alpha)$ in terms of the basis of outgoing waves, with amplitudes $\xi_\beta$:

\begin{align}
  \Psi_\alpha^+ (\vec{k}_\alpha) = \sum_{\beta} \xi_\beta(\vec{r}_\beta) \psi_\beta(\vec{x}_\beta). \label{incoming_wave_in_outgoing_basis}
\end{align}

The total Hamiltonian for either channel can be expressed in terms of the internal Hamiltonians for nucleus $b$ and $B$, collectively denoted $H_\beta = H_b + H_B$, in addition to the kinetic and potential energies of the relative positions of the nuclei:
\begin{align}
  \hat{H} = \hat{H}_\beta + \hat{K}_\beta + \hat{V}_\beta
\end{align}

Examining the form of the time-independent Schr\"odinger equation applied to a given outgoing channel:
\begin{align}
  \hat{H} \ket{\Psi_\alpha^+} &= E \ket{\Psi_\alpha^+} \notag \\
  0 & = \left[E - \hat{H}_\beta - \hat{K}_\beta - \hat{V}_\beta \right] \ket{\Psi_\alpha^+}.
  \label{schrodinger_for_scattering}
\end{align}
Owing to the orthonormality of the $\left\{\ket{\psi_\beta}\right\}$, the independence of the kinetic energy with respect to the positional coordinates $\vec{x}_\beta$, and Eq. \eqref{incoming_wave_in_outgoing_basis}, we find that upon multiplying $\bra{\psi_\beta}$ to either side:

\begin{align}
  0 & = \bra{\psi_\beta} \left[E - \hat{H}_\beta - \hat{K}_\beta - \hat{V}_\beta \right] \ket{\Psi_\alpha^+} \notag \\
  \bra{\psi_\beta} \hat{V}_\beta \ket{\Psi_\alpha^+} &= \left[(E-\hat{H}_\beta) - \hat{K}_\beta \right] \ket{\xi_\beta} \notag \\
  \bra{\psi_\beta} \hat{V}_\beta \ket{\Psi_\alpha^+} &= \left[E_\beta - \hat{K}_\beta \right] \ket{\xi_\beta}
  \label{inhomogenous eq.}
\end{align}

The interaction potential $V_\beta$ within the outgoing channel $\beta$ is separable into two terms of the form
\begin{align}
  V_\beta \left(\vec{x}_\beta,\vec{r}_\beta\right) = U_\beta(r_\beta) + W_\beta(\vec{x}_\beta, \vec{r}_\beta).
\end{align}
The first term, $U_\beta(r_\beta)$, is an average potential (in practice, the optical potential) which depends only on the relative inter-nuclear positioning of the nucleons participating in the reaction, whereas the second term, $W_\beta(\vec{x}_\beta,\vec{r}_\beta)$ can depend explicitly upon the internal nucleon coordinates (in practice, serving as the transition potential); in other words, the $W_\beta$ term allows for internal rearrangement within the channel and is, within this formalism, typically small in relation to $U_\beta$. In contrast, the potential $U_\beta$ is unable to induce transitions during the interaction.

This assumption is the premise of the \emph{Distorted Wave Born Approximation} (DWBA), and can be understood in the context that the elastic channel in the scattering process (directly modeled by $U_\beta$) dominates over inelastic channels, charge-exchange channels, \emph{etc.} which are each modeled by $W_\beta$. The general prescription is to treat $W_\beta$ then as a weak perturbation on the elastic channel that can only induce rearrangement or excitation of the participating nucleons, as evidenced by the choice of definition for each of the terms:
\begin{align}
  \hat{U}_\beta &= \braketop{\psi_\beta}{\hat{V}_\beta}{\psi_\beta} \notag \\
  \hat{W}_\beta &= \hat{V}_\beta (\vec{x}_\beta, \vec{r}_\beta) - \braketop{\psi_\beta}{\hat{V}_\beta}{\psi_\beta}.
  \label{transition_potential_defined}
\end{align}
The DWBA framework essentially models direct nuclear reactions as one-step processes; its validity is predicated on the transition amplitudes (equivalently, the cross sections) being small in relation to those of the incoming elastic channel. The significance of this is that Eq. \eqref{inhomogenous eq.} can be rewritten as an inhomogeneous equation:
\begin{align}
  \bra{\psi_\beta} \left[ \hat{U}_\beta + \hat{W}_\beta \right] \ket{\Psi_\alpha^+} &= \left[E_\beta - \hat{K}_\beta \right] \ket{\xi_\beta} \notag \\
  \bra{\psi_\beta} \hat{W}_\beta \ket{\Psi_\alpha^+} & = \left[ E_\beta - \hat{K}_\beta - \hat{U}_\beta \right] \ket{\xi_\beta}.
\end{align}
By projecting a complete set of states and resolving the identity with $\left\{ \ket{\psi_{\beta^\prime}}\bra{\psi_{\beta^\prime}}\right\}$, and further employing that the diagonal elements $\bra{\psi_\beta} \hat{W}_\beta  \ket{\psi_\beta}$ vanish identically owing to its definition in Eq. \eqref{transition_potential_defined}, one finds that
\begin{align}
  \left[ E_\beta - \hat{K}_\beta - \hat{U}_\beta \right] \ket{\xi_\beta} & = \sum_{\beta^\prime} \bra{\psi_\beta} \hat{W}_\beta \ket{\psi_{\beta^\prime}} \big<\psi_\beta^\prime \big| \Psi_\alpha^+ \big> \notag \\
  &= \sum_{\beta^\prime \neq \beta } \bra{\psi_\beta} \hat{W}_\beta \ket{\psi_{\beta^\prime}} \ket{\xi_{\beta^\prime}}. \label{distorted_waves_equation}
\end{align}
Under the aforementioned presumption that $W_\beta$ can be treated as a perturbation, then the solutions to the homogeneous equations can be used for a basis in the expansion of the inhomogeneous solutions of Eq. \eqref{distorted_waves_equation} and therefore used in the calculation of the transition matrix elements. These homogenous solutions are the eponymous \emph{distorted waves} $\ket{\chi_\beta^\pm}$, and in the case of first-order coupling only between the elastic and inelastic channels:
\begin{align}
  \left(E_\alpha - \hat{K}_\alpha - \hat{U}_{\alpha} \right) \ket{\chi_\alpha^+} & = 0 \notag \\
  \left(E_\beta - \hat{K}_\beta - \hat{U}_{\beta} \right) \ket{\chi_\beta^-} & = 0
\end{align}
In the asymptotic regime, Green's-function solutions for Eq. \eqref{distorted_waves_equation} exist in terms of the $\chi_\beta^+$ and its time-reversed solutions $\chi_\beta^-$ \cite{Satchler_direct_nuclear_reactions}. Using this result, the transition matrix element for the $\alpha\rightarrow \beta$ reaction is given in the DWBA framework, for the specific case of inelastic scattering (wherein $\beta \mapsto \alpha^\prime)$:
\begin{align}
  T_{\alpha^\prime,\alpha} &= \bra{\chi_{\alpha^\prime}^-} \hat{V}_{\alpha} \ket{\chi_{\alpha}^+} \notag \\
  f_{\alpha^\prime,\alpha} (\theta_{\alpha^\prime}) &= - \frac{\mu_{\alpha^\prime}}{2\pi \hbar^2} \bra{\chi_{\alpha^\prime}^-} \hat{V}_{\alpha} \ket{\chi_{\alpha}^+}  \label{transition_potential_in_terms_of_distorted_waves}
\end{align}

One should note that due to the definitions of the optical potential $\hat{U}_\alpha$ and the transition potential $\hat{W}_\beta$, in the case of inelastic scattering, only the latter transition term contributes to the inner products of Eq. \eqref{transition_potential_in_terms_of_distorted_waves}.

As the transition matrix elements and scattering amplitudes are directly related, this argumentation provides a road-map for the calculation of inelastic angular distributions given an optical potential, $U_\beta$. Upon acquisition of such a potential, providing that the elastic channel is comparatively strong in relation to the inelastic channels which one desires to model, Eq. \eqref{transition_potential_in_terms_of_distorted_waves} provides a framework within the DWBA method to calculate the transition matrix elements given a transition potential that then acts upon the readily-calculable elastic scattering solutions. As it so happens --- as we will describe in great detail in Chapter \ref{Data Analysis} --- the features of $U_\beta$ can be well-modeled with a correct choice of ansatz for its functional form based on the limiting behaviors of the nuclear force, and subsequently fitted to experimental elastic scattering data.

As we will see, $U_\beta$ not only provides the set of scattering solutions which constitute the set of distorted waves on which the DWBA theory is built, but also provides within the collective model a mechanism for calculating the transition potential itself. Thus, the problem of calculating the transition matrix elements and equivalently, the angular distributions for the inelastic channels of interest in this work, is reduced to determining an adequate characterization of the average optical potential that reproduces the observables from the elastic channel \cite{Satchler_direct_nuclear_reactions}. \interfootnotelinepenalty=10000 \footnote{Reference \cite{Satchler_direct_nuclear_reactions} makes the distinction between the DWBA method and the method of distorted-waves. The former directly calculates the potential by explicit treatment of the coupled channels problem under the assumption that the off-diagonal terms are fairly small in comparison to the diagonal terms. The method of distorted-waves, in contrast, \emph{fits} the optical potential $U_\beta$ such that experimentally-measured angular distributions are well-described in the elastic channel, and then utilizes that optical potential in the calculation of the transition amplitudes. Our methodology technically uses the latter methodology, but the distinction made by Ref. \cite{Satchler_direct_nuclear_reactions} is hardly adhered to in common literature, and so we will instead frequently refer to them interchangeably.}

\subsection{Transition potentials}

The task of calculating the angular distributions for an inelastic-scattering channel for which the DWBA is valid is therefore reduced to the calculation of the matrix element of Eq. \eqref{transition_potential_in_terms_of_distorted_waves}, which is readily implemented by various DWBA and coupled-channels codes (in this work, we have primarily used \texttt{PTOLEMY} \cite{ptolemy_manual}), which handle the solution for the distorted waves themselves and the calculations of the matrix elements. The input to these codes are essentially the chosen ansatz for the functional form of the optical potential (the form of that which was specifically used for the analysis of this work is discussed in Chapter \ref{Data Analysis}), the transition amplitudes as developed in Subsection \ref{transition_amplitudes_section}, and the transition potentials which are directly calculable from the optical potential and the transition densities of Subsection \ref{theory:transition_densities}.

Owing to the short-range nature of the effective nucleon-nucleon interaction (the Coulomb potential is considered separately), the transition potentials $W_\lambda$ which connect the elastic channel --- modeled by the optical model $U_\text{OM}$ --- to the inelastic channel of multipolarity $\lambda$ are well-approximated as having the same functional form as the transition densities \cite{harakeh_book,Satchler_direct_nuclear_reactions,satchler_isospin,khoa_satchler_single_folding}:
\begin{align}
  W_\lambda (r) & \propto \delta \rho_\lambda (r).
\end{align}

This assumption, combined with Eqs. \eqref{monopole transition density} --- \eqref{IVGDR transition density}, yields the radial component of the transition potentials:
\begin{align}
  W_0 (r) & \propto  - \beta_0^\text{OM} \left(3 + r \frac{\diff}{\diff r} \right) U_\text{OM} (r) , & (\lambda = 0) \notag \\
  W_1 (r) & \propto  - \frac{\beta_1^\text{OM} }{c } \left[3 r^2 \frac{\diff}{\diff r} + 10 r - \frac{5}{3} \expect{r^2} \frac{\diff}{\diff r} + \epsilon \left (r \frac{\diff^2}{\diff r^2} + 4 \frac{\diff}{\diff r} \right)\right] U_\text{OM} (r), &(\lambda =1) \notag \\
  W_\lambda (r) & \propto - \beta_\lambda^\text{OM} R \frac{\diff}{\diff r}  U_\text{OM}(r) ,  & (\lambda \geq 2) \notag \\
  W_\text{IVGDR}(r) & \propto - \beta_\text{IVGDR}^\text{OM} \gamma \left(\frac{N-Z}{A} \right) \left[ \frac{\diff }{\diff r} + \frac{1}{3} c \frac{\diff^2 }{\diff r^2} \right] U_\text{OM}(r). & (\text{IVGDR}) \label{transition_potentials_multipolarities}
\end{align}

This assumption of proportionality between the transition densities and potentials is logically equivalent to the stance that the interaction potential adopts the same deformation as that which is assumed by the nuclear density distribution during a transition \cite{khoa_satchler_single_folding, horen_deformation_optical_model}. In practice, this means that the deformation length of the density distribution, $\delta_\lambda$:
\begin{align}
  \delta_\lambda & = \beta_\lambda c,
\end{align}
is equal to the deformation length of the optical model potential, $\delta_\lambda^\text{OM}$ \cite{bernstein_deformation}:
\begin{align}
  \delta_\lambda^\text{OM} & = \beta_\lambda^\text{OM} R,
\end{align}
in which $c$ and $R$ are respectively the half-radii of the density distribution and optical potential. In practice, this is done separately for the real and imaginary components of the optical potential.

To this point, one can begin to characterize the behaviors of the calculated differential cross sections in terms of the transition potentials. Under these assumptions, the shapes of the transition densities and transition potentials are essentially independent of the strength of the amplitudes $\beta_\lambda$. Equations \eqref{cs_defined}, \eqref{transition_potential_in_terms_of_distorted_waves}, and \eqref{transition_potentials_multipolarities} suggest that the magnitudes of the transition amplitudes $\beta_\lambda$ realized in a transition directly scale the magnitudes of the measured cross sections, without influencing the structure of the angular distributions. These facts constitute a prelude to the multipole decomposition analysis that will be discussed in Chapter \ref{Data Analysis}.

%

%
%
%
%
%
%
%
%
%
%

%
%

\chapter{Experimental details and data reduction}
\label{experimental}

\section{ISGMR studies in stable nuclei}

From an experimental point of view, there are a number of pathways available for one who wishes to experimentally isolate the ISGMR in stable nuclei. The most overwhelming hurdle to be crossed in these experimental studies is the simultaneous excitations of different giant resonances which can then give rise to a structureless continuum in the detected spectra \cite{harakeh_book}. The purpose of this chapter is to both motivate and describe the actions undertaken by modern-day experimental campaigns --- and indeed, this thesis work --- to reconcile this issue and extract features of individual giant resonances (namely, the ISGMR) through both careful experimental planning and instrumentation.

\subsection{The importance of forward angle measurements}


The first experimental evidence for the ISGMR came in the 1970s from the experimental efforts of Harakeh \emph{et al.} \cite{harakeh_ISGMR_discovery_77,harakeh_ISGMR_discovery_79}, wherein \nuc{208}{Pb}($\alpha,\alpha^\prime$) spectra suggested that there was a peak at $14^\circ$ which was separate from that of the ISGQR (which was discovered and characterized several years previously) that was comparatively stronger than the same peak measured at $12^\circ$. The suggested explanation for the discrepant angular character of the peaks was that each carried different multipolarities. Ultimately, a definitive assignment of the monopole character of the peak was later made on the basis of comparison of experimental data with the characteristic $\lambda = 0$ angular distribution at extremely forward angles \cite{youngblood_ISGMR_discovery_77}. Just a few years later, an independent experimental effort probed the giant resonance region using inelastic deuterium scattering off of \nuc{40}{Ca}, \nuc{58}{Ni}, \nuc{90}{Zr}, \nuc{120}{Sn}, and \nuc{208}{Pb}, in which monopole strength was again assigned for each nucleus by inspection of the measured angular distributions \cite{willis_d_dprime}. As will be discussed shortly, in a sense, these angular-distribution analyses were progenitors of present-day ISGMR studies.

\begin{figure}[t!]
  \centering
  \includegraphics[width=\linewidth]{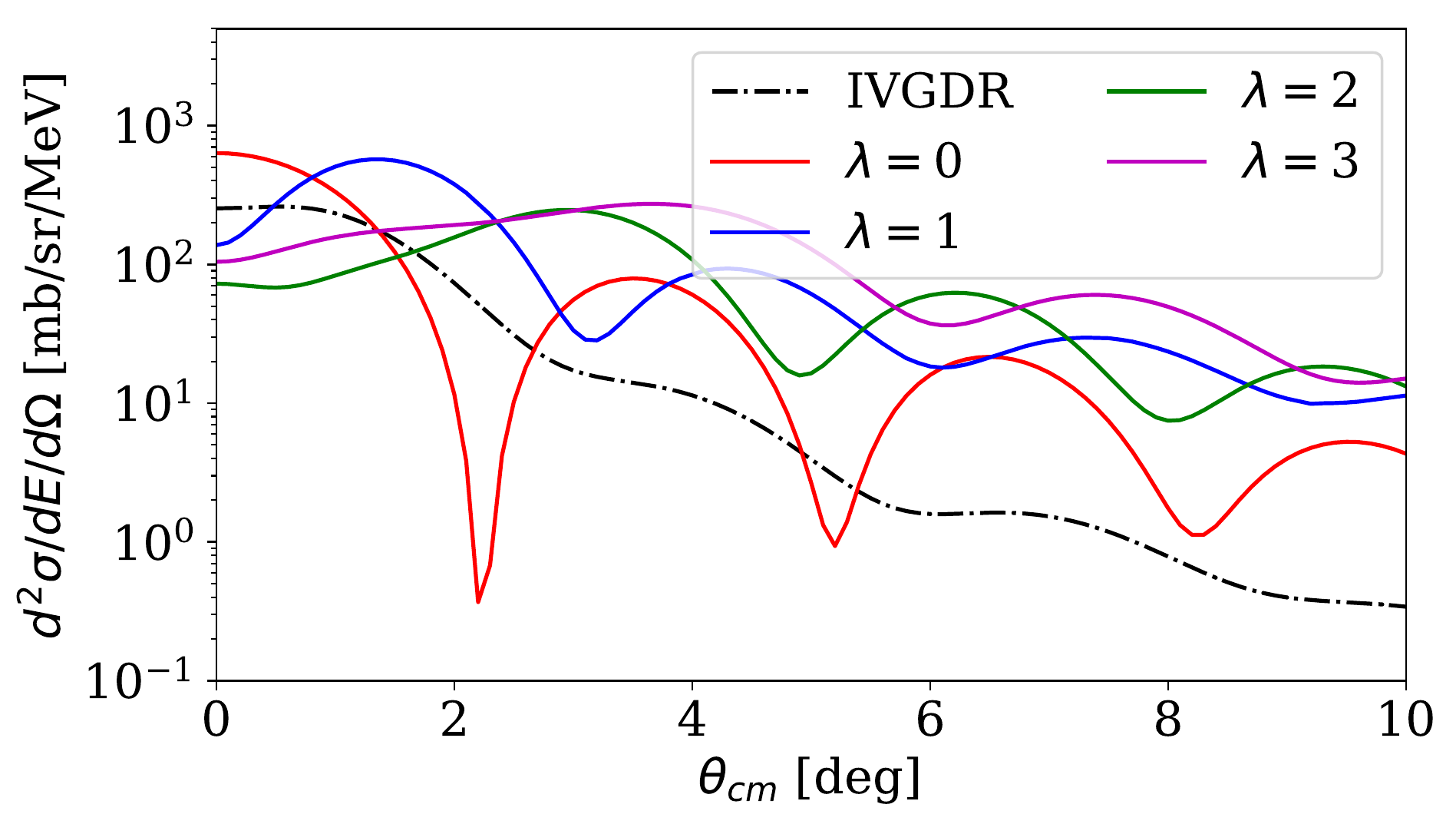}
  \caption[Sample DWBA angular distributions for \nuc{94}{Mo} at $E_x = 15$ MeV.]{Sample DWBA angular distributions for \nuc{94}{Mo} at $E_x = 15$ MeV. The angular distributions corresponding to isoscalar momentum transfers $\lambda=0$ (red), $\lambda = 1$ (blue), $\lambda = 2$ (green) and $\lambda=3$ (purple) are shown, in addition to the contributions of the isovector giant dipole resonance (black, dot-dashed).}
  \label{experimental:angular_distributions}
\end{figure}

In any event, this chain of events illustrates the main experimental difficulty with experimentally isolating the ISGMR response of a nucleus. Shown in Fig. \ref{experimental:angular_distributions} are some characteristic angular distributions for \nuc{94}{Mo}($\alpha,\alpha^\prime$) (with $E_\text{beam} = 386$ MeV), for the isoscalar monopole, dipole, quadrupole, and octupole resonances ($\lambda = 0, 1,2, 3$ respectively), as well as the isovector dipole resonance, at 15 MeV. It is clearly the case that the ISGMR angular distribution peaks at $0^\circ$, whereas for the ISGDR and ISGQR distributions, the maxima occur at larger angles. Notably, the angular distributions for angular momentum transfers which carry the same natural parities ($\pi = (-1)^\lambda$, with $\lambda=0,1,2,\ldots$) are very nearly in phase beyond their corresponding first maxima. To put it simply, although the ISGMR does technically have a measurable response at larger angles, it is nearly intractable to definitively isolate those measured features from higher multipolarities which rapidly begin to overlap at larger angles.


The predominant means by which modern-day experiments quantify the isoscalar giant resonance strength distributions is by measuring angular distribution data to decompose the responses of the giant resonances over a range of excitation energies using a multipole decomposition analysis (discussed further in Chapter \ref{Data Analysis}) \cite{Li_PRL,Li_PRC,itoh_sm_PRC,patel_cd,patel_MEM,gupta_A90_PLB,gupta_A90_PRC,KBH_EPJA}. In order to optimally constrain the ISGMR response on this basis, it is further required to acquire forward-angle angular distribution to mitigate the aforementioned difficulties arising from the overlap of the ISGMR and ISGQR.

\subsection{Choice of probe for ISGMR studies}
As discussed in Chapters \ref{intro} and \ref{theory}, there exist myriad giant resonances (cf. Fig. \ref{multipole figure}) which can be excited in an experiment. Much of this chapter and  Chapter \ref{Data Analysis} discuss in great detail, the instrumental and data-analysis techniques which allow for isolating a single mode, the ISGMR, among all of the possible oscillations shown in Fig. \ref{multipole figure}. However, with the aid of selection rules and conservation laws, it is possible to execute a well-planned experiment which precludes the excitation of certain resonances altogether to optimize the constraints on the ISGMR that can be determined from the experimental data.

In this lies the reasoning for using $\alpha$-particles as the primary probe for experiments on the ISGMR in stable nuclei. As the $\alpha$-particle carries neither an isospin nor a spin projection, it primarily excites the isoscalar and electric giant resonances.\footnote{Due to Coulomb excitation and the intrinsic angular momentum carried by photons, it is possible for $\alpha$-particles to couple to the IVGDR with measurable effect. This is discussed in greater detail in Chapters \ref{theory} and \ref{Data Analysis}.} Due to this fact, in this thesis work, $\alpha$-particles were our choice of probe for each of the experiments.

\section{ISGMR measurements at RCNP}

\begin{figure}[t!]
\centering
\includegraphics[width=0.75\linewidth]{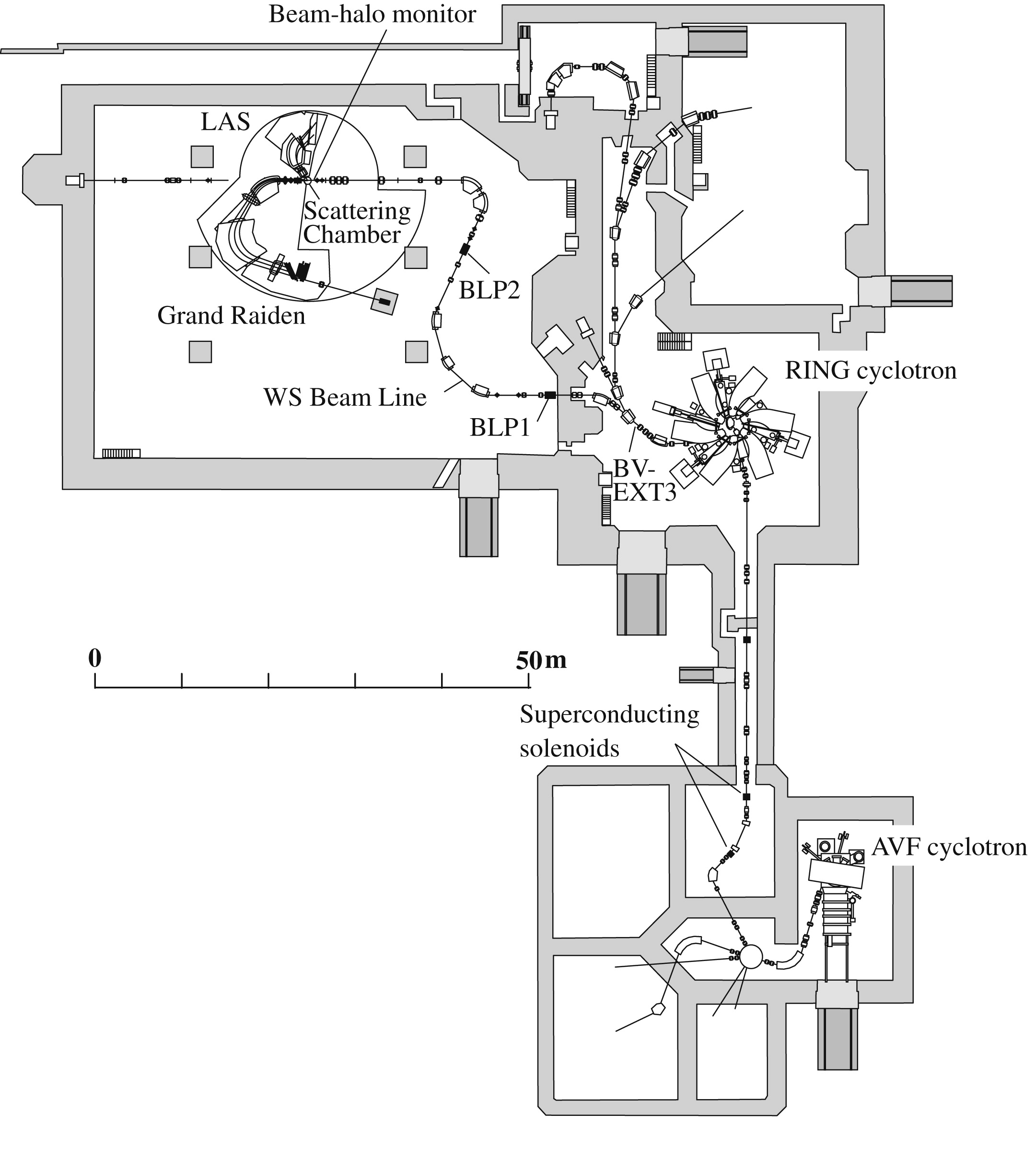}
\caption[Schematic of the coupled AVF and Ring Cyclotrons, the WS beamline which transported the beam, and the Grand Raiden magnetic spectrometer.]{Schematic of the coupled AVF and Ring Cyclotrons, the WS beamline which transported the beam, and the Grand Raiden magnetic spectrometer. Figure originally from Ref. \cite{tamii_grand_raiden}.}
\label{GR_beamline}
\end{figure}

\begin{table}[t!]
\centering
\caption[Areal densities of the target foils used in this work]{Areal densities of the target foils used in this work}
\resizebox{\textwidth}{!}{%
\begin{tabular}{@{}cccccccccc@{}}
\toprule
Nucleus                   & \nuc{40}{Ca} & \nuc{42}{Ca} & \nuc{44}{Ca} & \nuc{48}{Ca} & \nuc{94}{Mo} & \nuc{96}{Mo} & \nuc{97}{Mo} & \nuc{98}{Mo} & \nuc{100}{Mo} \\ \midrule
Areal Density [mg/cm$^2$] & 1.63         & 1.78         & 1.83         & 2.20         & 4.10         & 4.5          & 3.2          & 6.3          & 3.4           \\ \bottomrule
\end{tabular}%
}

\label{target thicknesses}
\end{table}

The work undertaken in this thesis was completed at the Research Center for Nuclear Physics (RCNP). A pair of experiments (E462 and E495) were conducted on, respectively, the \nuc{94,96,97,98,100}{Mo} and \nuc{40,42,44,48}{Ca} isotopic chains. The areal densities of the target foils used in the two experiments are reported in Table \ref{target thicknesses}. The $\alpha$-particles, which were generated by an electron-cyclotron resonance ion source \cite{RCNP_ion_source}, were first injected into the Azimuthally-Varying-Field (AVF) cyclotron and then transported to the Ring Cyclotron as shown in Fig. \ref{GR_beamline}. The Ring Cyclotron was operated such that only single-turn $386$ MeV $\alpha$-particles were extracted to ensure a high-quality beam, with a typical energy resolution of $\sim 150-200$ keV --- this is well below the characteristic energy scales of any giant resonances \cite{harakeh_book} and thus proved sufficient for our experimental purposes. The ability for the coupled cyclotrons to deliver high-quality beams of this energy is critical, for the ISGMR excitation is a direct reaction and therefore its associated cross sections scale directly with beam energy \cite{harakeh_book,Satchler_direct_nuclear_reactions}.

The accelerated $\alpha$-particles were transported by the West-South (WS) beamline
\cite{wakasa_beamline,wakasa_beamline_2} into the target chamber, where the target foils were bombarded and the scattered particles were accepted into the Grand Raiden high precision magnetic spectrometer \cite{Fujiwara_Grand_Raiden,wakasa_beamline,wakasa_beamline_2,tamii_grand_raiden}. Grand Raiden has a design resolving power of $p/\Delta p = 37000$; in our own experiments, this was not realized owing to limits in the energy resolution of the beam transport injected into the spectrograph. Design specifications of the spectrograph are given in Table \ref{GR_design_specs}.

\begin{figure}[t]
\centering
\includegraphics[width=0.85\linewidth]{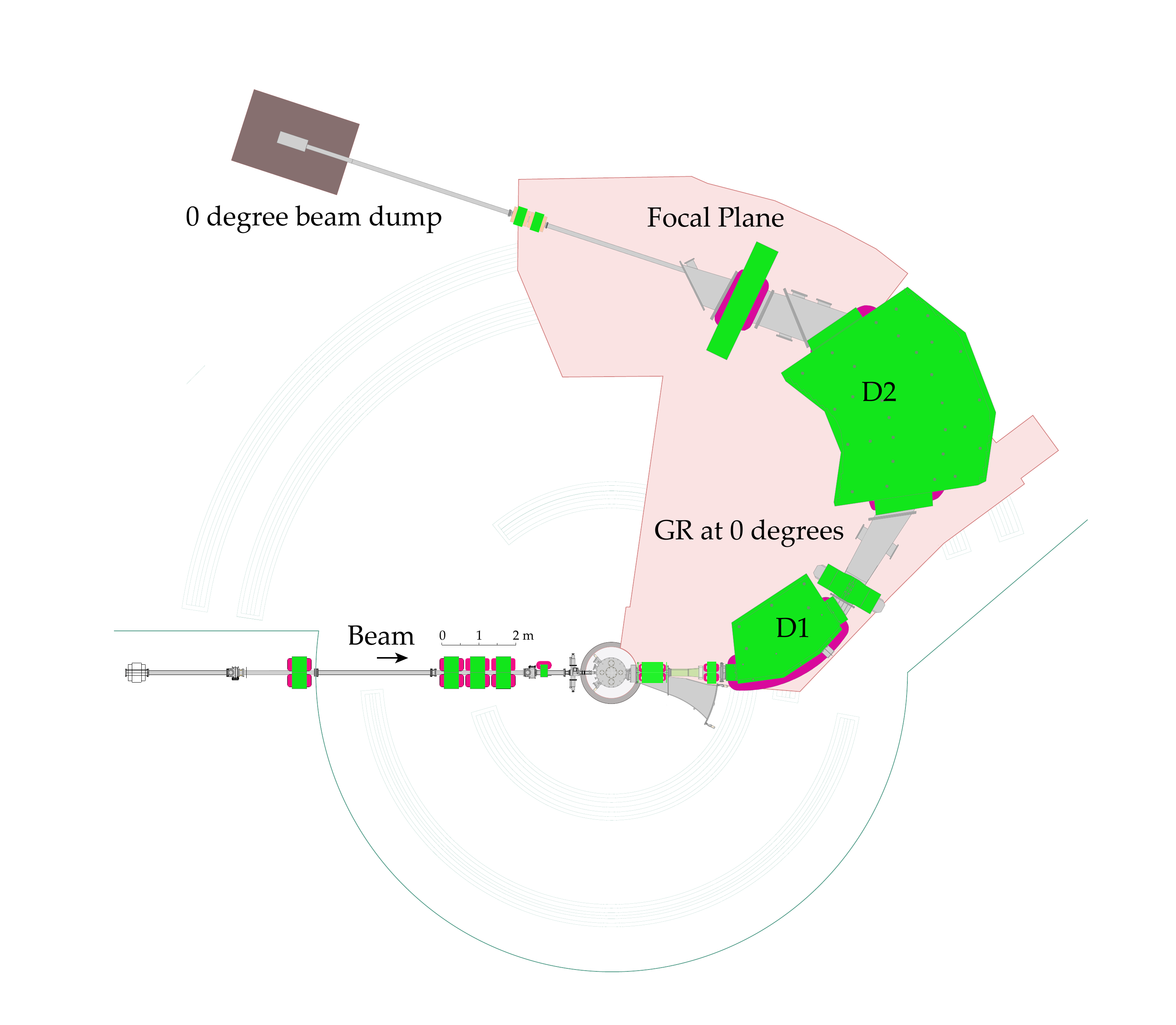}
\caption[Scale drawing of the Grand Raiden spectrometer in the zero-degree arrangement.]{Scale drawing of the Grand Raiden spectrometer in the zero-degree arrangement. Shown in green are the magnetic quadrupoles and dipoles; we have labelled the momentum-analyzing magnets D1 and D2. Figure courtesy of Prof. A. Tamii.}
\label{GR}
\end{figure}

\begin{table}[t!]
  \centering
  \caption{Design specifications of the Grand Raiden spectrograph \cite{Fujiwara_Grand_Raiden}}
  \begin{tabular}{@{}ll@{}}
  \toprule
  Mean Orbit Radius                                & 3 m           \\
  Focal Plane Horizontal Length         & 1.5 m         \\
  Maximum Bending Dipole Field          & 1.8 T         \\
  Maximum Magnetic Rigidity             & 5.4 T m       \\
  Horizontal Magnification $(x|x)$      & -0.419         \\
  Vertical Magnification $(y|y)$        & 5.98          \\
  Momentum Dispersion $(x|\delta)$                  & 15.45 m       \\
  Momentum Byte                         & 5\%            \\
  Resolving Power ($p/\Delta p$)        & 37 000        \\
  Maximum Horizontal Angular Acceptance & $\pm 20$ mrad \\
  Maximum Vertical Angular Acceptance   & $\pm 70$ mrad \\
  \bottomrule
  \end{tabular}
  \label{GR_design_specs}
\end{table}

A detailed schematic of the spectrograph in the zero-degree arrangement is shown in Fig. \ref{GR}. The zero-degree measurements require the beam to be transported through the spectograph alongside the inelastically scattered $\alpha$-particles; after the dipole fields laterally disperse the inelastically scattered particles according to their reduced momentum along the horizontal focal plane axis, the minimally-dispersed, unreacted beam was transported through a pipe in the high-energy side of the focal plane detector, and into a special Faraday Cup located downstream from the focal plane in the beam dump. For the $2.5^\circ$ data, a Faraday Cup was located just outside of the scattering chamber, as the unreacted beam is still very close to the scattered beam. For higher angle data, a Faraday Cup inside of the scattering chamber was used for stopping the beam. The focal plane itself was aligned at a $\Psi_x = 45^\circ$  angle to the incident beam axis to minimize the effects of second-order ion-optical requirements that induce abberations that couple the detected focal-plane angle to the focal plane position, \emph{i.e.} $(x|x\theta) + \tan \Psi_x \approx 0$ \cite{Fujiwara_Grand_Raiden}. Any abberations which were present in the resulting spectra were corrected for in the offline-analysis in the styles of Refs. \cite{tamii_grand_raiden,abberations_spectrographs}.

The focal plane detector system was comprised of a pair of vertical and horizontal position-sensitive multiwire drift chambers (MWDCs), separated by $250$ mm, each with a plastic scintillator backing which provided a signal to photomultiplier tubes for the purposes of triggering, timing reference, and particle identification \cite{tamii_grand_raiden}. The MWDCs themselves were comprised of a pair of anode wire planes, denoted $X$ and $U$, each of which was bounded by a single cathode plane constructed from a polymeric aramid film. The cathode-anode spacing was approximately $10$ mm. The anode wires are comprised of two types of wires:

\begin{enumerate}
  \item \emph{Sense wires} which are made from 20 micron gold-plated tungsten, and
  \item \emph{Potential wires} which are made from 50 micron gold-plated beryllium copper.
\end{enumerate}

\begin{figure}[t]
  \includegraphics[width=\linewidth]{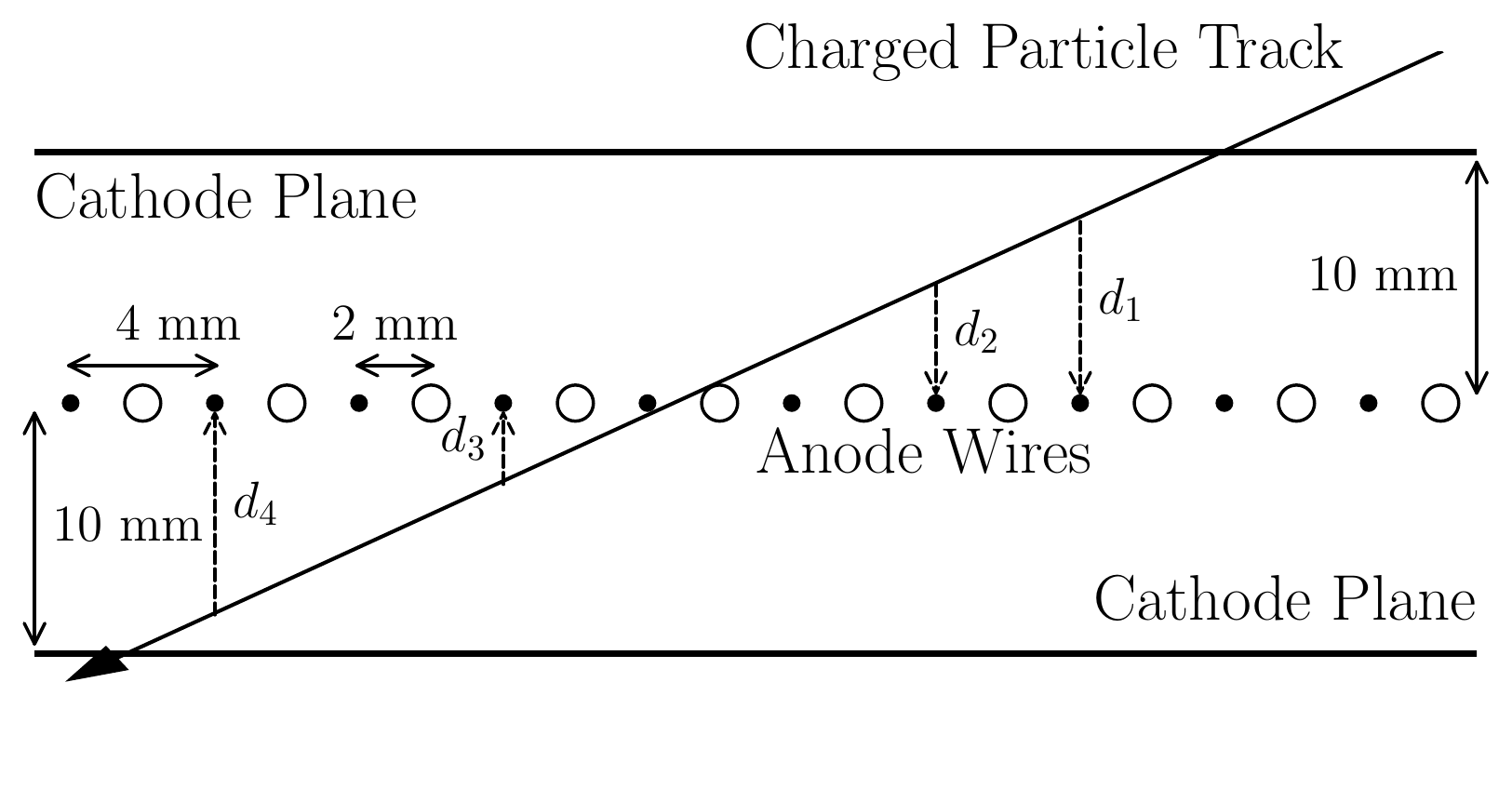}
  \caption[Schematic of the $X$-plane of the MWDCs.]{Schematic of an $X$-plane in the MWDCs, as viewed from above. A sample charged particle trajectory is shown with a set of possible ionization drift lengths. In the figure, filled circles denote the sense wires, while open circles denote the potential wires.}
  \label{MWDC_schematic}
\end{figure}

Each MWDC was filled with an admixture of argon and isobutane gasses. The role played by the potential wires is to generate a well-defined and very-nearly uniform electric field through this medium, with the goal of inducing a drift of the ionization electrons which are then detected by the sense wires. Due to the uniformity of the electric field sufficiently far from the potential wires, the ionized electrons then drift and are detected by the sense wires which then provide a signal indicating the position of the electron and thus the trajectory of the $\alpha$-particle. The aforementioned $X$ and $U$ anode planes consist of wires stretched, respectively, vertically and $\sim 48.2$ degrees from the vertical axis. Owing to this combination of wire orientations, the horizontal hit position at the focal plane can be calculated with a high precision on the order of essentially the sense-wire spacing.

Figure \ref{MWDC_schematic} shows a possible trajectory of a charged particle moving through an $X$ plane of the focal plane detection system. As charged particles move through the gas admixture, they induce ionizations in which the newly-freed electrons drift along the electric field generated by the cathode plane and anode potential wires, causing a near-constant-speed drift ($\sim$ 50 $\mu$m/ns) directly toward the anode wire plane. Using the time signals from the plastic scintillator as reference, the TDC readouts which yielded the times characterizing the transport from the ionization loci to the sense wires were recorded. With the drift speed being well-characterized with a given voltage difference between the cathode and anode planes, these drift times were readily converted into drift lengths, $d_j$, for a hit on the $j^\text{th}$ anode sense wire.

As the horizontal position, $p_j$ of each wire is known precisely, for a given charged particle trajectory, a set of tuples was generated with an entry for each sense wire hit, $P=(p_j,d_j)$. Events were only considered for which the set $P$ had three or greater elements and for which the magnitude of the drift lengths $d$ reaches a global minimum within the interior of the set $P$. The collection of these data then allowed for a least-$\chi^2$ minimization using a linear model function. The determination of this model function permitted inference of the exact location at which the given trajectory crosses the anode wire plane, which was then recorded as the true position of the charged particle at that plane of the MWDC.


Moreover, the extraction of the $x_1$ and $x_2$ positions at the first and second MWDC anode plans allows for a straightforward calculation of the focal-plane detected angle:
\begin{align}
  \tan \left(\theta_\text{horiz}^\text{fp}\right) = \frac{x_2-x_1}{L}, \label{angle_focal_plane}
\end{align}
wherein $L=250$ mm is the inter-MWDC spacing. The horizontal magnification from Table \ref{GR_design_specs} allows for the ready calculation of the scattering angle as
\begin{align}
  \theta_\text{horiz}^\text{fp} = (x|x) \theta_\text{horiz}^\text{scat}.
\end{align}

\begin{figure}[t]
  \centering
  \includegraphics[width=\linewidth]{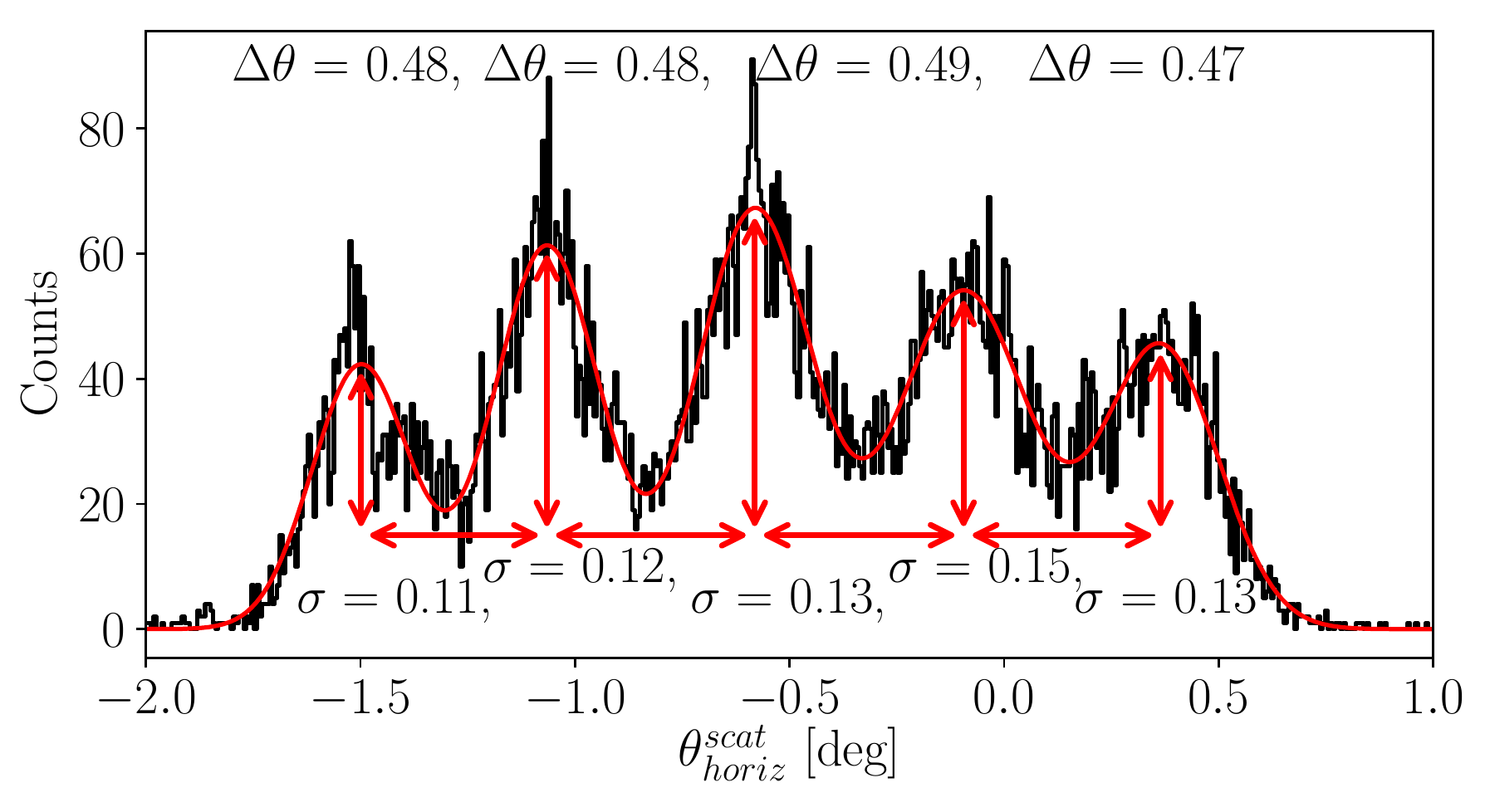}
  \caption[Sample $\theta_\text{horiz}^\text{scat}$ histogram obtained from a run using a sieve slit.]{Sample $\theta_\text{horiz}^\text{scat}$ histogram obtained from a run using the sieve slit mentioned in the main text. The slit is comprised of a grid of points separated by horizontal distances of $5$ mm at a distance of 585 mm from the focal plane itself. This constitutes a difference in scattering angle of approximately $0.489^\circ$ between the holes. Shown also in the figure is a $5$-peak fit to the data using Gaussian distributions, with the differences between the peak centers shown as well as the standard deviation of the distributions. The latter yields a measurement of the experimental angular resolution for extractions of the scattering angle, which is approximately $0.13^\circ$.}
  \label{sieve_slit_fig}
\end{figure}

Equation \eqref{angle_focal_plane} allowed for extraction of angular distribution data from the measured focal plane angles. To measure the experimental angular resolution as well as the transfer matrix element $(x|x)$, a sieve slit (a grid with collimated holes $5$ mm horizontally spaced and $12$ mm vertically spaced, located at the acceptance of the spectrograph) was utilized. The measured angular resolution was thus obtained to be approximately $0.13$ degrees for the scattering angle. A $\theta_\text{horiz}^\text{scat}$ histogram so obtained is shown in Fig. \ref{sieve_slit_fig}, alongside the multi-peak fit that allowed for a precise extraction of experimental scattering angle resolution.

The combination of the plastic scintillator signal with the MWDC signals served two purposes: the trigger signal was first generated by a coincidence between each pair of scintillators; later in the offline analysis, the energy deposited into the detector was utilized to characterize the particle identity (cf. Fig. \ref{background_excitation_spec}(a)). The timing signals and energy signals from the scintillators were digitized using a LeCroy FERA ({\bf F}ast {\bf E}ncoding and {\bf R}eadout {\bf A}DC) system and then fed to a LeCroy 1190 dual-port storage module within a VME crate \cite{tamii_DAQ,von-neumann-cosel_tamii_RCNP}.

The signals recorded by the MWDCs were pre-amplified and discriminated by REPIC RPA 220 cards and the signals were then digitized by CAEN V1190A multi-hit TDCs in a distinct VME crate; these signals were then stored within the memory buffer before being transmitted along with the signals from the scintillators to a local server via an Ethernet connection. The dead times associated with the hardware described here are typically $\sim 30$ $\mu$s/event. Further details on the electronics setup can be found in Ref. \cite{tamii_DAQ,von-neumann-cosel_tamii_RCNP}.

A unique feature of Grand Raiden which is of particular note is its composition of ion-optical capabilities which allow for the so-called \emph{vertical focusing} mode. With the spectrograph operating in such a setting, scattering events with momentum transfer occurring within the target chamber are coherently focused in the vertical direction, whereas events scattering elsewhere in the beamline --- thus constituting the instrumental background, to be discussed further shortly --- are over- or under-focused in the vertical direction at the focal plane. This feature, coupled with a vertical position sensitivity of the focal-plane detection system, lends itself usefully to accounting for the instrumental background in the offline analysis of the data.

To account for fluctuations in beam-intensity over the course of the experimental run, the integrated charge was utilized in the calculation of the experimental cross section data. Data were taken at each angle until sufficient statistics were acquired to reliably extract angular distributions to within $\sim 5\%$ uncertainty; in many cases, this goal was well-exceeded due to greater beam intensities (on the order of $>10$ nA), and statistical uncertainties were on the order of a few percent.

\section{Energy calibration} \label{experimental:energy_calibration}

The scattered particles were accepted into Grand Raiden and subsequently dispersed laterally according to their magnetic rigidity due to the presence of the dipole fields of the D1 and D2 elements of the spectrometer. The charged particle with mass $m$ and charge $q$ moving  relativistically ($\gamma = 1/\sqrt{1-v^2}$) through and perpendicular to a magnetic field $B$ will be bent through a circular arc of radius $\rho$. With the magnetic fields and velocity being perpendicular to one another, the Lorentz force results in uniform circular motion for a magnetic rigidity $[B \rho]$:
\begin{align}
  \frac{\gamma m v^2}{\rho} &= q v B \notag \\
  [B \rho] &= \frac{\gamma m v}{q} \notag \\
  &= \frac{p}{q}.
  \label{mag_rigidity}
\end{align}
Bearing this in mind as well as the geometry of Grand Raiden (cf. Fig. \ref{GR}) suggests that $\alpha$-particles which have higher momenta --- thereby having transferred less momentum to the recoiling target nucleus --- will have greater radii of curvature in their trajectories through the spectrometer. Inversely, $\alpha$-particles which have lower momenta will have smaller radii of curvature in their trajectory. The observable implication of this is that higher-momentum $\alpha$-particles, which have left the recoiling nucleus with a smaller excitation energy, will be detected further to the outside of the focal plane, whereas lower-momentum $\alpha$-particles will be detected closer to the interior edge of the focal plane. In this way, the position at which the incident $\alpha$-particles were detected at the focal plane is directly related to the particle momentum, and further, the excitation energy of the recoiling target nucleus.

\begin{figure}[t!]
  \centering
\includegraphics[width=\linewidth]{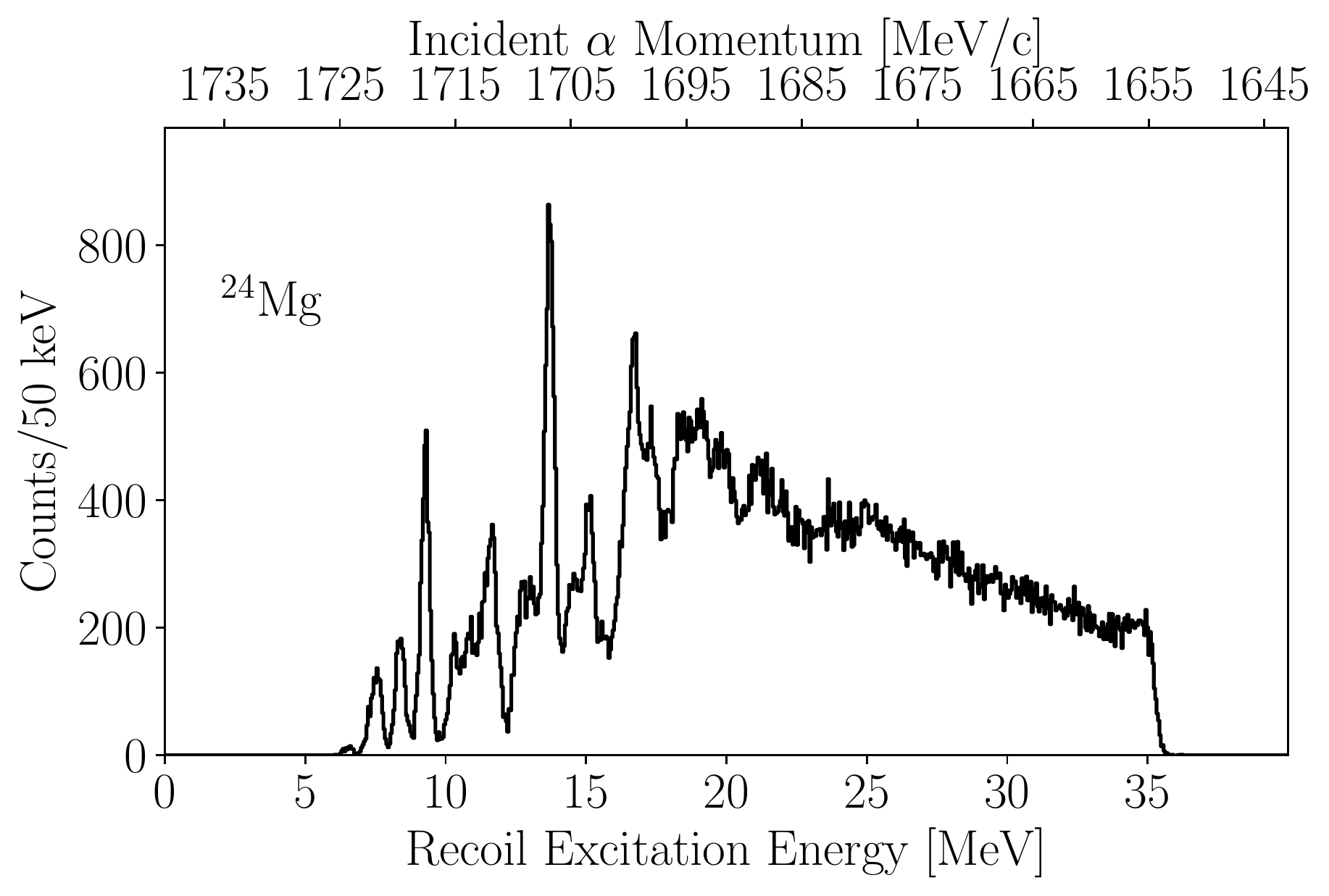}
\caption[Sample $0^\circ$ calibration spectra for \nuc{24}{Mg}($\alpha,\alpha^\prime$).]{Sample $0^\circ$ calibration spectra for \nuc{24}{Mg}($\alpha,\alpha^\prime$), showing both the calibrated excitation energy as well as the detected $\alpha$-particle momentum after calibration. }
\label{24Mg_calibration}
\end{figure}


The general procedure for the momentum calibration of the focal plane is to take a species with known excitations, and to calculate the corresponding $\alpha$-momenta which would leave the recoiling calibrant nucleus in its excited states. Upon doing so, one can then empirically determine, with great precision, the functional dependence of $p(X_\text{fp})$, which describes the dispersion of the focal plane momentum across the lateral dispersion plane. This is a property \emph{only} of the magnetic field settings of the spectrograph, and indeed scales directly with the field strength as seen in Eq. \eqref{mag_rigidity}.

For each experiment, \nuc{24}{Mg} was used as the calibration nucleus owing to high-quality reference spectra obtained from Ref. \cite{Kawabata_private}. Figure \ref{24Mg_calibration} shows a sample post-calibration spectrum of \nuc{24}{Mg} at the zero-degree setting of Grand Raiden.

\begin{figure}[t!]
  \centering
\includegraphics[width=0.75\linewidth]{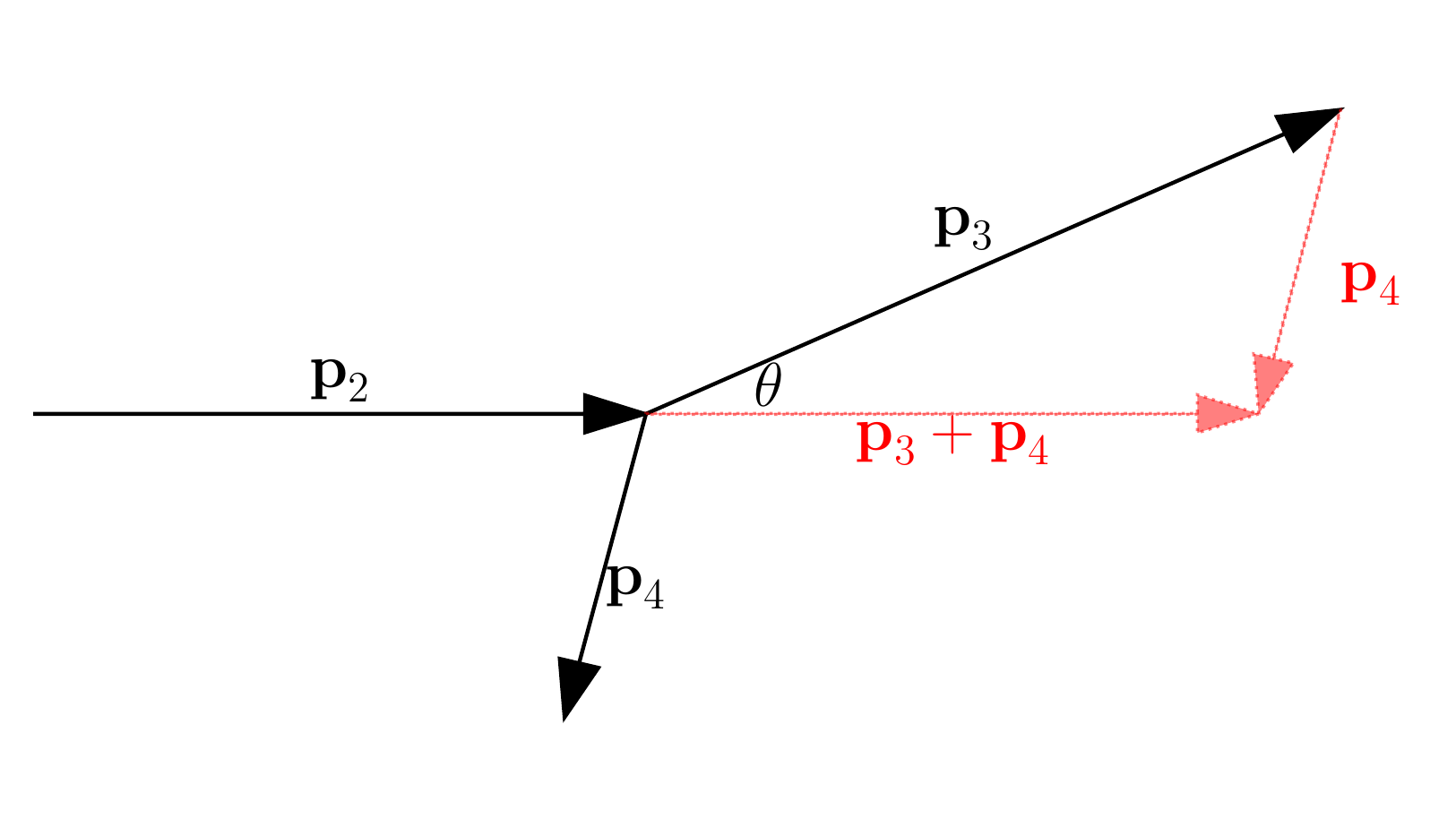}
\caption[Sketch of the forward-kinematics scattering process.]{Sketch of the forward-kinematics scattering process in the laboratory frame, with the scattering angle $\theta$ shown.}
\label{kinematics_figure}
\end{figure}

For a two-body reaction $A_1 + A_2 \rightarrow A_3 + A_4$ with the target nucleus $A_1$ starting from rest, the kinematics are sketched in Fig. \ref{kinematics_figure}.  With the incident projectile leaving the target nucleus with a recoil 4-momentum $p_4$ and leaving with itself a 4-momentum of $p_3$, the relativistic conservation of 4-momentum holds. The initial momentum of the system is:
\begin{align}
  p_i &= p_1 + p_2 \notag \\
  &= (m_1,\vec{0}) + (E_2,\vec{p}_2).
\end{align}
Similarly, the final momentum of the system is:
\begin{align}
  p_f & = p_3 + p_4 \notag \\
  &= (E_3,\vec{p}_3) + (E_4,\vec{p}_4).
\end{align}

The initial momentum of the incident $\alpha$ beam is known from the beam energy at the scattering site:
\begin{align}
  E_1^2 &= \vec{p}_1^2 + m_1^2 \notag \\
  (E_\text{beam}+m_1)^2 &= \vec{p}_1^2 + m_1^2 \notag \\
  |\vec{p}_1| & = \sqrt{E_\text{beam}^2 + 2 m_1 E_\text{beam}}
\end{align}

With the aforementioned calibration reference spectra of \nuc{24}{Mg}, the momentum $\vec{p}_3$ is directly measured and in so doing, the total energy $E_3$ of the ejectile is also measured:
\begin{align}
  E_3 = \sqrt{\vec{p}_3^2 + m_3^2}.
  \label{projectile_interval}
\end{align}

Furthermore, the total 3-momentum of the ejectile and recoiling target is conserved relative to that of the incident projectile and stationary target. Inspection of Fig. \ref{kinematics_figure} shows that the 3-momenta $\vec{p}_3$, $\vec{p}_4$, and $\vec{p}_2 = \vec{p}_3+\vec{p}_4$ constitute a triangle for which the law of cosines defines the recoil momentum in terms of the laboratory-frame scattering angle of $\vec{p}_3$ relative to the beam axis:
\begin{align}
  \vec{p}_4^2 &= \left(\vec{p}_3 + \vec{p}_4 \right)^2 + \vec{p}_3^2 - 2 \left(\vec{p}_3 + \vec{p}_4 \right) \cdot \vec{p}_3 \notag \\
  &= \vec{p}_2^2 + \vec{p}_3^2 - 2 | \vec{p}_2 | | \vec{p}_3| \cos \theta. \label{recoil_momentum}
\end{align}

Inspection of the energy component of the 4-momentum conservation equations yields:
\begin{align}
  m_1 + (m_2 + E_\text{beam}) &= E_3 + E_4, \label{recoil_mass}
\end{align}
and the corresponding invariant interval for $p_4$ yields
\begin{align}
  E_4^2 & = \vec{p}_4^2 + m_4^2. \label{recoil_interval}
\end{align}

In terms of the excitation energy $E_x$, $m_4 = m_1+E_x$, and Eq. \eqref{recoil_interval} yields the excitation in terms of the previously derived quantities in Eqs. \eqref{recoil_momentum} and \eqref{recoil_mass}:
\begin{align}
  E_x = \sqrt{E_4^2 - \vec{p}_4^2} - m_2.
  \label{excitation_energy_formula}
\end{align}

Equations \eqref{projectile_interval} -- \eqref{excitation_energy_formula} were applied event-wise to each data set using the $p_3(X_\text{fp})$ function constrained by the reference \nuc{24}{Mg} spectra taken at the same magnetic field and angular settings of the spectrograph. Using Eq. \eqref{mag_rigidity}, any fluctuation in the dipole field strengths between runs were accounted for in the extracted $p_3(X_\text{fp})$ function via proportional scaling of the fit functions by the dipole strengths. Finally, despite the energy losses through the target foil being small ($\sim 10$s of keV), they were nonetheless accounted for using the statistical SRIM framework \cite{SRIM}.

\section{Data acquisition and reduction}

\begin{figure}[t!]
  \centering
\includegraphics[width=0.90\linewidth]{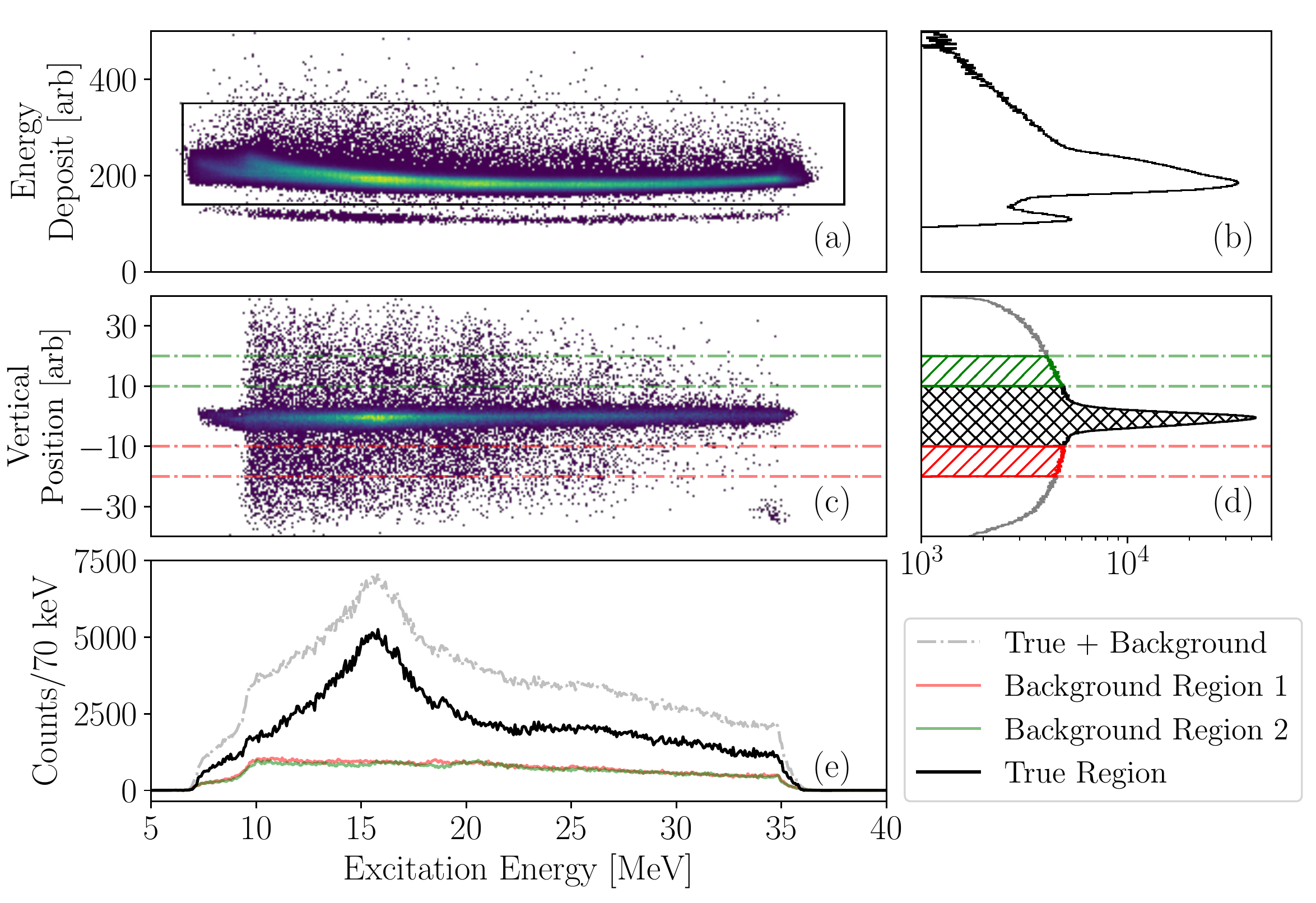}
\caption[Delineation of the offline data reduction process and gates utilized in this work.]{\textbf{(a)} Particle identification spectrum, showing the energy deposited into the plastic scintillator against excitation energy. The enclosed, strong line shown corresponds to $\alpha$ events, which were gated on in offline analysis, while the excluded weaker line is comprised of events corresponding to proton, deuteron, and triton detections. \textbf{(b)} Projection of the scintillator energy deposition histogram onto the vertical axis. \textbf{(c)} Two-dimensional histogram displaying the correlation between the energy-calibrated horizontal focal-plane position versus the vertical focal-plane position after application of the particle-ID gate of (a). \textbf{(d)} Vertical focal-plane position of (c) projected onto the vertical axis. \textbf{(e)} Excitation-energy spectra for each of the hatched regions in (d), as well as the subtracted spectrum which is comprised essentially of instrumental-background-free $\alpha$ events. Figure adapted from Ref. \cite{KBH_EPJA}.}
\label{background_excitation_spec}
\end{figure}

Figure \ref{background_excitation_spec} shows a series of plots which delineate the steps taken in the data reduction for these nuclei. The particle identification was completed via examination of the energy deposited into scintillators located at focal plane. Figures \ref{background_excitation_spec}(a) and \ref{background_excitation_spec}(b) show, respectively, the correlation between energy-deposition and excitation energy as well as the one-dimensional energy-loss histogram. The enclosed region in (a) corresponds to $\alpha$ events which were gated on in the offline analysis discussed hereafter, while the excluded events correspond to proton, deuteron, and triton detections.

Figure \ref{background_excitation_spec}(c) and \ref{background_excitation_spec}(d) show typical vertical focal-plane position spectra encountered during the data reduction. Operation of Grand Raiden in vertical focusing mode allows for true events, which originate from scattering off of the target, to be coherently focused along the vertical plane. In contrast, events originating up- or down-stream relative to the scattering chamber due to, for example, scattering off of the beamline, collimator, or the entrance slit and walls of the spectrograph, are over- or under-focused in the vertical direction. In Fig. \ref{background_excitation_spec}(d), the black doubly-hatched region corresponds to events which are focused to the median of the vertical focal-plane position and thus correspond to a combination of ``true'' events and those arising from instrumental background effects. The red and green singly-hatched regions correspond to gates on the off-median focal-plane positions in the spectra, which arise purely from instrumental background. This property of the measurement allows for a nearly complete and unambiguous subtraction of instrumental background.

The background contribution to the spectra is largest near $\theta_\text{GR} = 0^\circ$, as the elastic cross sections are high and thus, elastically scattered particles which subsequently scatter off elements in the beamline can contribute to the background at this spectrometer setting. Further, we make the point that the various background gates shown in Fig. \ref{background_excitation_spec}(d) result in nearly identical background contributions to the excitation-energy spectra, as evidenced in Fig. \ref{background_excitation_spec}(e).

\section{Angular distribution extraction} \label{experimental:angular_distribution_extraction}

\begin{figure}[t!]
  \centering
  \includegraphics[width=\linewidth]{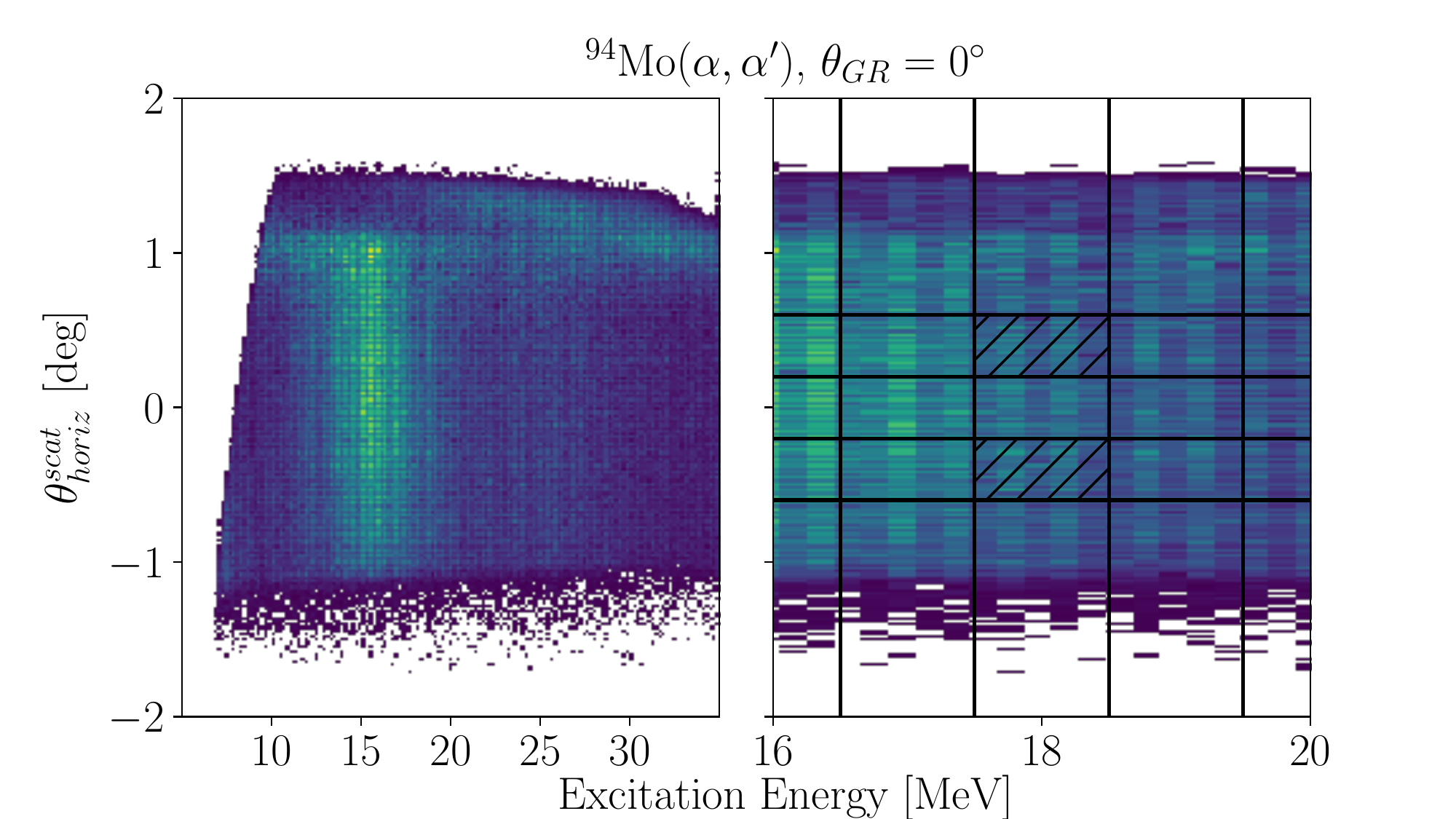}
  \caption[Sample two-dimensional $\theta_\text{horiz}^\text{scat}$ v. $E_x$ histogram, with typical regions-of-interest for count integration shown.]{Left: sample background-subtracted histogram showing the dependence of $\theta_\text{horiz}$, relative to $\theta_{GR} = 0^\circ$, on the calibrated excitation energy for \nuc{94}{Mo}. Right: sample region-of-interest wherein the bins used in the counts extraction are shown. Shaded is a typical bin region for the $0^\circ$ dataset; for $0^\circ$, the region was restricted to $\pm 0.6^\circ$ relative to the Grand Raiden angular setting with $3$ bins subdividing the angular range. For finite angle measurements, $\pm 0.8^\circ$ was used with $4$ angular bins.}
  \label{thscat_excitation_energy_zoomed}
\end{figure}

Figure \ref{thscat_excitation_energy_zoomed} shows a sample two-dimensional excitation energy spectrum extracted for \nuc{94}{Mo}, with the scattering angle $\theta_\text{horiz}^\text{scat}$ shown on the $y$-axis. The left panel shows this for the entire lateral acceptance of the spectrograph, whereas the right panel zooms in on sample regions-of-interest for a hypothetical $1$-MeV bin. After generating these histograms, each two-dimensional bin was integrated to determine the yield, $Y$, from which the experimentally-extracted double-differential cross section can be determined:
\begin{align}
  \frac{\diff^2 \sigma}{\diff E_x \diff \Omega} & = \frac{Y}{\Delta E_x \Delta \Omega N_\text{incident} n_\text{target} \varepsilon_\text{total} }
  \label{cross section defined}
\end{align}

The solid angle, $\Delta \Omega = \Delta \theta \, \Delta \phi$, is determined by the size of the horizontal acceptance, and the $\theta_\text{horiz}^\text{scat}$ bin width chosen in the offline analysis; the vertical acceptance of the collimator was fixed throughout the acquisition of any given data set. The energy bin width is also chosen manually in the offline analysis ($500$ keV for \nuc{94-100}{Mo}, $200$ keV for \nuc{40-44}{Ca}, and 1 MeV for \nuc{48}{Ca} owing to poorer statistical uncertainties). The number of incident particles, $N_\text{incident}$, is readily determined from the charge state of the beam and the integrated beam current which is measured during each experimental run, whereas the number of target particles per unit area, $n_\text{target}$, is determined from the areal density of the target and the molar mass of the species. Finally, the total efficiency $\varepsilon_\text{total}$ is calculated in two parts:
\begin{align}
  \varepsilon_\text{total} = \varepsilon_\text{DT} \, \varepsilon_\text{MWDC} ,
\end{align}

\begin{figure}[t!]
  \centering
  \includegraphics[width=\linewidth]{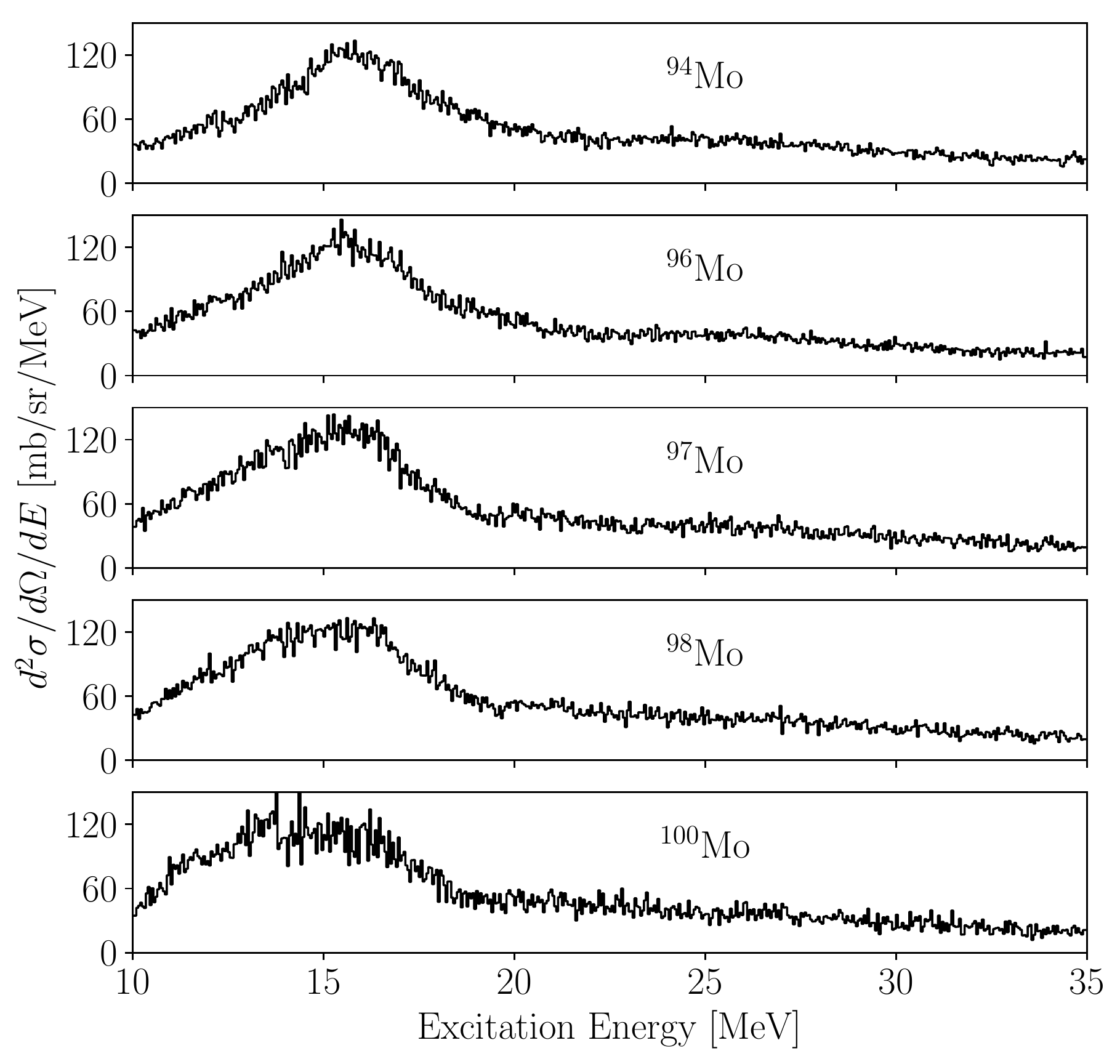}
  \caption{Background-subtracted energy spectra for \nuc{94-100}{Mo}($\alpha,\alpha^\prime)$ at $\Theta_\text{avg} = 0.69^\circ$, with $E_\alpha = 386$ MeV.}
  \label{molly_zeroDegSpec}
\end{figure}

\begin{figure}[t!]
  \centering
  \includegraphics[width=\linewidth]{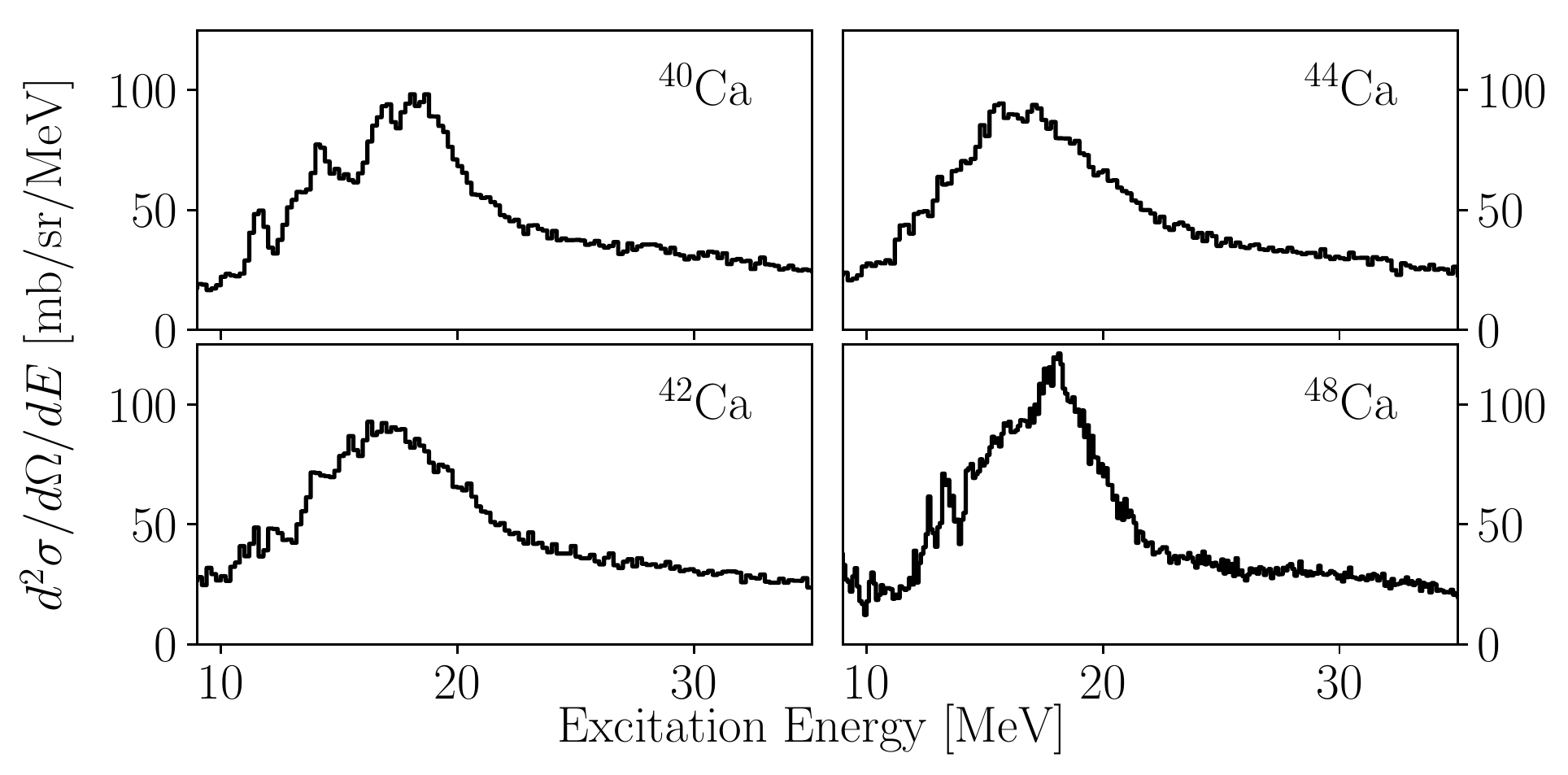}
  \caption{Background-subtracted energy spectra for \nuc{40-48}{Ca}($\alpha,\alpha^\prime)$ at $\Theta_\text{avg} = 0.69^\circ$, with $E_\alpha = 386$ MeV.}
  \label{calcium_zeroDegSpec}
\end{figure}

wherein $\varepsilon_\text{DT}$ arises due to \emph{dead time} of the data acquisition system. The efficiency $\varepsilon_\text{MWDC}$ arises due to occasional losses of scattered $\alpha$-particles in the middle of their trajectory through the MWDCs.

As each anode-wire plane serves essentially as an independent detector for the incident $\alpha$-particle, the total efficiency of the $P=\left\{X_1,X_2,U_1,U_2 \right\}$ system of planes is the product of the independent efficiencies of each element:

\begin{align}
  \varepsilon_\text{MWDC} & = \prod_{p \in P} \epsilon_{p}.
\end{align}


To calculate the individual $\epsilon_p$, each event was first enumerated according to the order in which it was registered by the data acquisition system. The set containing these numbers will be denoted $\mathcal{N}$. With this, one can define the set of events which registered signals in any given anode-wire plane via:
\begin{align}
  N_{p} = \left \{ n \in \mathcal{N} \; | \; \text{event \#$n$ is registered in $p$}   \right\}
\end{align}

The $\epsilon_p$ are defined in the following way:

\begin{align}
  \epsilon_p & =      \bigg | \bigcap_{p^\prime \in P} N_{p^\prime}  \bigg |  \bigg / \bigg|  \bigcap_{p^\prime \in \left[ P - \left\{p\right\} \right] } N_{p^\prime} \bigg |,
\end{align}
which is to say that it is the ratio of the number of events which hit \emph{all} anode planes, to the number of events which hit the \emph{other} anode planes (and \emph{possibly} the $p^\text{th}$ plane). The values for $\varepsilon_\text{MWDC}$ were calculated in this way after each run, and typical values were between $60\%-75\%$.

The procedure delineated here allowed for a precise extraction of cross section data over the excitation energy ranges of the spectrographic acceptance for each experiment. The assignment of angles for each extracted cross section was made after averaging over the finite horizontal and vertical acceptance of the collimator; the expression for $\Theta_\text{avg}$ is provided in Appendix \ref{averaging_appendix}, along with its corresponding derivation. The forward-angle energy spectra for \nuc{94-100}{Mo} and \nuc{40-48}{Ca} are respectively presented in Figs. \ref{molly_zeroDegSpec} and \ref{calcium_zeroDegSpec} for the forward-most angular bins which correspond to an average spectrographic scattering angle of $0.69^\circ$.

A comprehensive presentation of the extracted angular distributions (in addition to their multipole decompositions; this will be discussed in the following chapter) for each energy bin in each nucleus examined in this dissertation is made in Appendix \ref{MDA_results}.

\subsection{Hydrogen and oxygen contamination}

\begin{figure}[t!]
  \includegraphics[width=\linewidth]{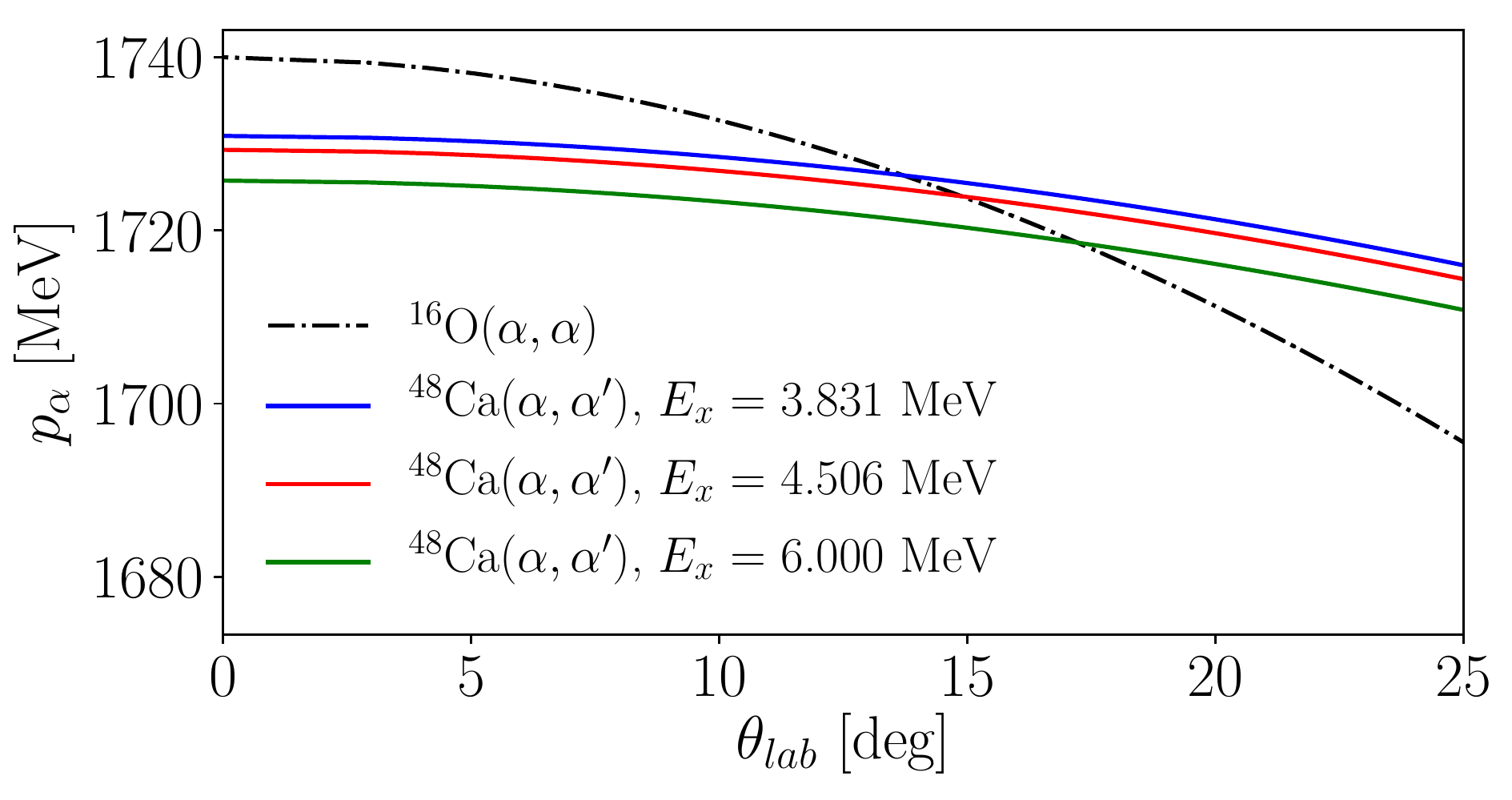}
  \caption[Momentum of scattered $\alpha$ particle as a function of laboratory-frame scattering angle for various reaction channels.]{Momentum of scattered $\alpha$ particle as a function of laboratory-frame scattering angle for various reaction channels. The dot-dashed line corresponds to the elastic scattering off of \nuc{16}{O}, whereas the solid lines correspond to inelastic scattering off of \nuc{48}{Ca} with different recoil excitation energies. One should note that with increasing scattering angle, the kinematics of the elastic scattering of \nuc{16}{O} overlap with the kinematics of increasingly-high excitation energy inelastic scattering channels.}
  \label{contaminant_figure}
\end{figure}

During each experiment, small amounts of impurities were found on the foils which demanded care for one to account for them correctly in the offline analysis. Many of the angular distributions presented in Appendix \ref{MDA_results} show conspicuous gaps in the experimental cross sections in certain panels which arose due to the incidence of \nuc{1}{H}($\alpha,\alpha$) elastic scattering channel onto the focal plane. These events are, of course, not filtered on the basis of the particle-identification signals, and originate from the scattering chamber; as a result, they are necessarily entangled within our measured spectra. As the ejectile momentum carries a different angular dependence owing to the kinematics of Section \ref{experimental:energy_calibration}, however, ejected $\alpha$ particles from lighter-mass contaminant channels are dispersed to the low-momentum side of the focal plane with increasing scattering angle, as shown in Fig. \ref{contaminant_figure}.

\begin{figure}[t!]
  \includegraphics[width=\linewidth]{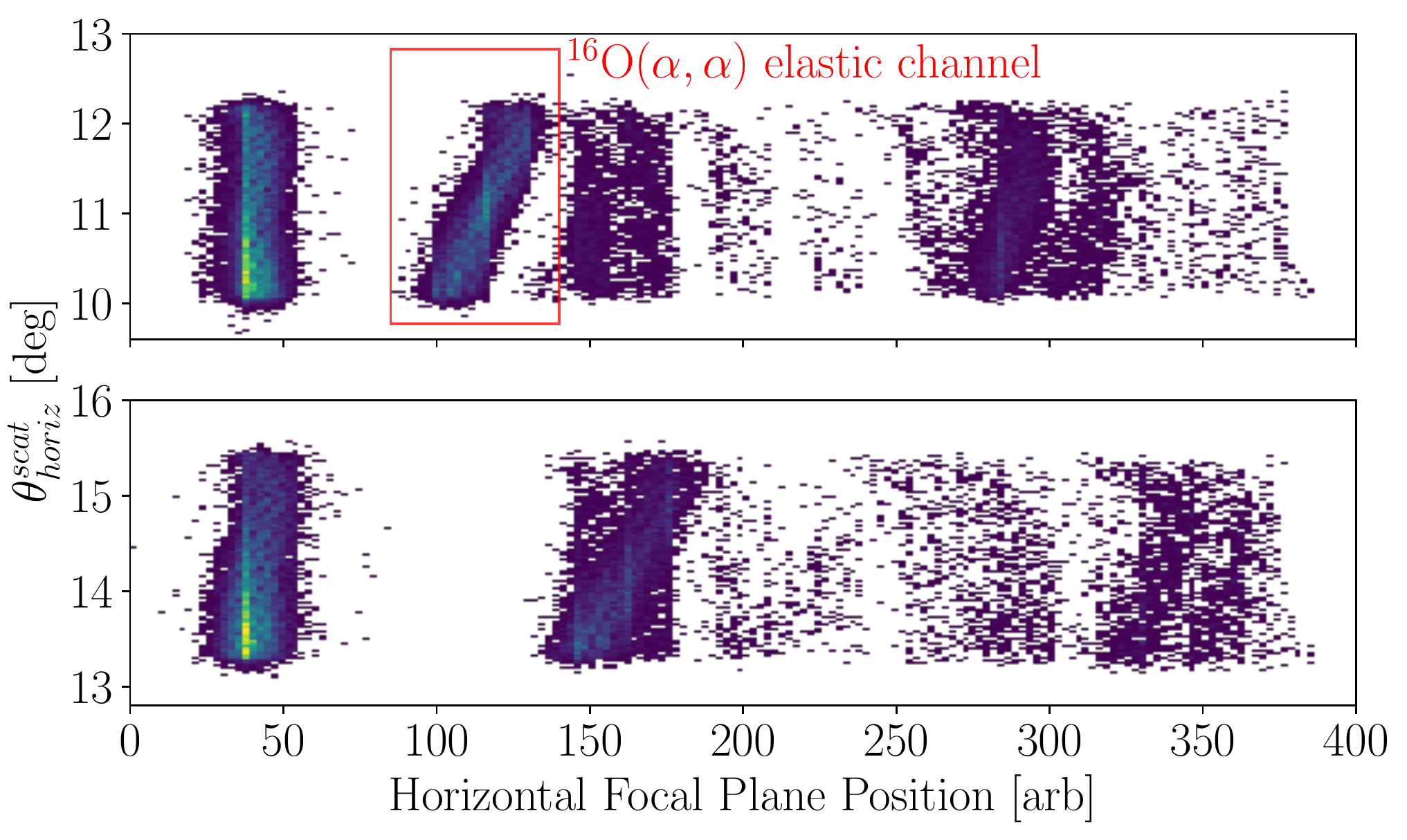}
  \caption[Two-dimensional ($\alpha,\alpha^\prime)$ scattering spectra measured from a partially oxidized \nuc{48}{Ca} target foil.]{Two-dimensional ($\alpha,\alpha^\prime)$ scattering spectra measured from a partially oxidized \nuc{48}{Ca} target foil. Top: scattering spectra at a central Grand Raiden setting of $11.2^\circ$. The excitation furthest to the left corresponds to the \nuc{48}{Ca}($\alpha,\alpha)$ elastic scattering channel.  The  measured line for the \nuc{16}{O} contamination is isolated, and the discrepant angular character relative to the kinematics of \nuc{48}{Ca}($\alpha,\alpha^\prime)$ is apparent. Bottom: same as above, but for a central Grand Raiden setting of $14.4^\circ$. One sees that the \nuc{16}{O}($\alpha,\alpha)$ elastic scattering channel has migrated to overlap with the \nuc{48}{Ca}($\alpha,\alpha^\prime$) channel that corresponds to the $0_1^+ \rightarrow 2_1^+$ transition of \nuc{48}{Ca} (see the calculations of Fig. \ref{contaminant_figure}).}
  \label{experimental_contaminant_figure}
\end{figure}

This predictable angular dependence for the presence of the contaminant allows for both identification and removal of each contaminant from the experimental angular distributions prior to the optical model or multipole decomposition analyses. In most cases, the removal of the contaminant can be implemented by simply omitting data points in small neighborhoods surrounding the loci of the contaminant within the inelastic angular distribution, as shown in Figs. \ref{contaminant_figure} and \ref{experimental_contaminant_figure}. In the case of hydrogen contamination, the elastic scattering off of the protons has a cross section which greatly dominates the inelastic excitation of the intended target nucleus, and so it is straightforward to identify and remove the affected data points. Further, in the case of hydrogen contamination, there is no collective structure to excite, and so the elastic channel is all that must be accounted for. The effects of this treatment are seen in many of the inelastic angular distributions which will be presented in the next section, and with completeness in Appendix \ref{MDA_results}.

In the case of \nuc{48}{Ca}, however, the \nuc{16}{O} contaminant was present with a substantial effective thickness. A subtraction of the contribution of \nuc{16}{O} to the experimental spectra was completed prior to further analysis. Figure \ref{experimental_contaminant_figure} shows the two-dimensional experimental spectra at various angles over the horizontal acceptance of the focal plane. The strong, vertical lines shown in each panel of this figure correspond to events which obey the kinematics of a recoiling \nuc{48}{Ca} nucleus; those which are askew originate from scattering off of a contaminant nucleus. A cross reference between the intersection of the enclosed and skewed line and the first excited state of \nuc{48}{Ca}, and the angular dependence of the ejected $\alpha$-momentum shown in Fig. \ref{contaminant_figure} indicates that the strong and skewed excitation that is shown in the former is indeed the result of elastic scattering off of \nuc{16}{O}.

This presence of \nuc{16}{O} is non-negligible and would, if it were to remain unaccounted for, pose major difficulty in the extraction of the ISGMR strength for \nuc{48}{Ca}. To mitigate this contaminant, the cross sections for elastic scattering of $100$ MeV/u $\alpha$-particles from \nuc{16}{O} were taken from Ref. \cite{wakasa_oxygen} and subsequently used to estimate the effective target thickness of \nuc{16}{O} using the measured counts from the \nuc{16}{O} elastic scattering channel at various angles (cf. Eq. \eqref{cross section defined}). This analysis indicated in an effective target thickness of $\sim 0.3$ mg/cm$^2$ present on the foil.

The \nuc{16}{O} spectra have substantial structure and excited states which extend into the giant resonance region, and therefore in addition to needing to account for the elastic scattering of the $\alpha$-particles off of \nuc{16}{O}, one needs to further account for the inelastic scattering channels. To this end, high-resolution \nuc{16}{O}($\alpha,\alpha^\prime$) cross-sections were acquired from Ref. \cite{itoh_private}.\interfootnotelinepenalty=10000 \footnote{We are very grateful for the measured \nuc{16}{O} spectra, which were very kindly provided by Prof. Masatoshi Itoh.}  These spectra were measured using the same beam energy and at the same scattering angles. After transformation of the kinematics to be consistent with the measured excitation energies for the \nuc{48}{Ca} inelastic spectra, these data --- combined with the effective target thickness extracted from the comparison of the measured elastic channel with the cross sections of Ref. \cite{wakasa_oxygen} --- allowed for one to calculate the contribution of \nuc{16}{O} to the inelastic cross sections at each of the angles.

\begin{figure}[t!]
  \includegraphics[width=\linewidth]{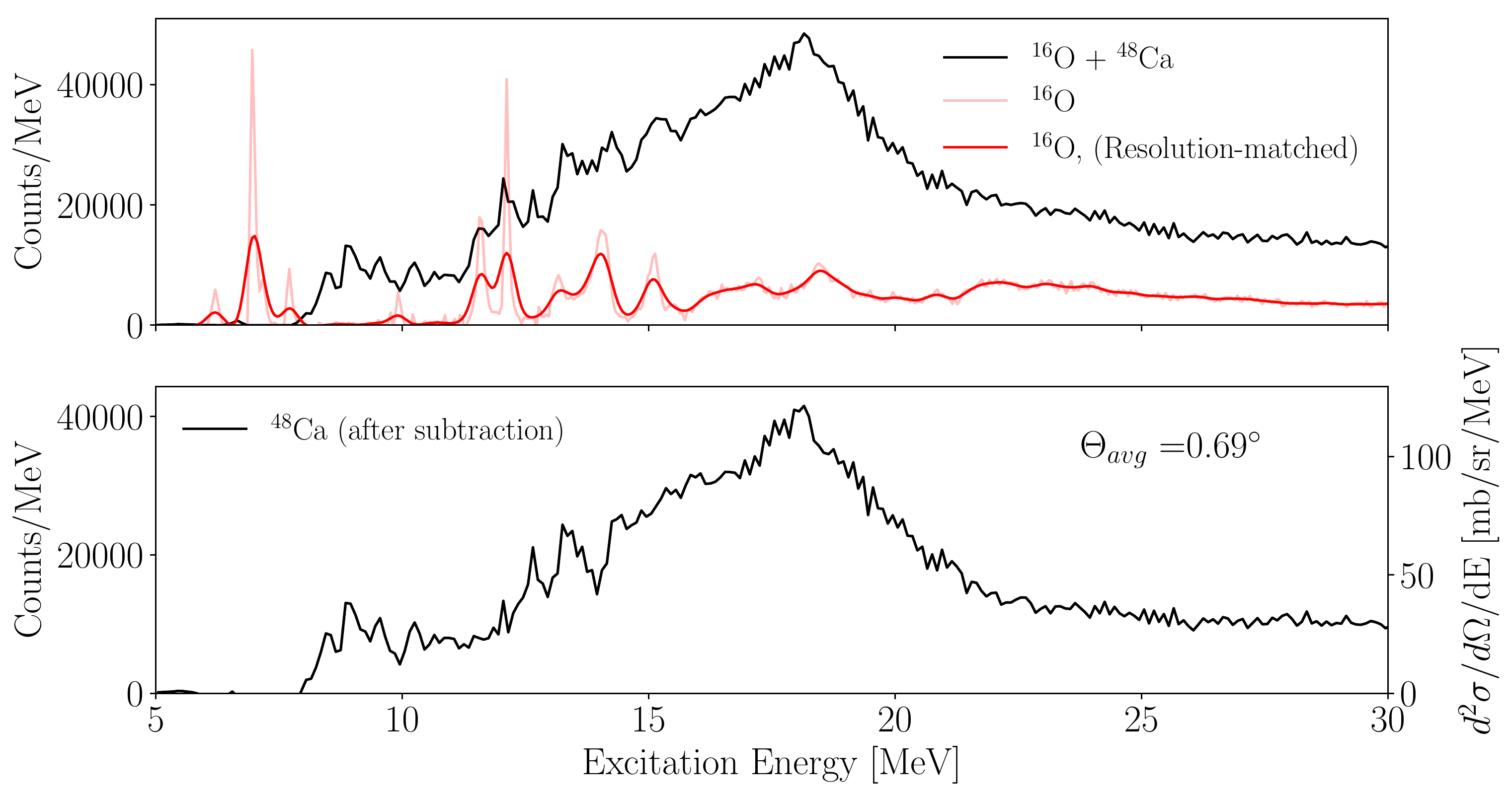}
  \caption[Background-subtracted experimental spectra for \nuc{48}{Ca} at an average spectrographic scattering angle of $\Theta_\text{avg}=0.69^\circ$.]{Background-subtracted experimental spectra for \nuc{48}{Ca} at an average spectrographic scattering angle of $\Theta_\text{avg}=0.69^\circ$. Top: total spectra, including the contribution from the \nuc{16}{O} contamination to the overall measured counts, as well as the calculated contribution by \nuc{16}{O} using the high-resolution inelastic \nuc{16}{O}($\alpha,\alpha^\prime$) cross-section data from Ref. \cite{itoh_private}, using the effective \nuc{16}{O} target thickness extracted from the measured \nuc{16}{O} elastic channel and the cross section data of Ref. \cite{wakasa_oxygen}. Also shown in the top panel is the resolution-matched spectra wherein a Gaussian filter of width $100$ keV was used to smooth the data from Ref. \cite{itoh_private} to match the present experimental energy resolution. Bottom: resulting post-subtraction spectrum for only \nuc{48}{Ca}($\alpha,\alpha^\prime)$ events.}
  \label{oxygen_contaminant_subtraction}
\end{figure}

Figure \ref{oxygen_contaminant_subtraction} shows the results of this subtraction for the spectrum corresponding to $\Theta_\text{avg}=0.69^\circ$. As the reference \nuc{16}{O} spectra were measured with higher resolution than the spectra acquired in the present experiment, it was necessary to convolve the present spectra with a Gaussian smearing function to match the resolutions before subtraction:

\begin{align}
  \widetilde{\sigma} (E_x,\Theta_\text{avg} ) & = \frac{1} { \sqrt{4\pi \omega^2} } \int_{-\infty}^{\infty}  \diff E^\prime \; \exp \left( \frac{(E_x-E^\prime)^2}{2 \omega^2} \right) \sigma(E^\prime,\Theta_\text{avg} ).
\end{align}

The values of the smearing width $\omega$ was tuned for each spectra comparison, but typically was on the order of $\sim 100$ keV. This facilitated a smooth subtraction of the \nuc{16}{O} contributions from the measured inelastic spectra. The additional uncertainties in this contaminant subtraction arising from, for example, the effective target thickness and the resolution matching between the experimental data, were accounted for by increasing the size of the energy bin-width --- from 200 keV (used for \nuc{40,42,44}{Ca}) to 1 MeV --- that was used for the extraction of the experimental cross sections. The culmination of this procedure resulted in smooth angular distributions for the \nuc{48}{Ca}($\alpha,\alpha^\prime$) reaction over the energy and angular ranges necessary to extract the giant resonance strength distributions, as shown in Appendix \ref{MDA_results}.

%

%
%
%
%
%
%
%
%
%
%

%
%

\chapter{Analysis of extracted angular distributions}
\label{Data Analysis}

The general post-reduction analysis procedures were fundamentally identical for all experiments. After extraction of the elastic and inelastic angular distributions using the data reduction methods of Section \ref{experimental:angular_distribution_extraction}, the analysis of the angular distributions can be partitioned into two branches which will be discussed in this chapter. The first branch deals with preparing for the multipole decomposition of the angular distributions by constraining the optical and transition potentials (discussed further in Section \ref{direct reaction theory}) that are used in the calculations of the characteristic angular distributions of the giant resonance transitions. The second branch involves employing the aforementioned characteristic angular distributions to isolate the contributions of specific multipole transitions to the experimentally-measured cross section data, using the so-called \emph{multipole decomposition analysis} \cite{first_MDA}.

In this chapter, the choice of ansatz for the functional form of the optical potential will be presented and the means by which its free parameters were constrained on the basis of experimental data will be described. The details of the multipole decomposition will then be presented, which heavily draws from the theory presented in Chapter \ref{theory}.

\section{Preparation for multipole decomposition}

\subsection{Optical model extraction}

As has been made clear, in order to characterize the features of the giant resonance strength distributions, it is necessary to have an optical model with which to perform Distorted Wave Born Approximation (DWBA) calculations. For these purposes, the angular distributions for the elastic scattering channels were extracted, as well as the inelastic scattering channels corresponding to pure transitions characterized by a unique value for the angular momentum transfer (\emph{e.g.} the ``low-lying'' excited states, $2_1^+$, $3_1^-$ \ldots).

The optical model code \texttt{PTOLEMY} was used for the optical model and DWBA calculations, using an optical model of the general form
\begin{align}
  U(r) = \mathcal{V}_\text{Coul}(r) - \mathcal{V}_\text{vol}(r) - i \mathcal{W}_\text{vol}(r),
  \label{optical_model_defined}
\end{align}
within which $\mathcal{V}_\text{Coul}$ is a point-sphere Coulomb potential, and $\mathcal{V}_\text{vol}$ and $\mathcal{W}_\text{vol}$ originate from a hybrid single-folding optical model with a modified density dependence \cite{khoa_satchler_single_folding}. For this model, the imaginary volume potential is taken to be the shape of a Woods-Saxon function:
\begin{align}
  \mathcal{W}_\text{vol} &= \frac{W_\text{vol}}{1+\exp\left(\frac{r-R_I}{a_I}\right)}.
\end{align}
The real volume potential takes the form of a realistic point-nucleon Gaussian interaction $\bar{v}_G$ which is then folded with the product of an  empirical model for the target nuclear density, $\rho(r^\prime)$, and a modified density dependence $f(\rho)$\interfootnotelinepenalty=10000\footnote{One should note that, taken out of context, the quantity $V_\text{vol}$ has a somewhat limited meaning. It is associated inextricably with the form of Eq. \eqref{real_shape_defined} used in the analysis procedure; if one has an additional prefactor, for example, in the integral of Eq. \eqref{real_shape_defined} or the definitions of Eq. \eqref{density_dependencies}, then the $V_\text{vol}$ found in the fitting procedure will be changed accordingly so as to keep the total potential depth --- the more meaningful quantity --- constant. In the present case, the values of the folding integral of Eq. \eqref{real_shape_defined} --- without the prefactor of $V_\text{vol}$ --- are typically of depth $\sim 2-2.5$ MeV, and so the total depth of the real volume potential ranges typically between 70-100 MeV in our analyses.}:
\begin{align}
  \mathcal{V}_\text{vol}(r) = V_\text{vol} \int \diff ^3 \vec{r^\prime} \rho(r^\prime) f(\rho^\prime) \bar{v}_G(s).
  \label{real_shape_defined}
\end{align}

Here, $V_\text{vol}$, $W_\text{vol}$, $R_I$ and $a_I$ are free parameters in the optical model parameter (OMP) set which were searched upon in the fitting procedure. The inter-particle separation $s= | \vec{r} - \vec{r^\prime}|$, and
\begin{align}
  f(\rho^\prime) &= 1 - \zeta \rho^\beta(r^\prime) \notag \\
  \bar{v}_G(s) &= \exp\left(-s^2/t^2\right).
  \label{density_dependencies}
\end{align}
The parameters $\zeta = 1.9$ fm$^2$, $\beta = 2/3$, and $t = 1.88$ fm were adopted from Ref. \cite{khoa_satchler_single_folding}, along with the extension to the calculations of transition form factors within this model framework. The target nuclear densities $\rho(r^\prime)$ were taken to be two-parameter Fermi distributions and are available from Ref. \cite{fricke_ADNDT}; those are shown, for completeness, in Table \ref{OMP_table}.

\begin{table}[t!]
\centering
\caption[Optical-model parameters for calcium and molybdenum nuclei]{Optical-model parameters for calcium and molybdenum nuclei}
\resizebox{\linewidth}{!}{
  \begin{tabular}{@{}ccccccccccccc@{}}
  \toprule
               & \multicolumn{4}{c}{Optical Model}                          &  & \multicolumn{2}{c}{Density \cite{fricke_ADNDT}} &  & \multicolumn{2}{c}{$2_{1}^+$ \cite{raman_ADNDT}}                         & \multicolumn{2}{c}{$3_{1}^-$ \cite{kibedi_ADNDT}}                         \\ \cmidrule(lr){2-5} \cmidrule(lr){7-8} \cmidrule(l){10-13}
        & $V_\text{vol}$  & $W_\text{vol}$ & $R_I$  & $a_I$  &  & $c$            & $a$           &  & $E_{x}$  & $B(E2)$  & $E_{x}$  & $B(E3)$  \\
        & [MeV]  &  [MeV] &  [fm] & [fm] &  & [fm]           &  [fm]          &  &[MeV] &  [$\text{e}^{2} \text{b}^{2}$] &  [MeV] &  [$\text{e}^{2} \text{b}^{3}$] \\ \cmidrule(r){1-5} \cmidrule(lr){7-8} \cmidrule(l){10-13}
  \nuc{42}{Ca} & 37.4                 & 31.6                 & 4.53       & 0.99      &  & 3.77              & 0.523             &  & 1.524         & 0.042                                 & 3.446         & 0.0081                                 \\ 
  \nuc{44}{Ca} & 37.4                 & 31.1                 & 4.64       & 0.99      &  & 3.75              & 0.523             &  & 1.116         & 0.047                                 & 3.307         & 0.0076                                 \\ 
  \nuc{48}{Ca} & 41.2                 & 32.7                 & 4.82       & 0.94      &  & 3.72              & 0.523             &  & 3.831         & 0.021                                 & 4.506         & 0.0063                                 \\ 
  \nuc{98}{Mo} & 30.5                 & 47.2                 & 5.19       & 1.09      &  & 5.11              & 0.523             &  & 0.787         & 0.267                                 & 2.017         & 0.133                                 \\ \bottomrule
  \end{tabular}
  }
\label{OMP_table}
\end{table}

\begin{figure}[t!]
  \centering
  \includegraphics[width=\linewidth]{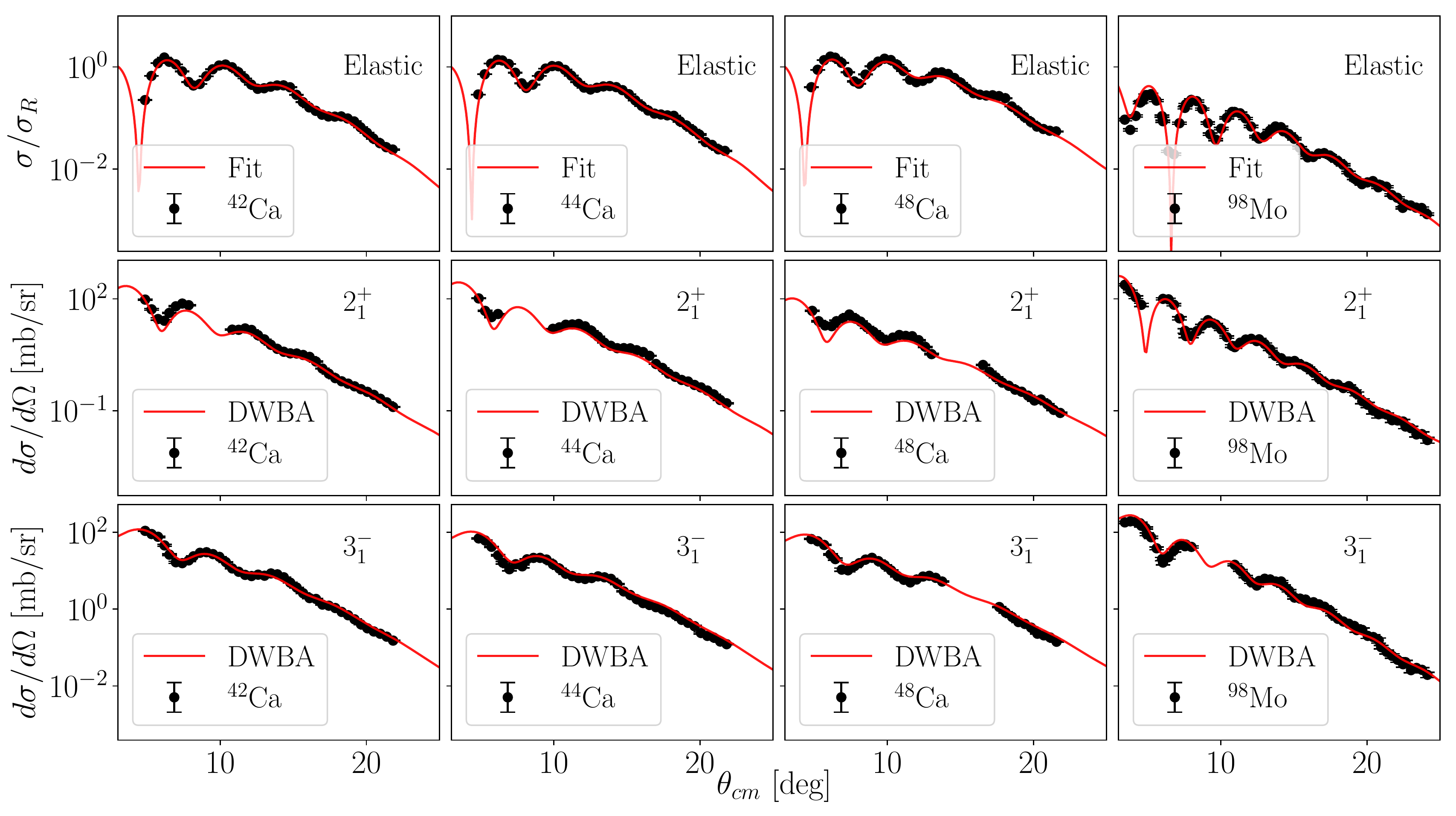}
  \caption[Results of optical model analyses for \nuc{42,44,48}{Ca} and \nuc{98}{Mo}]{Results of optical model analyses for \nuc{42,44,48}{Ca} and \nuc{98}{Mo}. Top: elastic fits to the elastic cross sections, normalized to the Rutherford cross section, along with optical model parameters. Middle: DWBA calculations shown along with experimentally extracted angular distributions for the pure $0_1^+ \rightarrow 2_1^+$ transitions for the nuclei. Bottom: same as middle, but for the $0_1^+ \rightarrow$ transitions.}
  \label{elastic_and_discrete_states}
\end{figure}

The top panels of Fig. \ref{elastic_and_discrete_states} depict the results of the least-$\chi^2$ fits to the extracted data, and the parameters are presented in Table \ref{OMP_table}.\footnote{Owing to constraints on available beamtime, it was not feasible to extract elastic angular distribution data for all nuclei over a sufficiently broad angular range. To mitigate this, optical model parameters for \nuc{42}{Ca} were used for all DWBA calculations for \nuc{40}{Ca}, and the optical model parameters for \nuc{98}{Mo} were used in the calculations for \nuc{94-100}{Mo}. One can see by inspection of the calcium nuclei in Table \ref{OMP_table} that the optical model parameters change very slightly over an isotopic range --- even when the change in nucleon number is substantial ---, and any effect on the resulting giant resonance strength extractions is certainly within the generous $20\%$ uncertainties which we quote as arising due to non-uniqueness of acceptable optical model parameter sets for modeling the data \cite{Li_PRC}.} Using the theory of distorted-waves developed in Chapter \ref{theory}, this optical model was then used as input to calculate the angular distributions for inelastic processes for each of the target nuclei.

\subsection{Comparison with the inelastic $0_1^+\rightarrow 2_1^+$ and $0_1^+\rightarrow 3_1^-$ reaction channels}

The optical model developed by \cite{khoa_satchler_single_folding} and defined by Eqs. \eqref{optical_model_defined}---\eqref{density_dependencies} has, in total, 4 free parameters in the fitting procedure. Utilizing the optical model constrained by the elastic scattering data is within the general methodology prescribed by Ref. \cite{Satchler_direct_nuclear_reactions}; it is  cautioned therein that the fitting of an optical potential to recalcitrant non-elastic data without imposing a constraint that the elastic angular distributions be simultaneously reproduced is both uncontrolled and arbitrary. Furthermore, it is stated explicitly and unequivocally that this methodology is unfounded \emph{even if it is the case that the non-elastic amplitudes and transition wavefunctions are indeed simply not effectively characterized by the same optical potential as that which well-models the elastic channel}. The requirement that the elastic angular distribution be accurately modeled by the choice of optical potential is somewhat intuitively obvious; in order to consider the one-step transitions from the ground to excited states as perturbations/distortions built upon the elastic scattering channel, the latter should be characterized as well as possible within the model framework. This procedure is described by Ref. \cite{Satchler_direct_nuclear_reactions} as being justified \emph{post facto}, in the sense that despite a lack of a particularly stable theoretical foundation, it has met with success in reproducing angular distribution data and further, in the extraction of nuclear structure properties that are in agreement with other forms of experiment.

We thus endeavored to mitigate any issues presented by this lack of rigorous underpinning of the distorted-waves method by strengthening the condition on the optical model parameters we employed in subsequent analysis. We imposed an extension of the aforementioned constraint on the optical model parameters reproduction of the elastic scattering channel. In addition to the accurate reproduction of those angular distributions, it was required that the low-lying discrete state angular distributions be in excellent agreement with the corresponding DWBA calculations using the transition potentials calculated from the ground-state optical potential. The degree of agreement between the angular distribution shapes for the $0_1^+\rightarrow2_1^+$ and $0_1^+\rightarrow 3_1^-$ transitions permits one to have confidence in the \emph{structure} of the extracted giant resonance strength distributions.


Figure \ref{elastic_and_discrete_states} shows, in the middle and bottom rows, the corresponding experimental data for these transitions in juxtaposition with the DWBA calculations using the optical model parameters presented in Table \ref{OMP_table}. One should note that there is no fitting performed in these panels; the degree of agreement between the DWBA and experimental data is excellent. Furthermore, Table \ref{OMP_table} shows the adopted $B(E\lambda;0_1^+\rightarrow \lambda_1^{+,-})$ reported in \cite{raman_ADNDT,kibedi_ADNDT}.\footnote{For \nuc{48}{Ca} $0_1^+\rightarrow 2_1^+$, it is well-established that the isoscalar transition probability extracted from isoscalar probes is different from the electromagnetic transition probabilities \cite{bernstein_isoscalar_electric_rates}. Thus, the value shown in the Fig. \ref{elastic_and_discrete_states} corresponds not to the adopted electromagnetic transition probability, but that which models the presented angular distributions.} These coupling parameters were used in the DWBA and essentially determine the magnitude of the cross section, as discussed in Section \ref{direct reaction theory}. Having the capacity to reproduce the structure of the transitions as well as the magnitude of the transitions with the adopted coupling parameters allows one to ascribe validity to the \emph{magnitude} of the extracted giant resonance strength distributions (as measured by the fraction of the EWSR extracted within each energy bin).

\section{Extraction of the giant resonance strength distributions}

Modern-day extractions of the giant resonance strength distributions ubiquitously rely upon decomposing experimental angular distributions into relative contributions from each individual multipolarity. As discussed previously, the giant resonances lie in a region in which the response functions of many multipolarities overlap heavily. With the aid of the optical model described in the previous section and the collective-model transition densities and potentials of Chapter \ref{theory}, one can begin to disentangle the measured distribution into responses of individual angular momentum transfers.

\subsection{Definition of the MDA} \label{MDA_definition_section}

The multipole-decomposition analysis (MDA) we shall describe here is the presently-accepted method for facilitating this disentanglement. The methodology endeavors to take experimental angular distributions as input, which are then decomposed into a superpositions of angular distributions corresponding to pure angular momentum transfers of $\lambda = 0$, $1$, $2$, $\ldots$. The set of angular momenta included in the fitting procedure is truncated at a reasonable value (in this work, $\lambda_\text{max} = 8$ or $10$). The functional form of the MDA is defined as follows:

\begin{align}
  \frac{\diff^2\sigma^\text{exp}(\theta_\text{c.m.},E_x)}{\diff \Omega\, \diff E} & = \sum_\lambda A_\lambda (E_x) \frac{\diff^2\sigma^\text{DWBA}_\lambda(\theta_\text{c.m.},E_x)}{\diff\Omega \, \diff E}.
  \label{MDA_defined}
\end{align}

If the DWBA calculations are completed using coupling parameters which correspond to $100\%$ of the EWSR, then $A_\lambda$ corresponds to the fraction of the corresponding EWSR exhausted within that particular energy bin \cite{harakeh_book,satchler_isospin,Li_PRL,Li_PRC,uchida_208Pb_ISGDR}. The isovector giant dipole resonance (IVGDR) strengths for these nuclei are known from Refs. \cite{berman_GDR,plujko_GDR}, and those, in combination with DWBA calculations incorporating the Goldhaber-Teller model \cite{satchler_isospin}, allow for the IVGDR strengths to be explicitly accounted for in the MDA procedure. Although multipolarities were included up to $\lambda_\text{max} = 10$, we report extracted strengths only for $\lambda \leq 2$; the extractions of these multipole strengths are insensitive to increasing values of $\lambda_\text{max}$ beyond the values employed here \cite{itoh_sm_nucA,itoh_sm_PRC}.

In this technique, the spectra underlying the giant resonance responses can include physical processes which do not arise from the inelastic scattering channel (\emph{e.g.} proton or neutron knockout reactions and three-body channels). Additionally, the nuclear continuum is comprised of highly overlapping excitations which themselves have small transition probabilities for population from the ground state, but which can be populated in multi-step processes that are theoretically modeled as complicated $n-$particle, $n$-hole transitions \cite{harakeh_book}. Some of these processes can contribute to the strength distributions at higher excitation energies for the similarly-forward-peaked multipoles (\emph{i.e.} the ISGMR and ISGDR), and have been shown to disappear in particle-decay coincidence measurements which investigated the origin of the observed extra high-energy strength \cite{hunyadi_208Pb_ISGDR,nayak_208Pb_ISGDR} and can be neglected in the considerations of the extracted strength. The events arising from the nuclear continuum, however, do not exhibit any coherence in the multipolarity of their angular distributions and so are then absorbed into the multipoles included within the MDA basis without favoring any particular multipolarity \cite{itoh_sm_PRC}.

\subsection{Implementation of the MDA}
This subsection will, in a brief interlude, describe the spirit of Markov-Chain Monte Carlo (MCMC) sampling and its extension to model-fitting in the manner employed in this work. The aforementioned Monte Carlo algorithm of Goodman and Weare \cite{goodman_weare,foreman_mackey} was chosen for implementation of the MDA; its sophistication precludes a tractable first-sketch of MCMC methods and so we will first begin by presenting a simpler algorithm, known as the Metropolis Algorithm \cite{metropolis_seminal} --- see Algorithm \ref{metropolis_sketch}. From here, we will expound upon the extension of MCMC with Bayes' Theorem to model data; changing the MCMC algorithm itself is trivial as the outputs of each algorithm are functionally identical, with the differences largely arising due to the computational timescales of each method. Further development and specialized discussions of applications of MCMC methodology to scientific computing and nuclear physics are available in recent literature \cite{computational_nuclear_physics_notes,monte_carlo_statistical_methods,monte_carlo_strategies_in_scientific_computing}; the interested reader is referred there for the presentation of details outside the scope of this thesis work.

In the coarsest sense, MCMC methods are simply those which allow one to systematically make draws from the domain, $\mathcal{X}$, of a known probability distribution function (PDF) with probabilities defined by the PDF range, $\mathcal{Z}$. The Markov-Chain Monte Carlo techniques are so-named due to the algorithm employing \emph{walkers}, which then take \emph{steps} around the parameter space subject to the PDF that is being sampled. A Markovian process is one in which the $j+1^\text{th}$ step depends \emph{only} on the $j^\text{th}$ position of the walker in the parameter space. Owing to this fact, each step of the walker through the parameter space is an essentially independent draw from the PDF, and providing that the algorithm runs for sufficiently long, one eventually finds that the histograms of the walker positions $\vec{x} \in \mathcal{X}$ in the parameter space, tabulated at each step, converge to the (unnormalized) PDF being sampled. In so doing, this allows for the statistical evaluation of observables $f(\vec{x})$ and thus the generation of probability distributions for those quantities. This treatment yields not only the expected values of $f$, but also its variance and spread --- thus yielding a statistical measure of the effect of uncertainties of components of $\vec{x}$ on the uncertainties in $f(\vec{x})$ (\emph{i.e.} propagating the uncertainties in $x_j$ through to $f(\vec{x})$ in a probabilistic formalism).

\begin{algorithm}[t!]
 \caption{Sample pseudocode for implementation of the Metropolis Algorithm.}
  \begin{algorithmic}[1]
    \STATE With $\vec{x}^0 = \left(x_1^0,x_2^0,\ldots,x_j^0,\ldots x_N^0\right)$, initialize randomly the initial values of each $x_j$ over an acceptable domain.
    \STATE Evaluate the PDF at $\vec{x}^0$, \emph{i.e.} evaluate $p_x = p(\vec{x}^0)$.
    \FOR {each $k$, $k=1,\ldots,N_\text{steps}$}
      \STATE $\vec{w} \gets \vec{x}^{k-1}$
      \FOR {each $j$, $j=1,\ldots,N$}
        \STATE Draw a random number $\Delta_j$ from a normal distribution of mean $\mu=0$ and width $\sigma_j$.
        \STATE $w_j \gets w_j + \Delta_j$
      \ENDFOR
      \STATE Evaluate the PDF at $\vec{w}$, \emph{i.e.} evaluate $p_w = p(\vec{w})$.
      \IF {$p_w/p_x \geq 1$}
        \STATE $\vec{x}^k \gets \vec{w}$
      \ELSE
        \STATE Draw random number $u$ from the uniform distribution over $[0,1]$.
        \IF{$u < p_w/p_x$}
          \STATE $\vec{x}^k \gets \vec{w}$
        \ELSE
          \STATE $\vec{x}^k \gets \vec{x}^{k-1}$
        \ENDIF
      \ENDIF
    \ENDFOR
  \end{algorithmic}
  \label{metropolis_sketch}
\end{algorithm}

Algorithm \ref{metropolis_sketch} delineates the general procedure for sampling from a known PDF $p:\mathcal{X}\rightarrow\mathcal{Z}$ utilizing the Metropolis Algorithm and a normal proposal distribution. The summary description of the algorithm is: for each iteration starting from the current walker position $\vec{x}$, to iteratively \emph{propose} an updated walker position $\vec{w}$ for each step\footnote{In the Metropolis Algorithm, the values of the various $\sigma_j$ for the values of the proposal step $\Delta_j$ are determined by the values which optimize the convergence of the algorithm as measured by the \emph{acceptance fraction} of the proposal steps. Further discussion of the acceptance fraction and its value as a diagnostic tool are found in Refs. \cite{computational_nuclear_physics_notes,monte_carlo_statistical_methods,monte_carlo_strategies_in_scientific_computing}.}, all the while accepting proposed steps that result in an increased probability density (\emph{i.e.} $p(\vec{w}) > p(\vec{x}^{k-1})$) and \emph{occasionally} accepting those which reduce the probability density. The latter allowance permits the walker to explore the entire parameter space and not to become trapped in the most highly probable regions of the domain, and is indeed crucial to the walker fully exploring the intended probability distribution.

The application of the aforementioned MCMC techniques to model fitting requires some additional formalism in order to precisely define the probability density that will govern the trajectory of the Markov Chain. The general problem posed is as follows: one has measured an $N$-tuple of data, $X = \left\{\vec{X}_j = (\vec{x}_j,y_j)\right\}_{j=1}^N$, --- perhaps with some uncertainties in the (here, a scalar) $y_j$, denoted $\delta y_j$ --- and wishes to determine the optimal parameters for modeling the data using a model function $f(\vec{x},\vec{a})$, wherein $\vec{a}$ is a set of \emph{free parameters} that constrain the function shape. Providing that there are more data points than free parameters, the problem posed is \emph{over-constrained} (in the sense that multiple choices of $\vec{a}$ will reproduce different partitions of the data with varying success), and thus one endeavors to find the optimal $\vec{a}$ that reproduces the entire data set as a whole.

One measure of goodness-of-fit that is frequently used is the weighted $\chi^2$, defined as follows:

\begin{align}
  \chi^2  = \sum_{j=1}^N \left( \frac{y_j - f(\vec{x}_j,\vec{a})}{\delta y_j} \right)^2.
\end{align}

Given a set of model parameters $\vec{a}$, one can define the \emph{likelihood distribution} of the data coordinates $\vec{X}$, denoted $P(\vec{X}|\vec{a})$, as:
\begin{align}
  P(\vec{X}|\vec{a}) &= \exp\left(-\chi^2/2\right) \notag \\
  &= \exp \left( -\frac{1}{2}\sum_{j=1}^N \left( \frac{y_j - f(\vec{x}_j,\vec{a})}{\delta y_j} \right)^2  \right). \label{likelihood_defined}
\end{align}

Equation \eqref{likelihood_defined} contains no information about the bounds of the parameters, and it is indeed the inverse conditional probability $P(\vec{a}|\vec{X})$, which denotes the probability distributions of the parameters of interest as constrained by the measured experimental data. The relationship between these two is known as Bayes' Theorem \cite{bayes_theorem_axiomatic}:
\begin{align}
  P(\vec{a}|\vec{X}) & = \frac{  P(\vec{X}|\vec{a}) P(\vec{a})  }{P(\vec{X})}.
  \label{bayes_defined}
\end{align}


In Eq. \eqref{bayes_defined}, the distribution $P(\vec{a})$ is referred to as the \emph{prior} distribution, which is constructed from whatever knowledge one has about the parameters absent the experimental data. As we lack previous information on the relative contributions from the EWSR over the experimental energy range, the convention we have employed is to simply choose a uniform distribution defined over the bounds of acceptable parameters. In this case, this is the support interval $[0,1]$ for the $A_\lambda$ coefficients denoting the fraction of the EWSR exhausted within any given energy bin):

\begin{align}
  P(\vec{a}) &= \prod_{j}\frac{1}{a_j^\text{max}-a_j^\text{min}} \left[ \Theta(a_j-a_j^\text{min}) - \Theta(a_j-a_j^\text{max}) \right],
  \label{priors_defined}
\end{align}
wherein $\Theta(x)$ is the Heaviside step function. The quantity $P(\vec{X})$ is the prior distribution of the data itself, which is generally unknown; examination of Algorithm \ref{metropolis_sketch}, however, suggests that one can conveniently omit the evaluation of the denominator, owing to the fact that only ratios of the $P(\vec{a}|\vec{X})$ are evaluated and the prior distribution of the data is constant throughout the MCMC analysis.

\begin{figure}[t!]
  \centering
  \includegraphics[width=\linewidth]{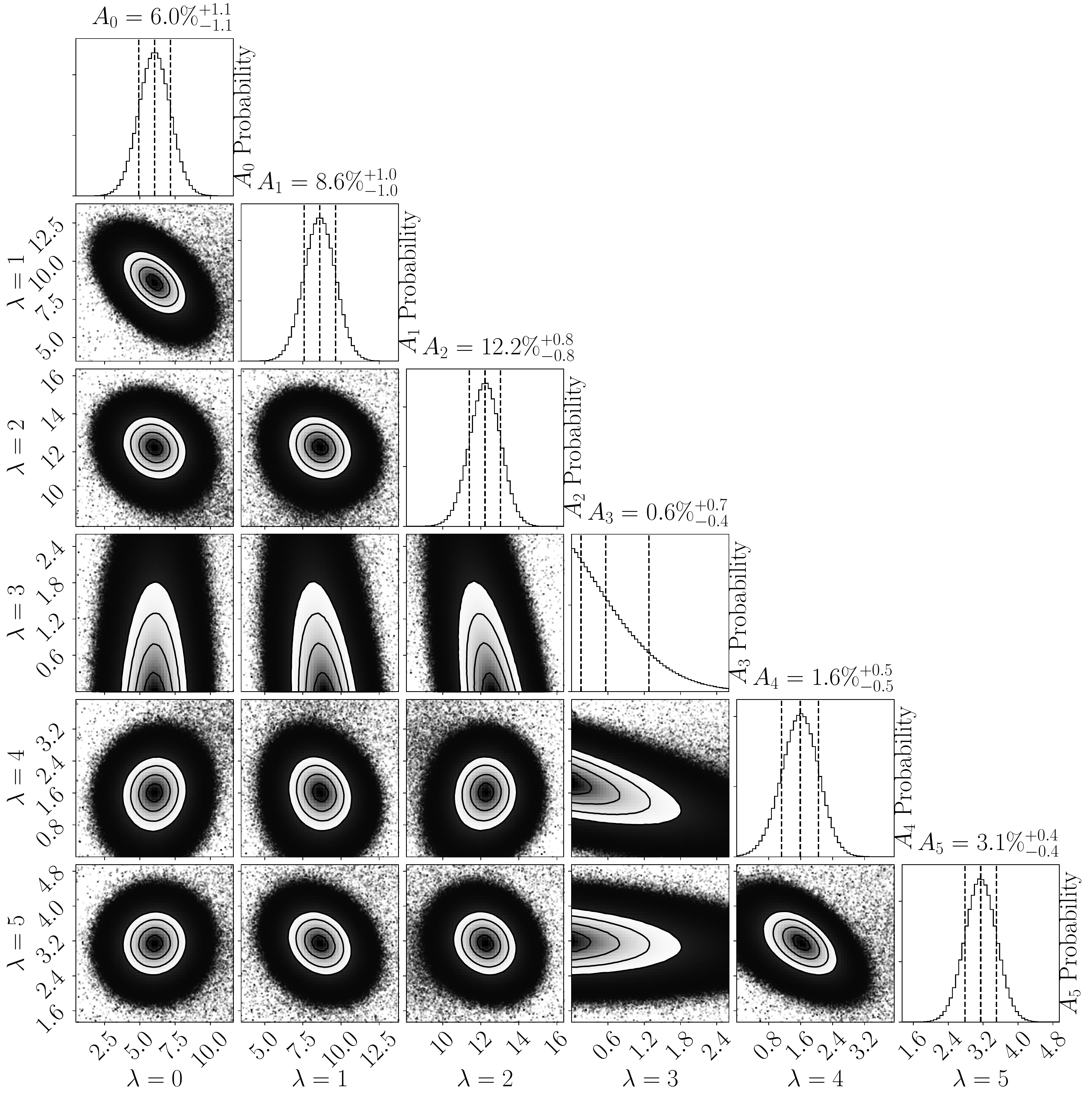}
  \caption[Matrix of probability distributions in $A_\lambda$ for \nuc{94}{Mo} at $E_x = $ 15 MeV.]{Matrix of probability distributions in $A_\lambda$ for \nuc{94}{Mo} at $E_x = $ 15 MeV for a restricted angular momentum space $\lambda=0,\ldots,5$ to facilitate a comprehensive visualization of the covariances.}
  \label{probability_distributions_sample}
\end{figure}

Thus, Eqs. \eqref{likelihood_defined} and \eqref{priors_defined}, when used as input to Bayes' Theorem (Eq. \eqref{bayes_defined}), construct the PDF that is sampled in Algorithm \ref{metropolis_sketch} --- or in any other MCMC algorithm. Indeed, as is described in Ref. \cite{goodman_weare}, the Metropolis algorithm is reasonably effective for sampling from distributions that are ``normal'', in the sense that they are not particularly skewed or anisotropic in the parameter spaces. It is further shown that the computational advantage of using the more sophisticated affine-invariant\footnote{\emph{Affine} transformations are a subset of linear transformations which preserve all parallel lines and the angles subtended between vectors. One example of an affine transformation is a rotation; one can actively \emph{rotate} an element of a large class of 2-dimensional PDFs (\emph{e.g.}, the $\lambda=0$, $\lambda=1$ panel of Fig. \ref{probability_distributions_sample}) until its principle axes are parallel with the component axes being sampled in the MCMC. The invariance of the algorithm to these transformations renders such a transformation unnecessary; the convergence of the MCMC is identical for correlated distributions providing an affine-transformation exists that maps the covariance to zero \cite{foreman_mackey}.}  MCMC algorithms presented therein lies with faster convergence and thus limits the computational encumbrance of the technique. The affine-invariance defined in Ref. \cite{foreman_mackey} and implemented within \texttt{emcee} has the striking feature that the covariances in the parameters do not affect the convergence of the MCMC run. So, for the purposes of the MDA, this was the algorithm of choice due to the significant covariances present in the probability densities of the $A_\lambda$ coefficients (see the off-diagonal panels of Fig. \ref{probability_distributions_sample}).

\begin{figure}[t]
  \centering
  \includegraphics[width=\linewidth]{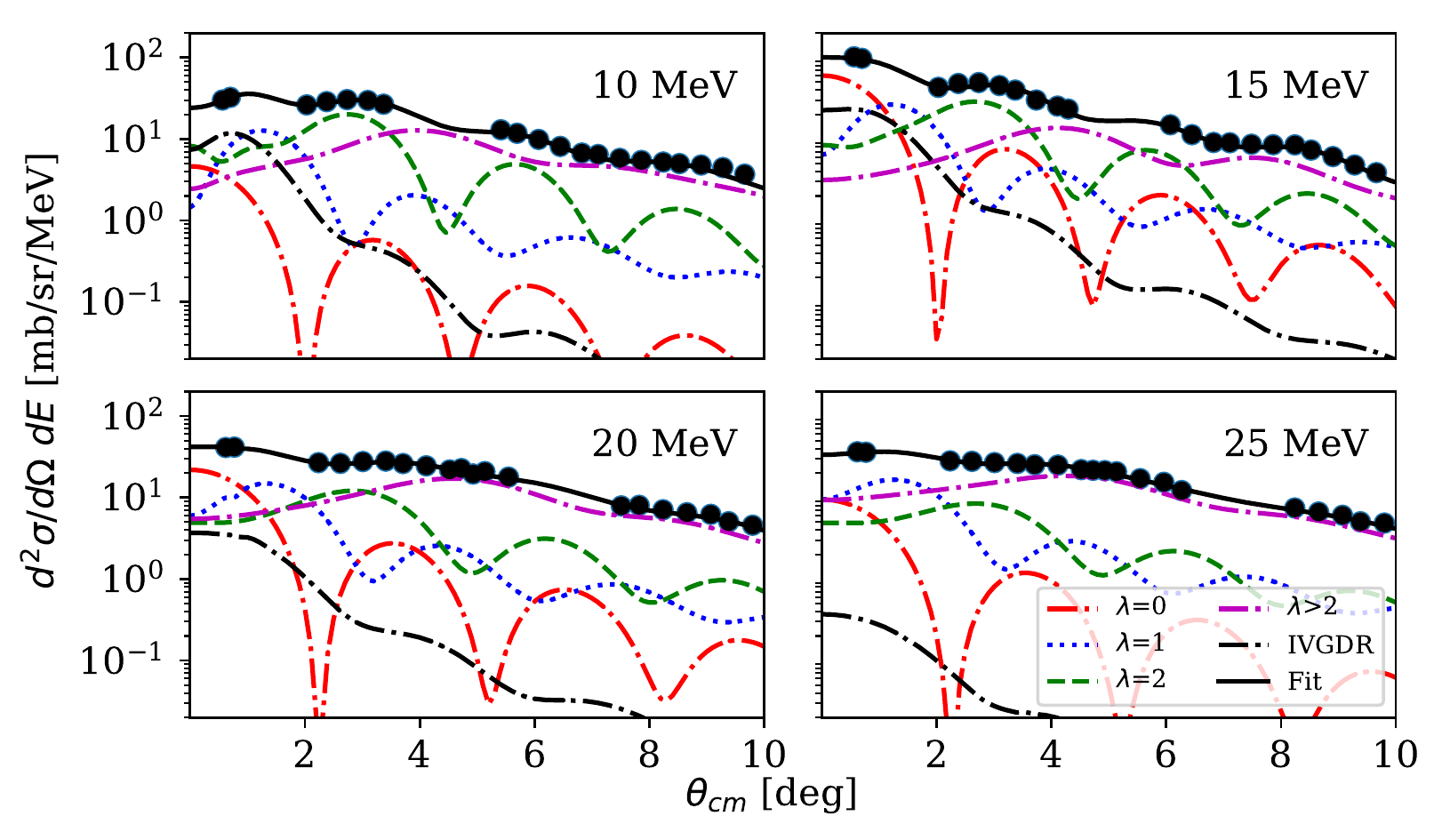}
  \caption[Sample MDA results for \nuc{94}{Mo}.]{Sample angular distribution decompositions for \nuc{94}{Mo} at $E_x = 10$, $15$, $20$, and $25$ MeV, with $\lambda = 0$ (red dot-dashed), $\lambda = 1$ (blue dotted), $\lambda = 2$ (green dashed) and $\lambda \geq 2$ (magenta dot-dashed) shown \cite{KBH_EPJA}. Also visible are the contributions from the IVGDR (black dot-dashed); these contributions are shown to drop off rapidly in intensity with increasing excitation energy.}
  \label{94Mo_mda_figure}
\end{figure}

As discussed previously, the strength distributions of the IVGDR were taken from previously extracted photoneutron distributions \cite{berman_GDR,plujko_GDR}. Here, the photoneutron cross sections are a direct measure of the IVGDR strength distribution. In these works, the distributions extracted from the aforementioned references were modeled with (at least) one Breit-Wigner distribution:
\begin{align}
  \sigma(E_x) &= \frac{\sigma_\text{max}}{1 + (E_x^2-E_\text{cent}^2)^2/E_x^2 \gamma^2}. \label{breit_wigner_dist}
\end{align}
The exact lineshape chosen for the modeling of the cross sections is somewhat irrelevant for our own purposes, providing that the experimental distribution is well-characterized by the fit. The integral of the lineshape over the excitation energy results yields the Thomas-Reiche-Kuhn (TRK) sum rule of Eq. \eqref{TRK_sum_rule}, and within the Goldhaber-Teller model, the fraction of this sum rule exhausted within any excitation energy bin can be related to the transition amplitude that exhausts the TRK sum rule (Eq. \eqref{TRK_amplitude}):
\begin{align}
  \beta_\text{IVGDR}(E_x) &= A_\text{IVGDR}(E_x) \, \beta_\text{IVGDR}^{100\% \text{ TRK}}(E_x) \notag \intertext{and, with $\diff E$ being the bin width,}
  A_\text{IVGDR}(E_x) &= \int_{E_x-\diff E /2}^{E_x+\diff E/2} \sigma(E_x^\prime) \; \diff E_x^\prime  \bigg /  \int_0^\infty \sigma(E_x^\prime) \; \diff E_x^\prime.
\end{align}

\begin{figure}[t]
  \centering
  \includegraphics[width=\linewidth]{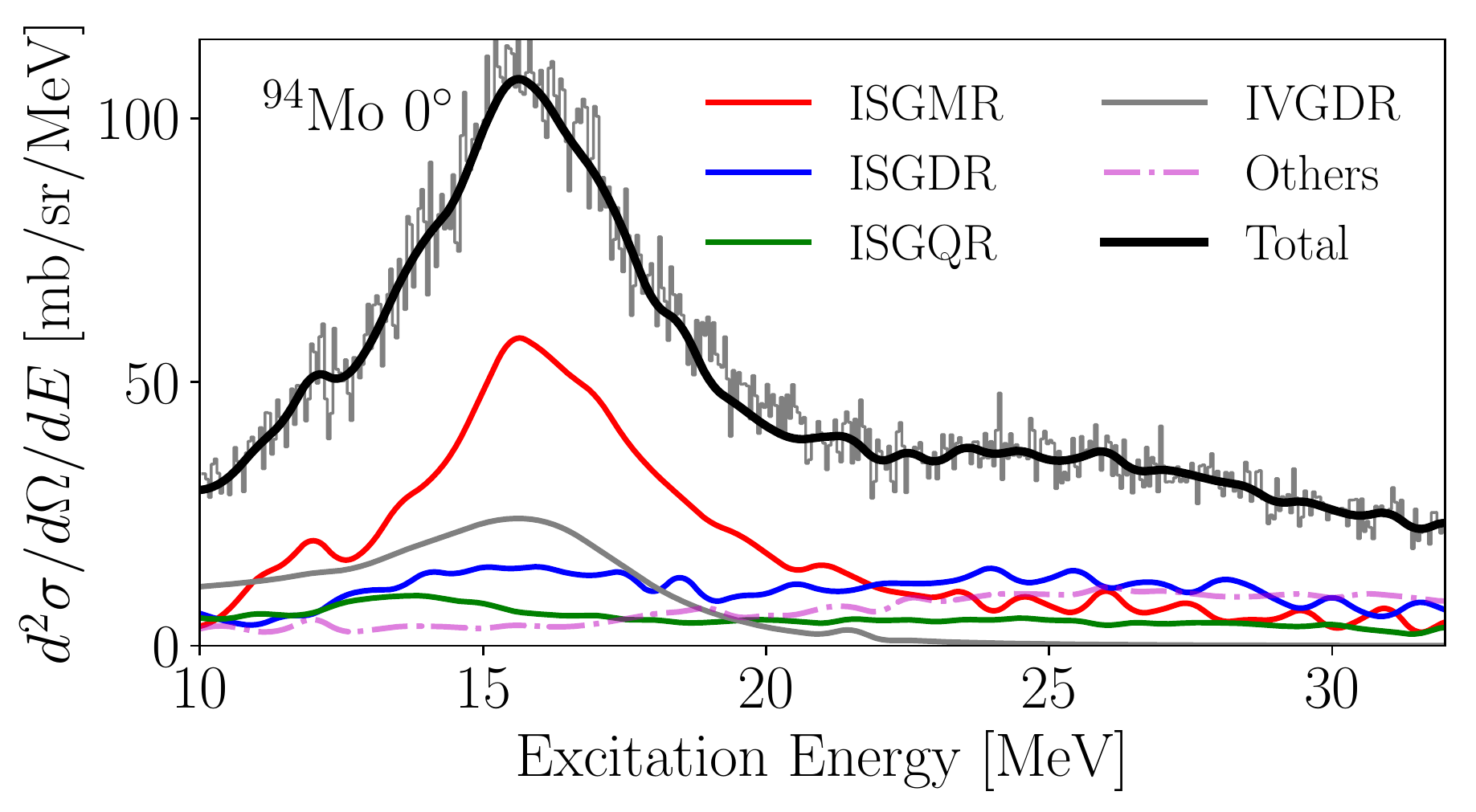}
  \caption[Experimental excitation energy spectra acquired with Grand Raiden set to $\theta^\text{GR}_\text{lab} = 0^\circ$.]{Experimental excitation energy spectra acquired with Grand Raiden set to $\theta^\text{GR}_\text{lab} = 0^\circ$, with a spectrographic averaged angle of $\theta_\text{lab}^\text{avg} = 0.6^\circ$. Shown are the contributions from the ISGMR (red), ISGDR (blue), ISGQR (green), IVGDR (gray), and higher multipolarities (purple) to the experimental spectrum, as determined from the multipole decomposition analysis.}
  \label{analysis:zeroDeg_decomposed}
\end{figure}

Upon conclusion of the MDA, the $68\%$ confidence interval of each $A_\lambda$ probability distribution --- centered at the distribution median --- was taken as the uncertainty in each parameter. For visualization purposes, sample probability distributions extracted for \nuc{94}{Mo} at 15 MeV are shown in Fig. \ref{probability_distributions_sample}.\footnote{As this is a restricted angular momentum space, these results will \emph{not} correspond exactly to the EWSR coefficients extracted in the true analysis. Owing to the high dimensionality ($\lambda_\text{max}=10$ for \nuc{94}{Mo}), it proved to be visually intractable to generate a plot including all correlations between the parameters for inclusion within this dissertation.}

This procedure was applied methodically to each nucleus and energy bin for which angular distributions were extracted. Some typical angular distributions and their associated decompositions are shown in Fig. \ref{94Mo_mda_figure}; a comprehensive presentation of the decomposed angular distributions analyzed in this work is shown in Appendix \ref{MDA_results}.

\begin{figure}[t]
  \centering
  \includegraphics[width=\linewidth]{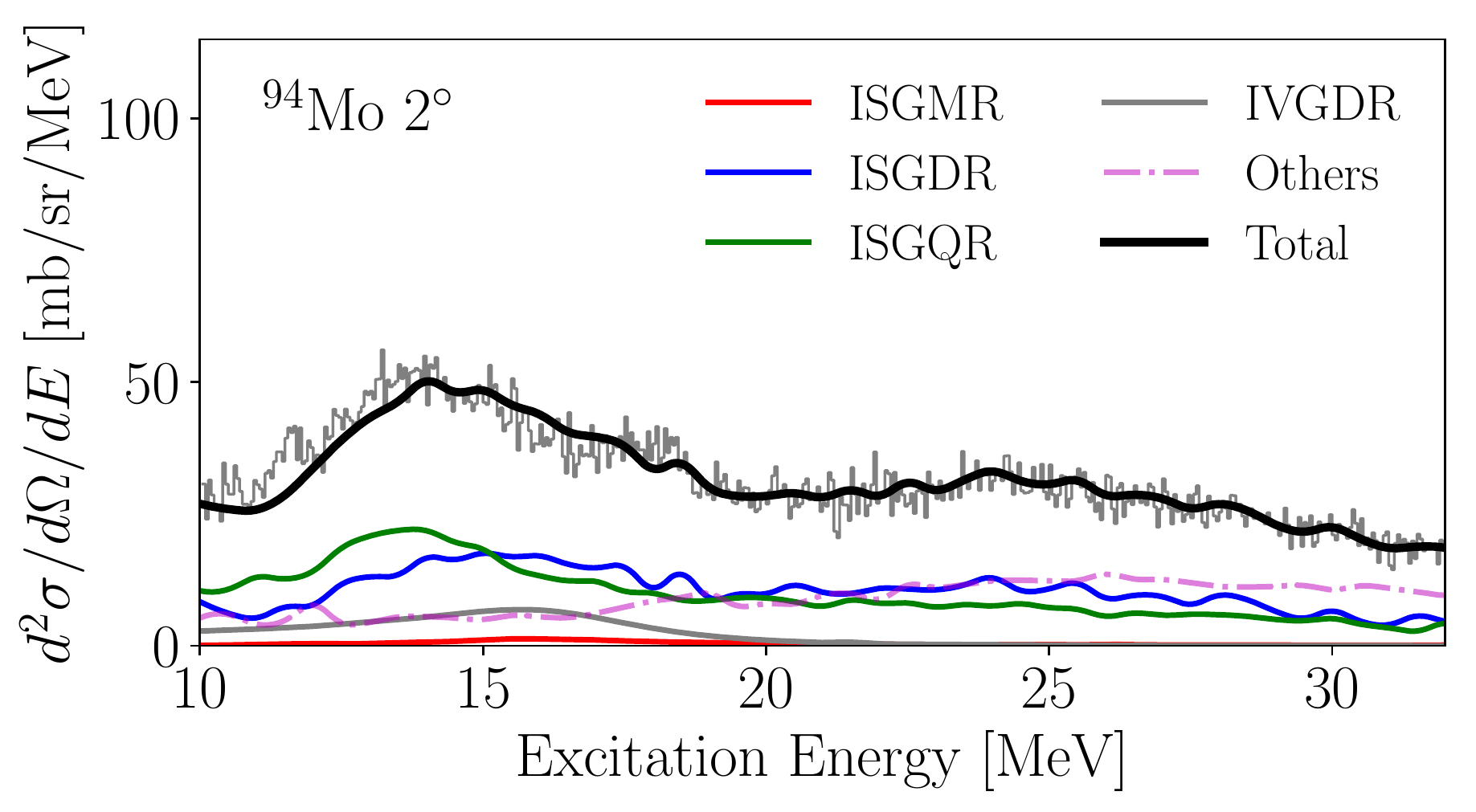}
  \caption[Experimental excitation energy spectra acquired with Grand Raiden set to $\theta^\text{GR}_\text{lab} = 2.0^\circ$.]{Experimental excitation energy spectra acquired with Grand Raiden set to $\theta^\text{GR}_\text{lab} = 2.0^\circ$, with a spectrographic averaged angle of $\theta_\text{lab}^\text{avg} = 2.2^\circ$. Shown are the contributions from the ISGMR (red), ISGDR (blue), ISGQR (green), IVGDR (gray), and higher multipolarities (purple) to the experimental spectrum, as determined from the multipole decomposition analysis.}
  \label{analysis:twoDeg_decomposed}
\end{figure}

Figures \ref{analysis:zeroDeg_decomposed} and \ref{analysis:twoDeg_decomposed} show two decompositions of post-reduction energy spectra of \nuc{94}{Mo} using the MDA framework discussed previously. In these figures, the corresponding DWBA calculations for $100\%$ of the EWSR at the corresponding laboratory-frame angles were then multiplied by the extracted $A_\lambda$ coefficients obtained within the MDA (or in the case of the IVGDR, as calculated by the TRK sum rule and the photoneutron distribution data from the literature). With these results, one can visualize the contributions from each multipolarity to the overall experimental energy spectra.

Figure \ref{analysis:zeroDeg_decomposed} shows this decomposition for the $0^\circ$ setting of Grand Raiden, at which the ISGMR response is maximal. In contrast, examination of Fig. \ref{analysis:twoDeg_decomposed} shows the same decomposition at $2^\circ$. Inspection of Fig. \ref{experimental:angular_distributions} alongside these figures allows for one to explain the features which are manifest within the spectra on the basis of angular distribution data. The characteristic angular distribution for the ISGMR is a sharp minimum, whereas the ISGQR is approaching a maximum, as the scattering angle approaches $2^\circ$. One sees that the contribution of the ISGMR to the experimental spectra at this angle is essentially negligible, whereas at $0^\circ$, the ISGMR contributes overwhelmingly to the detected response owing to its relatively high cross section at forward angles. Figs. \ref{analysis:zeroDeg_decomposed} and \ref{analysis:twoDeg_decomposed} demonstrate even further the significance that angular distribution data hold in attempts to characterize the giant resonance responses due to the overlapping positions and spreads of the giant resonances in experimental spectra.

%
%
%
%
%

%

%
%
%
%
%
%
%
%
%
%

%
%

\chapter{Results and discussion}\label{results}

After the multipole decomposition, the strength distributions are determined from the EWSR fractions $A_\lambda$ and the full $m_1$ EWSR derived in Subsection \ref{theory:ewsr_derived}:

\begin{align}
  S_\lambda(E_x) & = A_\lambda(E_x) \frac{m_1^\text{100\% EWSR}}{E_x}. \label{conversion_from_ewsr_fraction}
\end{align}

Given the ISGMR strength distribution, one can calculate the nuclear incompressibility of a finite nucleus, $K_A$, from the energy of the compressional-mode electric isoscalar giant monopole resonance \cite{harakeh_book}:
\begin{align}
  E_\text{ISGMR} &= \hbar \sqrt{\frac{K_A}{m\expect{r_0^2}} } \label{energy_KA},
\end{align}

\noindent where $m$ is the free-nucleon mass, and $\expect{r_0^2}$ is the ground-state mean-square nuclear mass radius. Generally, the ISGMR energies would be associated with one of the experimental moment ratios $\sqrt{m_3/m_1}$, $m_1/m_0$, or $\sqrt{m_1/m_{-1}}$ \cite{jennings_jackson_constrained_scaling,treiner_incompressibility_models,harakeh_book,stringari_lipparini_sum_rules,bohigas_sum_rules,blaizot_nuclear_compressibilities}, where the moments $m_k$ of the strength function are defined as
\begin{align}
  m_k = \int \diff  E_{x} \,  S_\lambda(E_{x}) E_{x}^k. \label{moments_defined}
\end{align}
These moments are related to the constrained- and scaling-model energies and incompressibilities \cite{jennings_jackson_constrained_scaling,treiner_incompressibility_models,stringari_sum_rules,stringari_lipparini_sum_rules}, as well as the centroid (energy-weighted average) energies of the strength distributions via:
\begin{align}
  E_\text{constrained} &= \sqrt{\frac{m_1}{m_{-1}}}, \notag \\
  E_\text{centroid} & = \frac{m_1}{m_0}, \notag \\
  E_\text{scaling} &= \sqrt{\frac{m_3}{m_1}}. \label{moment_ratios}
\end{align}

\begin{figure}[t!]
  \centering
  \includegraphics[width=\linewidth]{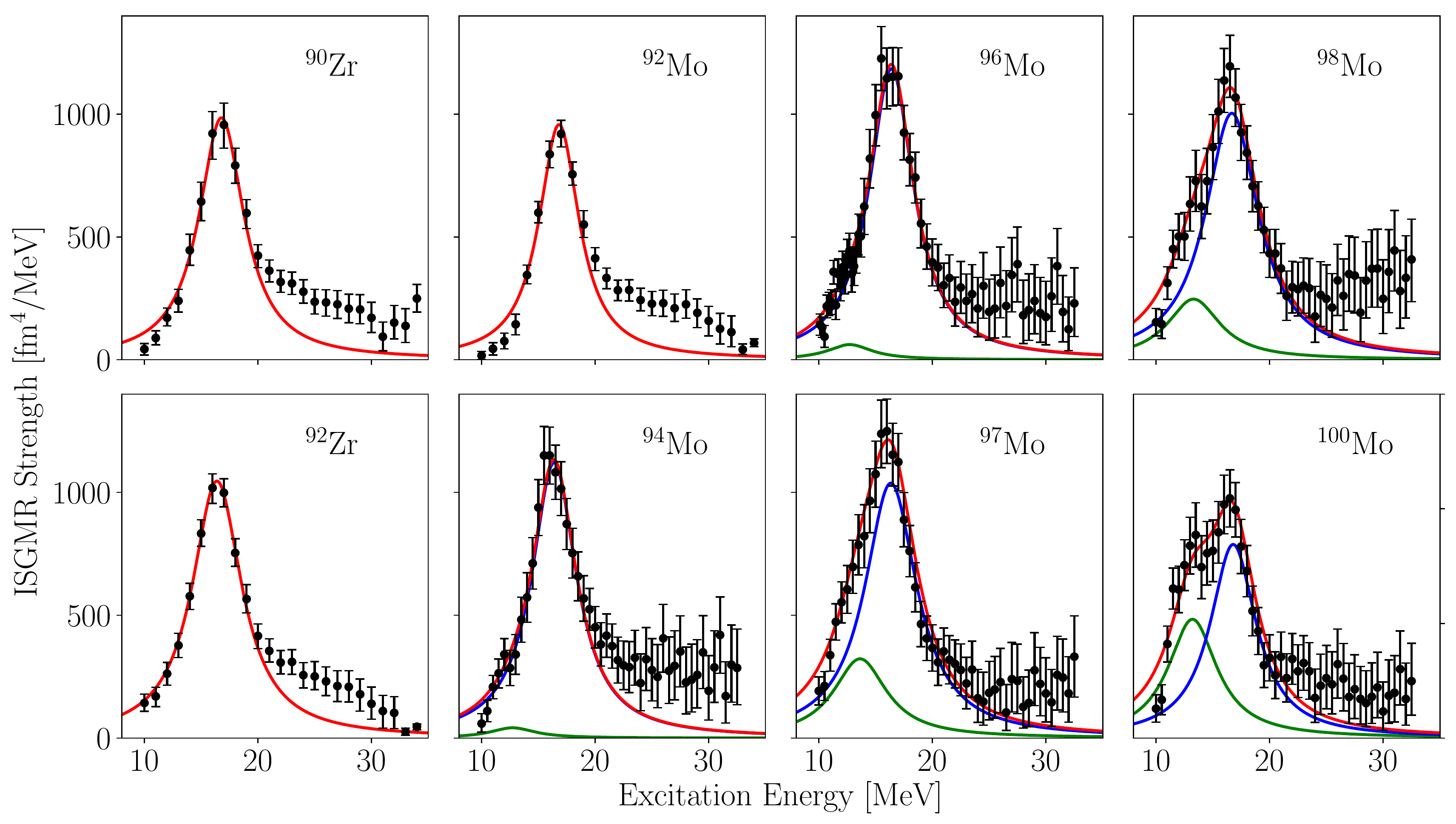}
  \caption[ISGMR strengths for \nuc{90,92}{Zr}, \nuc{92}{Mo}, and \nuc{94-100}{Mo}.]{ISGMR strengths for \nuc{90,92}{Zr}, \nuc{92}{Mo} (data originally from Ref. \cite{gupta_A90_PRC}) and \nuc{94-100}{Mo}. For \nuc{94-100}{Mo}, a two-peak Lorentzian distribution (Eq. \eqref{lorentz}) is also shown; else, a one-peak fit is plotted with the data. Fit parameters, EWSR percentages exhausted underneath the distributions, and the corresponding moment ratios are presented in Tables \ref{molly_moment_ratios_total_EWSRS} and \ref{molly_fit_parameters} \cite{KBH_EPJA}.}
  \label{molly_strengths_with_lorentz}
\end{figure}

Utilization of Eq. \eqref{energy_KA} with a corresponding ISGMR energy from Eq. \eqref{moment_ratios} is formally predicated on the assumption that the strength distribution of the resonance is contained within a single collective peak \cite{harakeh_book,KBH_EPJA} and that nearly 100\% of the EWSR is identified within the peak in question; in cases where the ISGMR strength distribution is fragmented, multiply-peaked, or found to under-represent the EWSR, the extraction of technical and meaningful $K_A$ values demands care and can in some cases become untenable. Nonetheless, extractions of $K_A$ in finite nuclei are useful in characterizing the bulk response of the nucleus to density oscillations in the ISGMR, and when appropriate, we will report the $K_A$ values associated with the scaling model energies $\sqrt{m_3/m_1}$ \cite{jennings_jackson_constrained_scaling} to be consistent with contemporary literature.

Strength distributions for the ISGDR and ISGQR have been extracted simultaneously from the MDA. Similar analyses of these distributions have been completed and the moment ratios have been likewise calculated for each nuclei investigated in this work. A comprehensive presentation and discussion of the results for these higher multipolarities are presented in Appendix \ref{strength_results}.

\begin{figure}[t!]
  \centering
  \includegraphics[width=\linewidth]{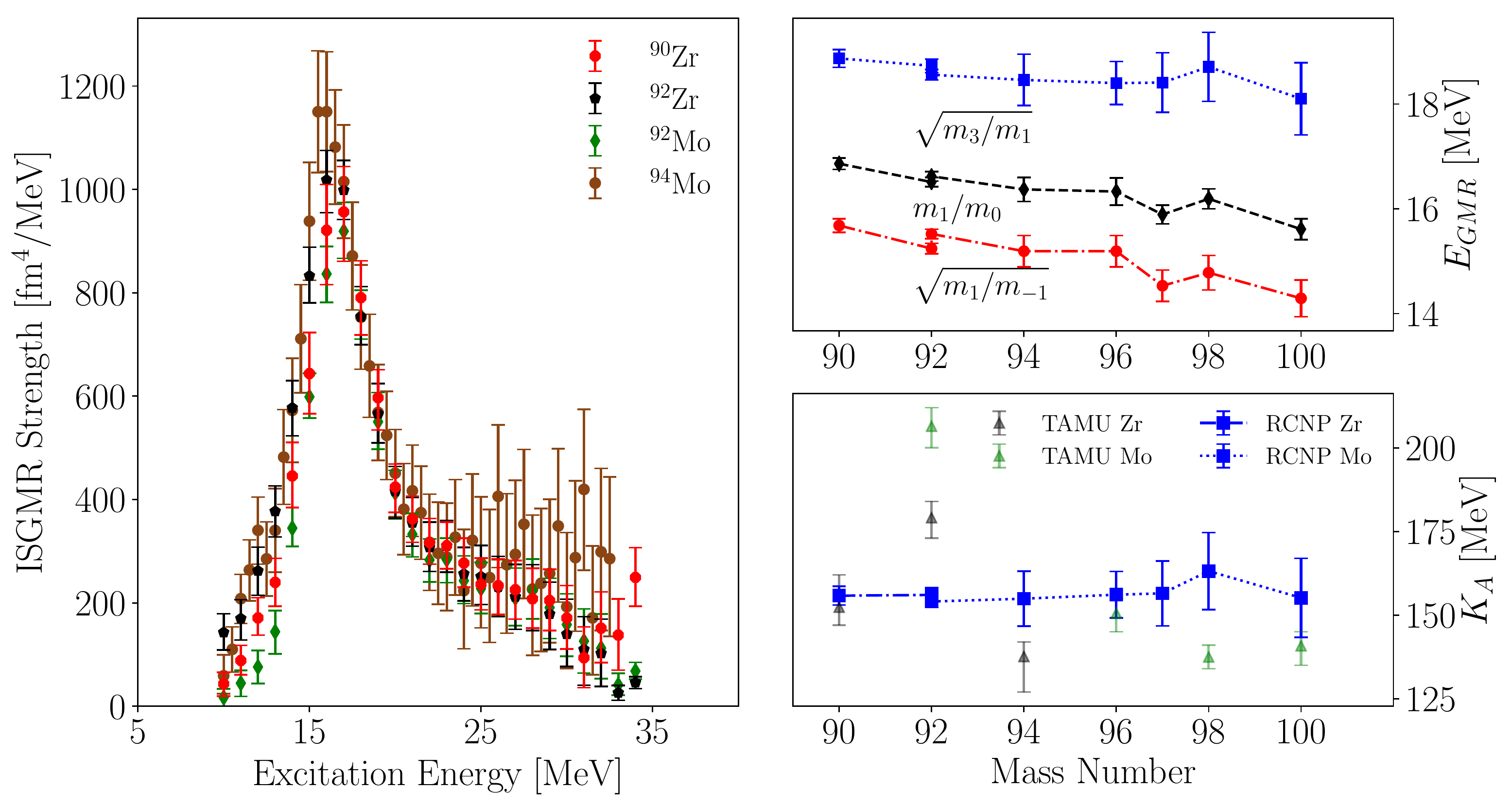}
  \caption[ISGMR strength distributions for \nuc{90}{Zr}-\nuc{94}{Mo}, and associated moment ratios and scaling-model $K_A$ values.]{Left: ISGMR strength distributions for the progression from \nuc{90}{Zr} --- \nuc{94}{Mo}. Evident is the structural and positional agreement of the distributions, especially for the $A=92$ isobars. Top right: Various moment ratios for the nuclei in the zirconium-molybdenum region. Lines connect \nuc{90,92}{Zr} and \nuc{92-100}{Mo}. Bottom right: TAMU extractions of $K_A$ shown previously in Fig. \ref{AM_KA} from Refs. \cite{youngblood_A90_unexpected,krishichayan_Zr,youngblood_92_100Mo,button_94Mo} (black and green triangles), juxtaposed with the finite nuclear incompressibilities $K_A$ (blue squares) measured for \nuc{90,92}{Zr}, \nuc{92}{Mo} \cite{gupta_A90_PLB,gupta_A90_PRC,KBH_EPJA} and \nuc{94-100}{Mo} within these works. Shown clearly is a consistent scaling-model nuclear incompressibility for the nuclei in this mass region.  In all cases, we have calculated $K_A$ from the resonance energies of Eq. \eqref{moment_ratios} over an energy range within which we have identified nearly $100\%$ of the EWSR, using the resonance energies listed in Table \ref{molly_moment_ratios_total_EWSRS}. Figure adapted from Ref. \cite{KBH_EPJA}.}
  \label{molly_strengths_KA}
\end{figure}

\section{ISGMR in the $A \approx 90$ region}

\begin{table}[t!]
  \centering
  \caption[Lorentzian-fit parameters for \nuc{90,92}{Z\lowercase{r}}, \nuc{92-100}{M\lowercase{o}}]{Lorentzian-fit parameters for \nuc{90,92}{Z\lowercase{r}}, \nuc{92-100}{M\lowercase{o}} \cite{KBH_EPJA} and integrated EWSR between 0 --- 35 M\lowercase{e}V}
  \label{molly_fit_parameters}
\resizebox{\linewidth}{!}{
\begin{tabular}{@{}ccccccccccc@{}}
\toprule
             &  & \multicolumn{3}{c}{Low Peak}                       &  & \multicolumn{3}{c}{High Peak} & & Total           \\ \cmidrule(lr){3-5} \cmidrule(l){7-9} \cmidrule(l){11-11}
      &  & $E_0$       & $\Gamma$   & $m_1$             &  & $E_0$       & $\Gamma$ & $m_1$   & &  $m_1$ \\

      &  & [MeV]      &  [MeV]  &  [\%]            &  &  [MeV]      & [MeV]  & [\%]  & & [\%]\\
\cmidrule(r){1-5} \cmidrule(l){7-9} \cmidrule(l){11-11}
\nuc{90}{Zr} &  & -                & -               & -                    &  & $16.8 \pm 0.2$ & $2.4 \pm 0.4$ & $84 \pm 2$ & & $84\pm2$ \\ 
\nuc{92}{Zr} &  & -                & -               & -                    &  & $16.4 \pm 0.1$ & $2.2 \pm 0.3$ & $91 \pm 2$ & & $91 \pm 2$ \\
\nuc{92}{Mo} &  & -                & -               & -                    &  & $16.5 \pm 0.1$ & $2.3 \pm 0.1$ & $73 \pm 2$ & & $73 \pm 2$\\ 
\nuc{94}{Mo} &  & $12.7 \pm 0.5$ & $2.4 \pm 0.4$ & \asymmerror{2}{3}{2} &  & $16.4 \pm 0.2$ & $2.4 \pm 0.4$  & $86 \pm 3$ & & $88 \pm 4$\\ 
\nuc{96}{Mo} &  & $12.7 \pm 0.5$  & $2.3 \pm 0.3$ & \asymmerror{4}{3}{4} &  & $16.4 \pm 0.2$ & $2.4 \pm 0.3$ & $89 \pm 3$  & & $93 \pm 4$\\
\nuc{97}{Mo} &  & $13.6 \pm 0.6$ & $2.8 \pm 0.5$ & $23\pm 4$ &  & $16.3 \pm 0.4$ & $2.8 \pm 0.4$  & $86 \pm 4$ & & $110 \pm 6$\\ 
\nuc{98}{Mo} &  & $13.3 \pm 0.5$ & $2.8 \pm 0.5$ & $16 \pm 4$ &  & $16.7 \pm 0.4$ & $2.8 \pm 0.4$  & $85 \pm 4$ & & $102 \pm 6$\\ 
\nuc{100}{Mo} &  & $13.2 \pm 0.4$ & $2.6 \pm 0.6$ & $32 \pm 4$ &  & $16.8 \pm 0.4$ & $2.5 \pm 0.5$  & $60 \pm 3$ & & $93 \pm 6$\\ \bottomrule
\end{tabular}
}
\end{table}

The extracted ISGMR strengths for \nuc{94-100}{Mo}, as well as \nuc{90,92}{Zr} and \nuc{92}{Mo} \cite{gupta_A90_PRC,KBH_EPJA}, are shown in Fig. \ref{molly_strengths_with_lorentz} along with with Lorentzian distributions which were fitted to the data:

\begin{align}
    S(E_x,S_0,E_{0},\Gamma) & = \frac{S_0 \Gamma}{\left(E_x - E_0\right)^2 + \Gamma^2}.
    \label{lorentz}
\end{align}

\begin{table}[t!]
\centering
\caption{Moment ratios for for \nuc{90,92}{Z\lowercase{r}} and \nuc{92-100}{M\lowercase{o}} calculated between $0$ --- $35$ M\lowercase{e}V from the fit distributions of Table \ref{molly_fit_parameters}}
\label{molly_moment_ratios_total_EWSRS}
\begin{tabular}{@{}cccc@{}}
\toprule
Nucleus      & $\sqrt{m_{1}/m_{-1}}$ & $m_1/m_0$   & $\sqrt{m_{3}/m_{1}}$   \\
     & [MeV] & [MeV]  &  [MeV]  \\ \midrule
\nuc{90}{Zr} & $15.7 \pm 0.1$            & $16.9 \pm 0.1$ & $18.9 \pm 0.2$          \\
\nuc{92}{Zr} & $15.2 \pm 0.1$            & $16.5 \pm 0.1$ & $18.7 \pm 0.1$            \\
\nuc{92}{Mo} & $15.5 \pm 0.1$            & $16.6 \pm 0.1$ & $18.6 \pm 0.1$           \\
\nuc{94}{Mo} & $15.2 \pm 0.3$            & $16.4 \pm 0.2$ & $18.5 \pm 0.5$           \\
\nuc{96}{Mo} & $15.2 \pm 0.3$            & $16.3 \pm 0.2$ & $18.4 \pm 0.4$            \\
\nuc{97}{Mo} & $14.5 \pm 0.3$            & $15.9 \pm 0.2$ & $18.4 \pm 0.6$            \\
\nuc{98}{Mo} & $14.8 \pm 0.3$            & $16.2 \pm 0.2$ & $18.7 \pm 0.7$            \\
\nuc{100}{Mo} & $14.3 \pm 0.4$            & $15.6 \pm 0.2$ & $18.1 \pm 0.7$            \\ \bottomrule
\end{tabular}
\end{table}

In further analyses of \nuc{94-100}{Mo}, it was found that deformation effects became manifest in the more neutron-rich nuclei. To account for this, the ISGMR strength distributions for those nuclei were fitted with a constrained combination of two peaks to account for possible coupling of the ISGMR strength with the $K=0$ component of the ISGQR \cite{garg_sm_PRL,kvi-def,itoh_sm_PRC,gupta_24Mg_plb,gupta_24Mg_prc,peach_28Si_prc}.  In the cases of \nuc{94,96}{Mo}, although a second peak was included in the modeling of the data, the extracted EWSR for the low-energy peak is consistent with $0\%$, as shown in Table \ref{molly_fit_parameters}. This would suggest that the deformation effects (and thus, any shifting of the ``main'' ISGMR peak) are negligible insofar as a comparison with the peak energies of \nuc{90,92}{Zr}, \nuc{92}{Mo} data is concerned for these nuclei, as pertains to the question posed in Subsection \ref{intro:A92_anomalous}.

The uncertainties in the parameters shown in Table \ref{molly_fit_parameters} are somewhat higher for \nuc{94-100}{Mo} due to the inclusion of a second, highly-correlated peak in the fitting procedure for these nuclei. The uncertainties in the quantities derived from the fit distributions (\emph{i.e.} the moment ratios and assigned EWSRs) were calculated using the probability distributions from outputs of Algorithm \ref{metropolis_sketch} and its surrounding model-fitting discussion. The uncertainties in the EWSRs themselves are only statistical; there is up to an additional $\sim$ 20\% uncertainty which is due to ambiguities in the choice of optical model and transition density input to the DWBA calculations \cite{Li_PRC,uchida_208Pb_ISGDR}; there is minimal effect on the extracted structure of the ISGMR, and so the features of the strength distribution are generally insensitive to choice of a given optical model from the subset of those which reproduce the elastic and low-lying channel angular distributions well.

In all cases, the total assigned EWSR underneath the modeled lineshape is found to be very-nearly 100\% over the experimental excitation energy range.  The moment ratios corresponding to the fit distributions are presented in Table \ref{molly_moment_ratios_total_EWSRS}.

\subsection{Comparison with theoretical calculations for the ISGMR responses and the question of softness in the molybdenum nuclei}

\begin{figure}[t!]
  \centering
  \includegraphics[width=\linewidth]{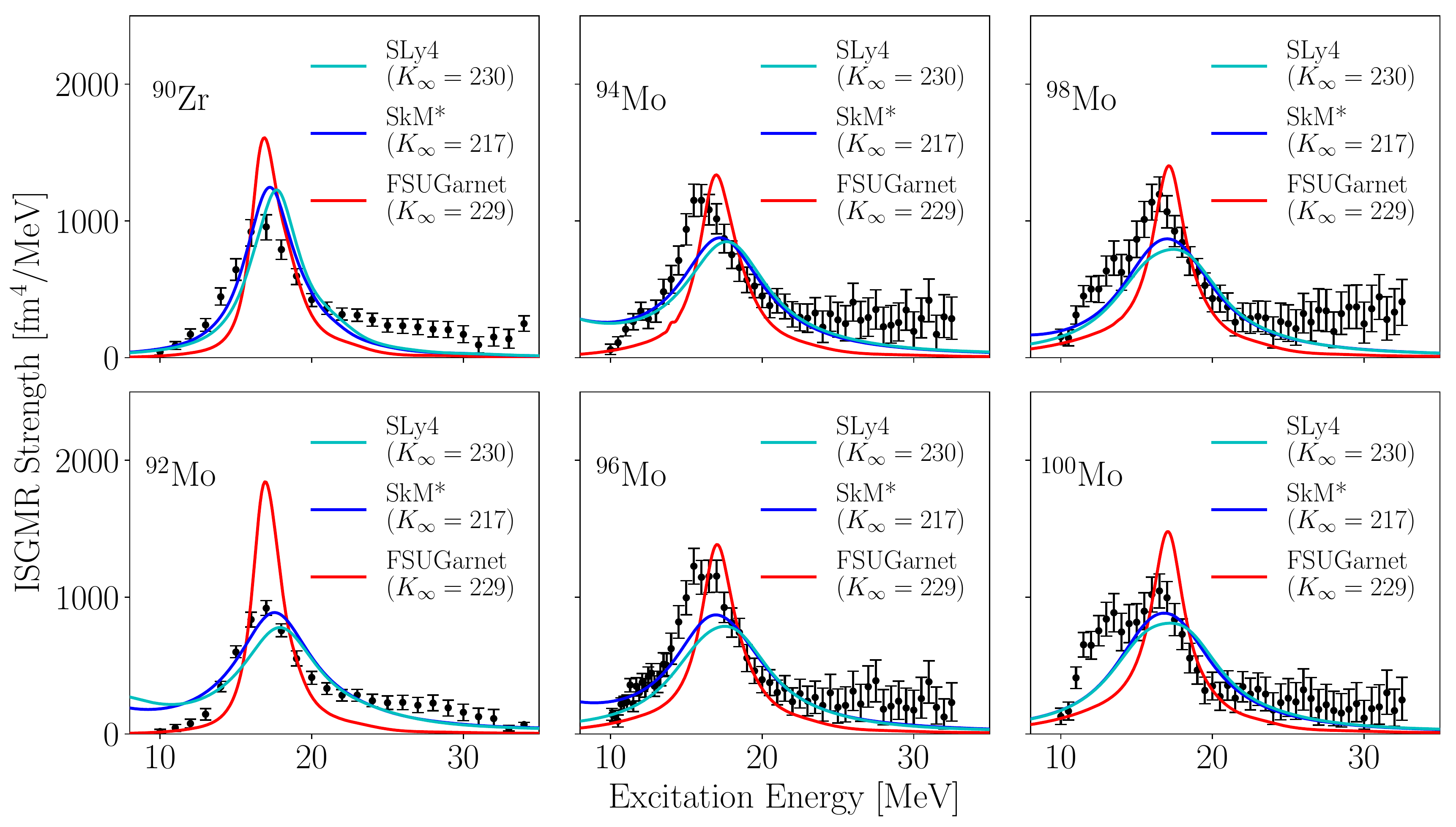}
  \caption[Experimental ISGMR strength distributions for \nuc{90}{Zr} and \nuc{92-100}{Mo} alongside nonrelativistic and relativistic (Q)RPA calculations.]{Experimental ISGMR strength distributions for \nuc{90}{Zr} and \nuc{92-100}{Mo} compared to those acquired in QRPA calculations with the SLy4 and SkM$^*$ interactions, from Ref. \cite{colo_private}, as well as RPA calculations with the FSUGarnet interaction from Ref. \cite{piekarewicz_private}.}
  \label{strengths_a90_only}
\end{figure}

Theoretical efforts to describe the ISGMR response in the zirconium and molybdenum nuclei are critical to elucidating the origins of the softness of the tin and cadmium nuclei, as described in Subsection \ref{intro:softness_question}. Nonrelativistic \cite{colo_private} spherical quasiparticle random phase approximation (QRPA) calculations (using the SLy4 \cite{SLYx_commissioned} and SkM$^*$ \cite{SkM_star_commissioned} interactions) and relativistic spherical RPA calculations using the FSUGarnet \cite{piekarewicz_calcium_isotopes,fsugarnet_commissioned} interaction are shown along with the ISGMR strengths of \nuc{90}{Zr} and \nuc{92}{Mo} reported in Ref. \cite{gupta_A90_PRC} and those of \nuc{94-100}{Mo} extracted in this work in Fig. \ref{strengths_a90_only}. \interfootnotelinepenalty=10000 \footnote{The theoretical calculations presented in this section were very graciously provided by Prof. Jorge Piekarewicz \cite{piekarewicz_private} (for the RPA results using FSUGarnet) and Prof. Gianluca Col\`o \cite{colo_private} (for the QRPA calculations employing Skyrme interactions).}

\begin{figure}[t!]
  \centering
  \includegraphics[width=\linewidth]{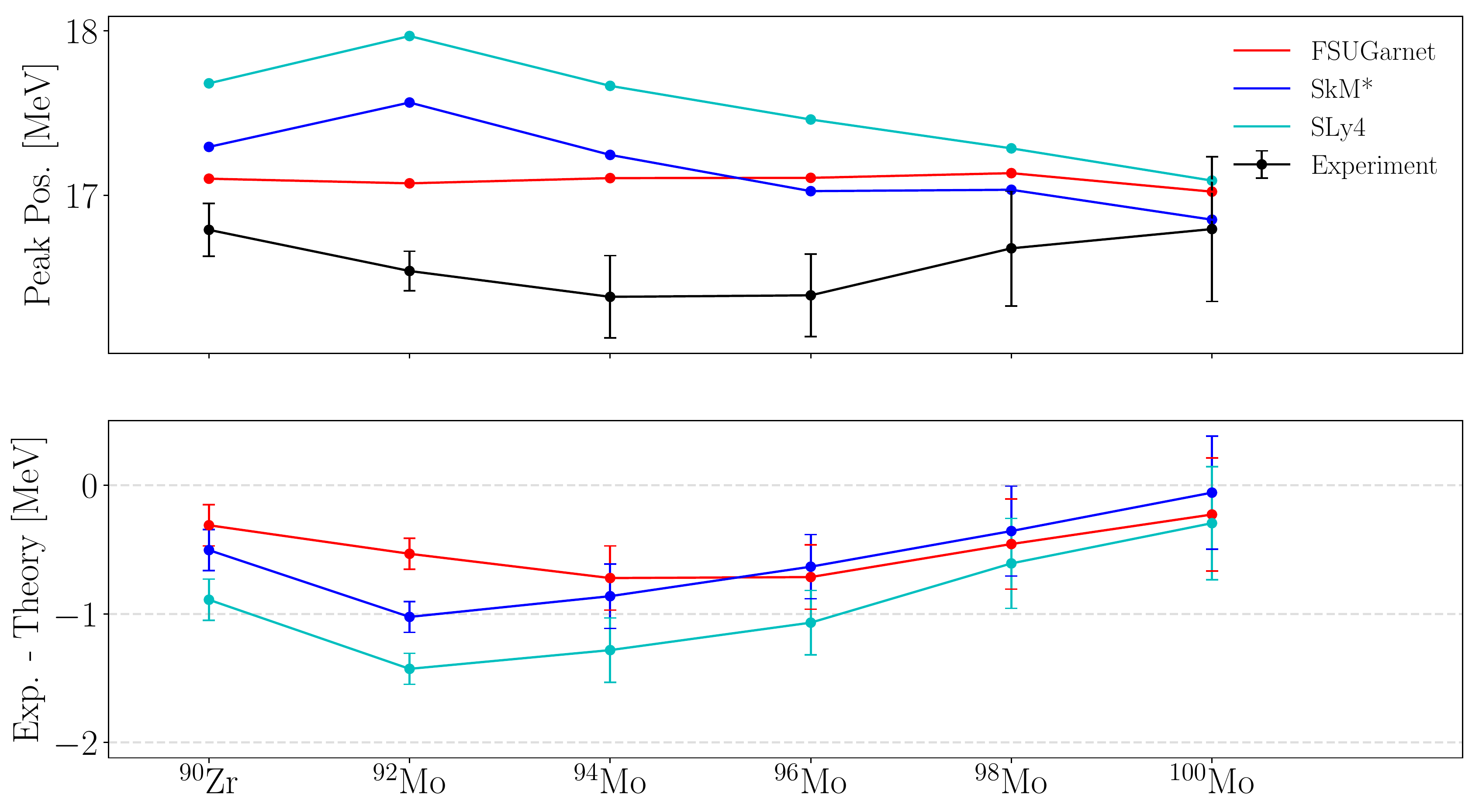}
  \caption[Comparison of the ISGMR peak centers for \nuc{90}{Zr} and \nuc{92-100}{Mo} with peak centers from theoretical calculations of the ISGMR strength.]{Comparison of the ISGMR peak centers for \nuc{90}{Zr} and \nuc{92-100}{Mo} with peak centers from theoretical calculations of the ISGMR strength using the FSUGarnet \cite{piekarewicz_private}, SLy4, and SkM$^*$ \cite{colo_private} interactions.}
  \label{a90_peak_positions}
\end{figure}

A close examination of Fig. \ref{strengths_a90_only} is illustrative in determining the local systematics of the ISGMR response as one moves away from \nuc{90}{Zr}. To quantify the agreement of the calculated ISGMR responses with the extracted strength distributions, Fig. \ref{a90_peak_positions} shows a comparison of the main ISGMR peak energy (as determined by a fit to a Lorentzian --- see Eq. \eqref{lorentz}) over the chain of nuclei \nuc{90}{Zr} and \nuc{92-100}{Mo} in the top panel, and the relative difference between the experimental peak centers and those of the theoretical calculations in the bottom panel. Bearing in mind that the effect of axial deformation on the ISGMR response is to moderately increase the position of the main ISGMR peak \cite{kvasil_deformation_analysis}, examination of Fig. \ref{a90_peak_positions} shows that there appears to be a mild softening of the molybdenum isotopes relative to \nuc{90}{Zr} which is later obfuscated by the manifesting effects of deformation in the higher-mass nuclei; this conclusion is drawn on the basis of the increase in the discrepancy between the theoretical and experimental peak positions in the \nuc{92-96}{Mo} isotopes, relative to those seen for \nuc{90}{Zr}.

The interpretation of this, following the prescription of Refs. \cite{blaizot,colo_2004a,garg_colo_review,cao_sagawa_colo}, is that the ISGMR of \nuc{92-96}{Mo} would suggest that a lower bulk nuclear incompressibility is appropriate for a microscopic description for the finite nuclei in this chain. Incidentally, the positioning of the ISGMR in all cases seems to prefer a slightly lower value of $K_\infty$ than one within the currently accepted range of $K_\infty = 240 \pm 20$ MeV; this is particularly evident within the nonrelativistic RPA calculations. SkM$^*$, with a relatively soft value for $K_\infty = 217$ MeV, is able to reproduce the ISGMR strength distributions reasonably well, as is FSUGarnet with $K_\infty = 230$ MeV; even still, the positioning of the extracted strength distributions seem to tend toward a lower $K_\infty$ than the presently adopted values.

At least a portion of this is conjectured to lie with the well-documented disagreement with the ISGMR energies of \nuc{90}{Zr} as determined by the present methodology \cite{gupta_A90_PLB,gupta_A90_PRC} and those of TAMU \cite{youngblood_90Zr_knm_240}. In short, previous constraints on the ISGMR were obtained on the basis of the latter \nuc{90}{Zr} data which placed the centroid energy $m_1/m_0 \approx 17.8$ MeV \cite{youngblood_90Zr_knm_240}, which is substantially higher than that reported in these analyses (see Table \ref{molly_moment_ratios_total_EWSRS}). Indeed, FSUGarnet directly used this \nuc{90}{Zr} centroid energy of Ref. \cite{youngblood_90Zr_knm_240} in its calibration \cite{piekarewicz_private}. It is therefore possible that the calibration of an interaction and associated $K_\infty$ value using the presently-described \nuc{90}{Zr} strength distribution would remedy a portion of this observed overestimation of the ISGMR strength position. However, one sees plainly that in a model-independent way, the ISGMR energy immediately drops in moving from \nuc{90}{Zr} to \nuc{92}{Mo}. To this end, we again point out that the molybdenum nuclei seem to be inconsistent with the value of $K_\infty$ which ought to best reproduce \nuc{90}{Zr}.

\begin{figure}[t!]
  \centering
  \includegraphics[width=\linewidth]{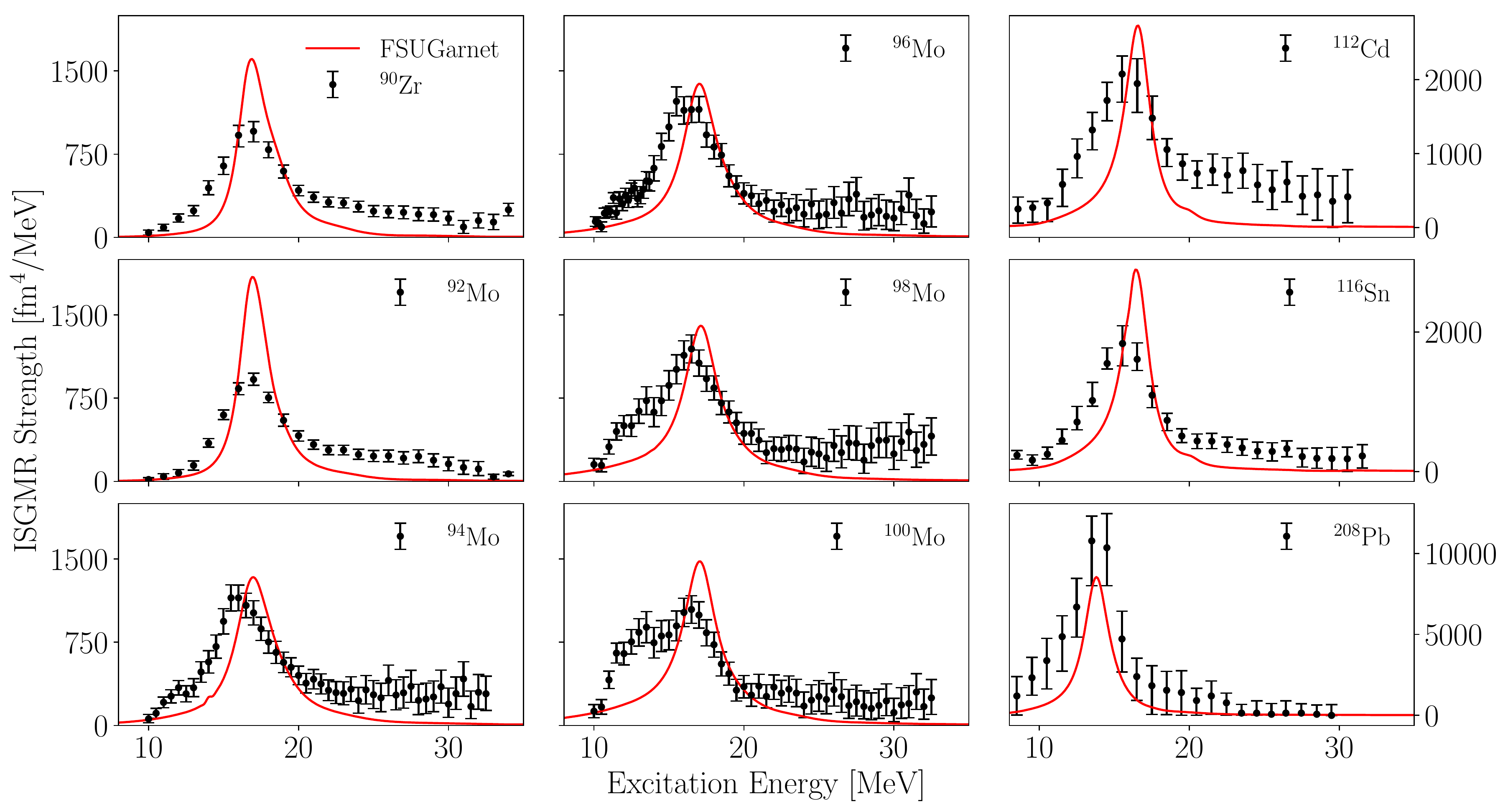}
  \caption[ISGMR strengths for \nuc{90}{Zr}, \nuc{92-100}{Mo}, \nuc{112}{Cd}, \nuc{116}{Sn}, and \nuc{208}{Pb}, and corresponding FSUGarnet RPA calculations.]{Extracted ISGMR strengths for \nuc{90}{Zr}, \nuc{92}{Mo} \cite{gupta_A90_PRC,KBH_EPJA}, \nuc{94-100}{Mo}, \nuc{112}{Cd} \cite{patel_cd}, \nuc{116}{Sn} \cite{Li_PRL,Li_PRC}, and \nuc{208}{Pb} \cite{patel_MEM}, in addition to relativistic RPA calculations with FSUGarnet \cite{piekarewicz_private,fsugarnet_commissioned}.}
  \label{FSUGarnet_strengths}
\end{figure}

If one endeavors to probe further the question as to whether the molybdenum nuclei are also soft in their ISGMR responses in the style of the tin and cadmium nuclei, as discussed in Subsection \ref{intro:softness_question}, it is necessary to examine the systematics of the ISGMR strength distributions of molybdenum in relation to those nuclei. To this end, Fig. \ref{FSUGarnet_strengths} shows the ISGMR strength distributions for the molybdenum nuclei of this work in comparison to those of \nuc{90}{Zr} \cite{gupta_A90_PRC,KBH_EPJA}, \nuc{112}{Cd} \cite{patel_cd}, \nuc{116}{Sn} \cite{Li_PRL,Li_PRC}, and \nuc{208}{Pb} \cite{patel_MEM}. RPA calculations using FSUGarnet \cite{fsugarnet_commissioned,piekarewicz_calcium_isotopes} are shown atop the experimental strength distributions for each of these nuclei across the nuclear chart; this interaction in particular is designed with a goal of constraining the bulk nuclear properties from the ISGMR response of doubly-magic or semi-magic nuclei, and so does not include the contributions due to pairing or deformation effects on the overall ISGMR response \cite{piekarewicz_private}; the effect of pairing on the ISGMR has been shown to generally decrease the peak energy of the ISGMR in the case of the tin isotopes \cite{junli_pairing}, but has been clearly shown to be insufficient to fully account for the discrepancy that is attributable to the softness of the tin and cadmium nuclei \cite{patel_MEM,patel_cd,khan_pairing}.

\begin{figure}[t!]
  \centering
  \includegraphics[width=\linewidth]{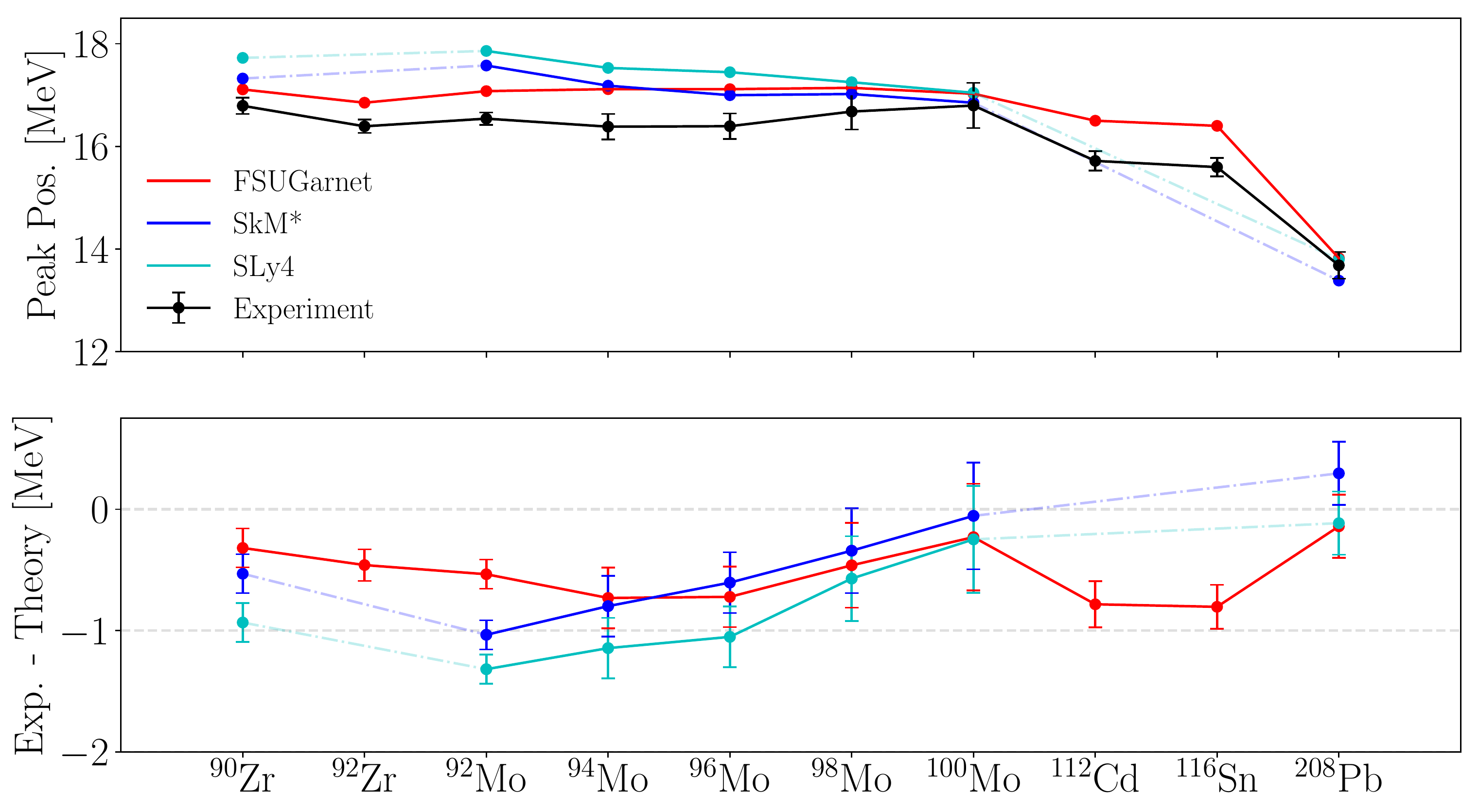}
  \caption[Comparison of the ISGMR peak centers for \nuc{90,92}{Zr}, \nuc{92-100}{Mo}, \nuc{112}{Cd}, \nuc{116}{Sn}, and \nuc{208}{Pb} with peak centers from theoretical calculations of the ISGMR strength.]{Comparison of the ISGMR peak centers for \nuc{90,92}{Zr}, \nuc{92}{Mo} \cite{gupta_A90_PRC,KBH_EPJA}, \nuc{94-100}{Mo}, \nuc{112}{Cd} \cite{patel_cd}, \nuc{116}{Sn} \cite{Li_PRL,Li_PRC}, and \nuc{208}{Pb} \cite{patel_MEM} using the FSUGarnet \cite{piekarewicz_private} interaction, SkM$^*$, and SLy4 \cite{colo_private}.}
  \label{FSUGarnet_peak_positions}
\end{figure}

\begin{figure}[t!]
  \centering
  \includegraphics[width=\linewidth]{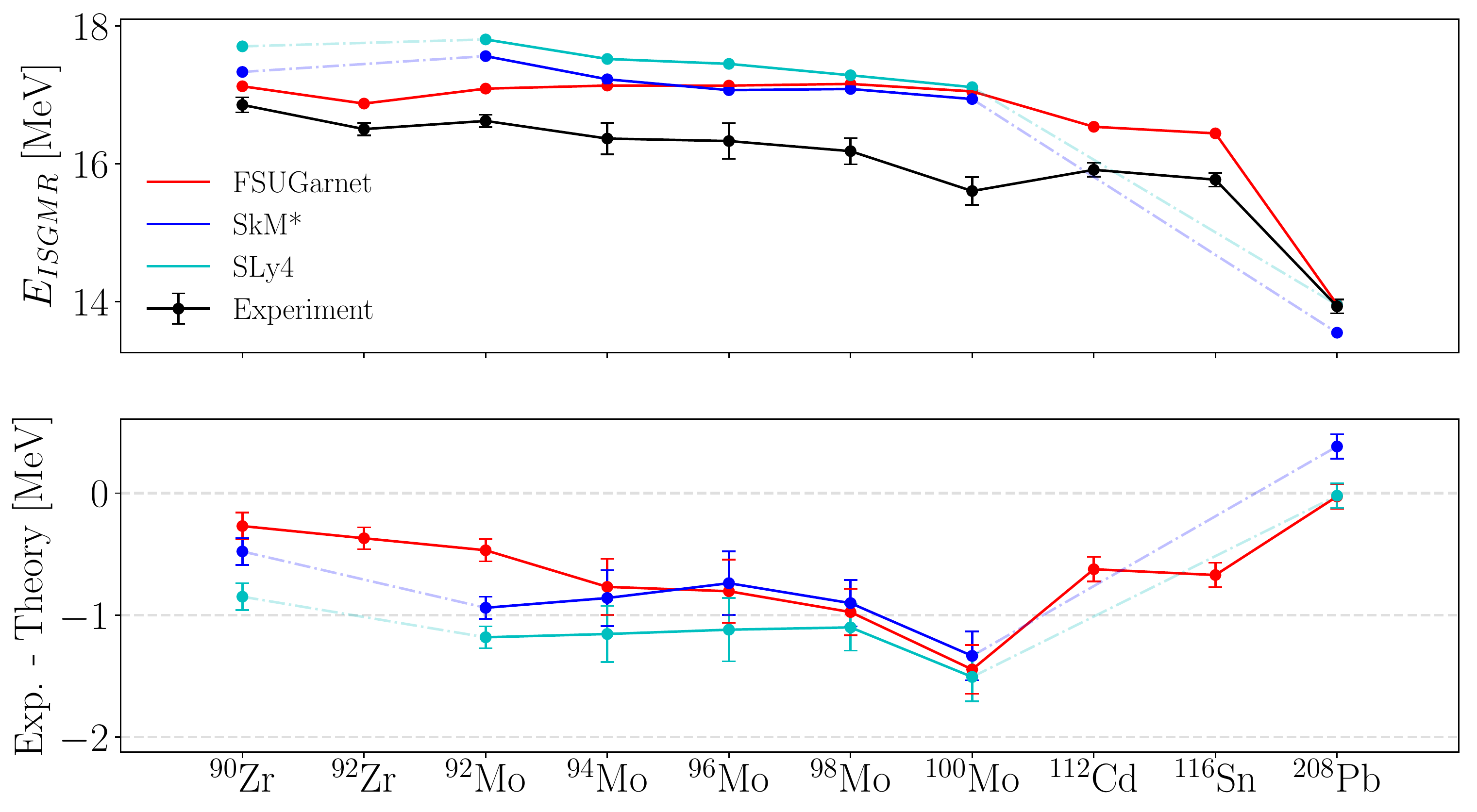}
  \caption[Comparison of the ISGMR centroid energies for \nuc{90,92}{Zr}, \nuc{92-100}{Mo}, \nuc{112}{Cd}, \nuc{116}{Sn}, and \nuc{208}{Pb} with centroids from theoretical calculations of the ISGMR strength.]{Comparison of the ISGMR centroid energies for \nuc{90,92}{Zr}, \nuc{92}{Mo} \cite{gupta_A90_PRC,KBH_EPJA}, \nuc{94-100}{Mo}, \nuc{112}{Cd} \cite{patel_cd}, \nuc{116}{Sn} \cite{Li_PRL,Li_PRC}, and \nuc{208}{Pb} \cite{patel_MEM} using the FSUGarnet \cite{piekarewicz_private} interaction, SkM$^*$, and SLy4 \cite{colo_private}.}
  \label{FSUGarnet_centroids}
\end{figure}

Bearing this fact in mind, a comparison of the theoretical strength distributions and those extracted from experiment suggests a general qualitative consistency between the peak positions across the chain of nuclei, particularly considering the estimate of typical theoretical uncertainties being approximately 100 keV \cite{piekarewicz_private}. Figures \ref{FSUGarnet_peak_positions} and \ref{FSUGarnet_centroids} show the corresponding ISGMR peak centers and centroid energies ($m_1/m_0$) extracted from the FSUGarnet, SkM$^*$, and SLy4 calculations in relation to the corresponding values extracted from the experimental data on \nuc{90,92}{Zr} \cite{gupta_A90_PRC,KBH_EPJA}, \nuc{92}{Mo} \cite{KBH_EPJA}, \nuc{94-100}{Mo}, \nuc{112}{Cd} \cite{patel_cd}, \nuc{116}{Sn} \cite{Li_PRL,Li_PRC}, and \nuc{208}{Pb} \cite{patel_MEM}.

In the case of \nuc{98,100}{Mo}, the peak positions increase whereas the centroid energies decrease as a result of the development of a low-energy peak in the strength distribution; this effect is believed to be accountable to the manifestation of the aforementioned deformation effects. The presence of axial deformation has been shown to result in a coupling between the monopole and $K=0$ component of the ISGQR as discussed previously; further theoretical work to examine the properties of the deformation of the molybdenum nuclei as pertains to the giant resonances and warrants future study outside of the scope of this dissertation. It is nonetheless our position that the changes in the ISGMR energies of \nuc{98-100}{Mo} as presented graphically in this work should be considered separately in light of this possibility so as to make a crisp judgment on the presence of softness.

We first point out that in both the peak-center plot and the centroid-energy plot, the ISGMR energy of \nuc{208}{Pb} is well-reproduced by FSUGarnet and SLy4, and underestimated by SkM$^*$. What is of note in each of these trends, however, is not only the relative softness that is observed in the experimental ISGMR response of \nuc{90}{Zr}, but the development of the softness in moving from \nuc{90}{Zr} to \nuc{96}{Mo} for interactions which are capable of reproducing the energies of \nuc{208}{Pb}. It is of even further interest that the difference between the theoretical energies and those extracted from experiment exhibit striking agreement between \nuc{94-96}{Mo} and the previously-studied nuclei \nuc{112}{Cd} and \nuc{116}{Sn} in the case of the FSUGarnet comparison. This is suggestive that the molybdenum nuclei exhibit precisely the same open-shell softness that has been previously documented in the cases of the cadmium and tin isotopic chains.



\section{ISGMR in the calcium nuclei}

\begin{figure}[t!]
\centering
\includegraphics[width=\linewidth]{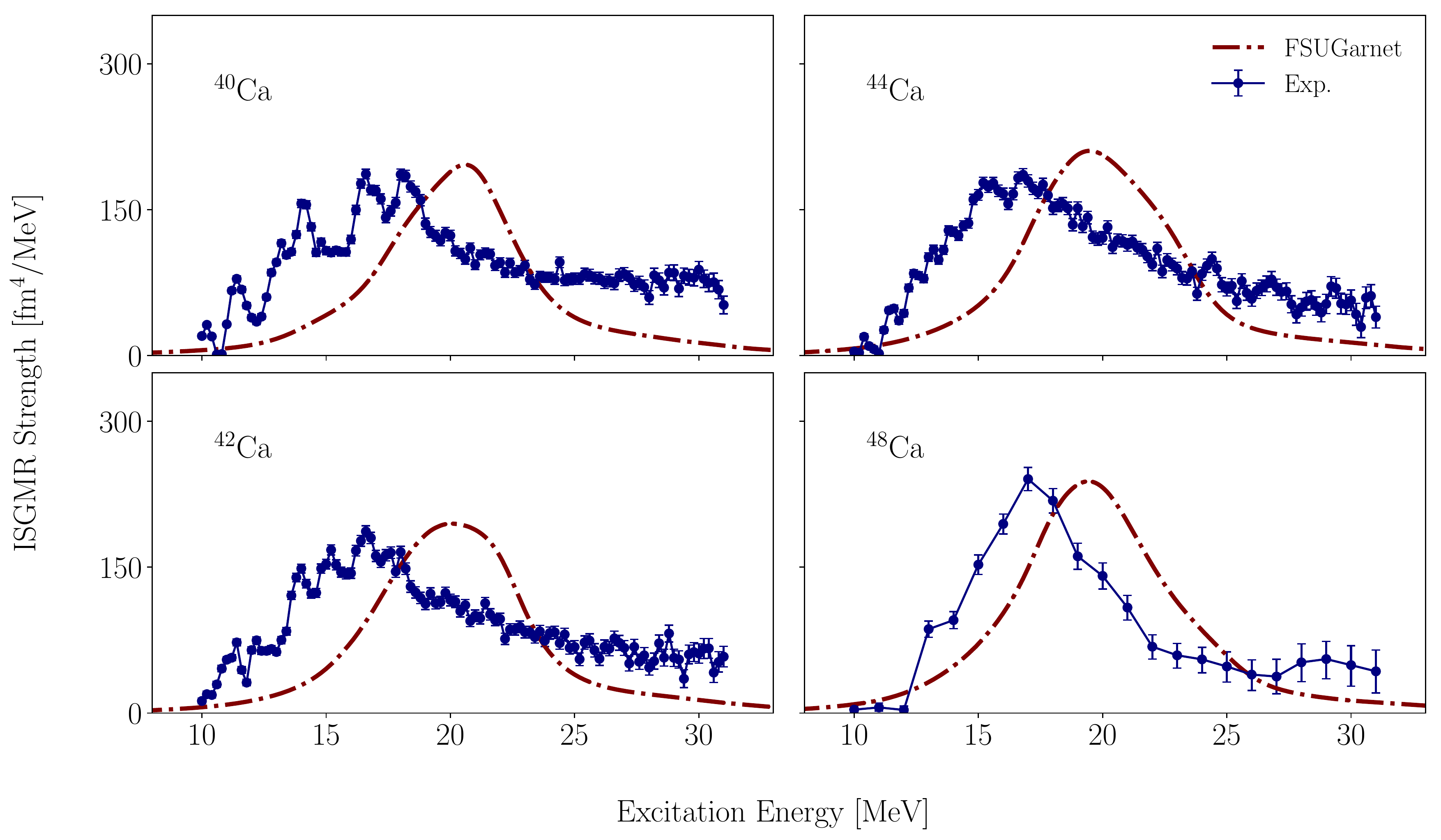}
\caption[ISGMR strength distributions for \nuc{40,42,44,48}{Ca} \cite{KBH_calcium_PLB}, along with relativistic RPA calculations with FSUGarnet.]{ISGMR strength distributions for \nuc{40,42,44,48}{Ca} \cite{KBH_calcium_PLB}, in addition to relativistic RPA calculations for the ISGMR response from Ref. \cite{piekarewicz_calcium_isotopes}.}
\label{calcium monopole strength}
\end{figure}

\begin{table}[t!]
\centering
\caption[Percentages of the EWSR exhausted and moment ratios for the ISGMR strength distribution in \nuc{40,42,44,48}{Ca}]{Percentages of the EWSR ($m_1$) for the ISGMR strength distributions and the corresponding moment ratios [in M\lowercase{e}V] }
\label{calcium_moment_ratios_total_EWSRS}
\begin{tabular}{@{}ccccccccc@{}}
\toprule
 & \multicolumn{4}{c}{RCNP} &  & \multicolumn{3}{c}{TAMU \cite{TAMU_40Ca,TAMU_44Ca,TAMU_48Ca}}  \\ \cmidrule(lr){2-5} \cmidrule(l){7-9}
 & \nuc{40}{Ca} & \nuc{42}{Ca} & \nuc{44}{Ca} & \nuc{48}{Ca} &  & \nuc{40}{Ca} & \nuc{44}{Ca} & \nuc{48}{Ca} \\
$m_1$ \% & $102^{+3}_{-4}$ & $89^{+3}_{-3}$ & $88^{+4}_{-4}$ & $78^{+4}_{-3}$ &  & $97^{+11}_{-11}$ & $75^{+11}_{-11}$ & $95^{+11}_{-15}$ \\
$\displaystyle \sqrt{\frac{m_1}{m_{-1}}}$  & $19.5^{+0.1}_{-0.1}$ & $19.0^{+0.1}_{-0.1}$ & $18.9^{+0.1}_{-0.1}$ & $19.0^{+0.2}_{-0.2}$ &  & $18.3^{+0.30}_{-0.30}$ & $18.73^{+0.29}_{-0.29}$ & $19.0^{+0.1}_{-0.1}$ \\
$\displaystyle {\frac{m_1}{m_{0}}}$ & $20.2^{+0.1}_{-0.1}$ & $19.7^{+0.1}_{-0.1}$ & $19.5^{+0.1}_{-0.1}$ & $19.5^{+0.1}_{-0.1}$ &  & $19.2^{+0.40}_{-0.40}$ & $19.5^{+0.35}_{-0.35}$ & $19.9^{+0.2}_{-0.2}$ \\
$\displaystyle \sqrt{\frac{m_3}{m_{1}}}$ & $22.3^{+0.1}_{-0.1}$ & $21.7^{+0.1}_{-0.1}$ & $21.5^{+0.1}_{-0.1}$ & $21.3^{+0.3}_{-0.3}$ &  & $20.6^{+0.40}_{-0.40}$ & $21.78^{+0.84}_{-0.72}$ & $22.6^{+0.3}_{-0.3}$ \\ \bottomrule
\end{tabular}
\end{table}

Figure \ref{calcium monopole strength} shows the extracted ISGMR strength distributions for each of the calcium nuclei investigated in E495 \cite{KBH_calcium_PLB}. Owing in part to the lighter mass of the nuclei, the resonances lie substantially higher in energy relative to the extracted molybdenum nuclei. Furthermore, there is a clear increase in the amount of fine structure present in the extracted strength owing to the well-documented fragmentation of the ISGMR in light-mass nuclei \cite{harakeh_book,von-nuemann-cosel_fine_structure}. Due to this fact, the ISGMR strength is not well-characterized by a single (or pair of) coherent peak(s) as were the molybdenum nuclei; the only means by which the moments of the strength distributions can be meaningfully characterized is by direct integration of the extracted strength over the full energy range. Furthermore, this is the same procedure which was followed by Refs. \cite{TAMU_40Ca,TAMU_44Ca,TAMU_48Ca}, and we proceeded accordingly while ensuring that all aspects of the $K_A$ extraction that followed were consistent with those of the aforementioned references.

Also shown in Fig. \ref{calcium monopole strength} are the corresponding ISGMR responses predicted by FSUGarnet \cite{piekarewicz_calcium_isotopes,fsugarnet_commissioned,piekarewicz_private}. It is worth noting that although FSUGarnet seems to over-predict the ISGMR energies of the calcium nuclei, it predicts a consistently decreasing ISGMR softening with increasing neutron excess in the stable calcium isotopes \cite{piekarewicz_calcium_isotopes} (see Table \ref{interactions_table}).


\subsection{Leptodermous analysis of the ISGMR in the calcium isotopes}

The moment ratios $\sqrt{m_1/m_{-1}}$, $m_1/m_0$, and $\sqrt{m_3/m_1}$ that are customarily used in characterizing the excitation energy of the ISGMR \cite{stringari_lipparini_sum_rules} are presented in Table \ref{calcium_moment_ratios_total_EWSRS} \cite{KBH_calcium_PLB}. The quoted uncertainties have been estimated using a Monte Carlo sampling (cf. Algorithm \ref{metropolis_sketch}) from the probability distributions of the individual $A_\lambda(E_x)$ and constitute a $68\%$ confidence interval. The pattern of moment ratios observed in the calcium isotopic chain (decreasing with $A$, as expected from the $A^{-1/3}$ rule)  is contrary to that reported in Ref. \cite{TAMU_44Ca} {\em viz.} increase in the moment ratios with increasing $A$.

\begin{figure}[t!]
  \centering
  \includegraphics[width=\linewidth]{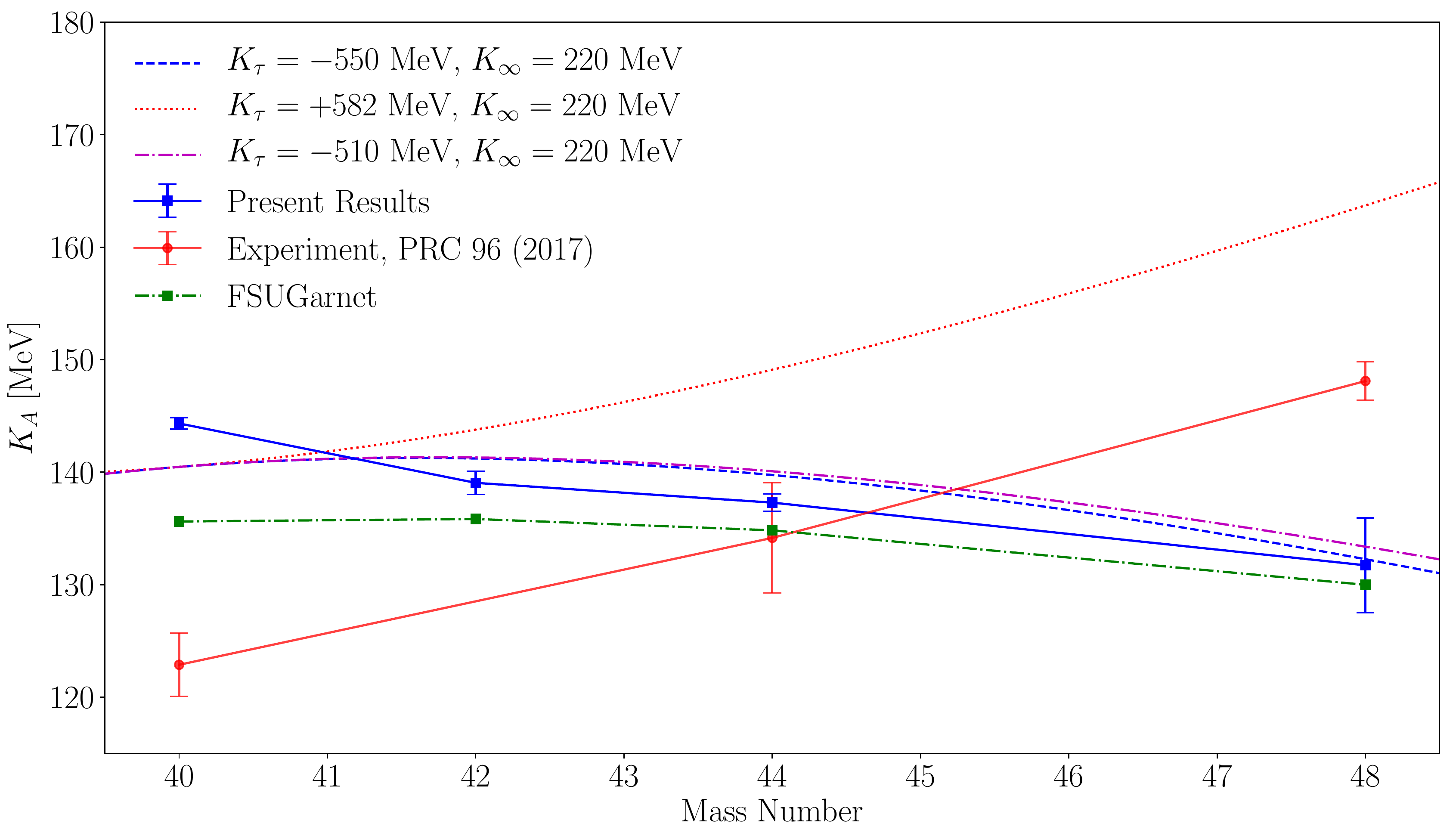}
  \caption[The extracted scaling-model incompressibilities for the calcium isotopes investigated in this work alongside various model predictions.]{The incompressibility, $K_A$, for the calcium isotopes investigated in this work (blue squares). These were calculated within the scaling model from the experimental data ($E_\text{ISGMR} = \sqrt{m_3/m_1}$, for consistency with the presentation of Ref. \cite{TAMU_44Ca}; see Table \ref{calcium_moment_ratios_total_EWSRS}). The expected trend for these values utilizing the previously documented central value for $K_\tau = -550$ MeV, and $K_\infty = 220$ MeV as input to Eq. \eqref{lepto} is presented (blue dashed line), along with the same calculation but with the value $K_\tau = +582$ MeV reported in Ref. \cite{TAMU_44Ca} (red dotted line). A fit to the data leads to a curve that is nearly identical to that shown above (blue dashed line) and leads to a value of $K_\tau = -510 \pm 115$ MeV. For comparison purposes, the data from Ref. \cite{TAMU_44Ca} are shown (red circles), as well as the $K_A$ values calculated from the ISGMR responses predicted by the relativistic FSUGarnet interaction (green squares) \cite{piekarewicz_private,piekarewicz_calcium_isotopes}. The solid lines through the data points are merely to guide the eye. Figure adapted from Ref. \cite{KBH_calcium_PLB}.}
  \label{RCNP KA}
\end{figure}

In addition to the moment ratios exhibiting the expected behavior over the isotopic chain, the demonstrated trend for the extracted finite incompressibilities, $K_A$, is even more illustrative (see Fig. \ref{RCNP KA}): The agreement of the extracted $K_A$ values with the behavior modeled by the leptodermous expansion of Eq. \eqref{lepto} using the accepted values for $K_\tau$ and $K_\infty$ is rather good, and stands in stark contrast to the results from Ref. \cite{TAMU_44Ca}. While the extracted $K_A$ for \nuc{44}{Ca} is consistent with that which was measured in Ref. \cite{TAMU_44Ca}, the $K_A$ for the extrema of \nuc{40}{Ca} and \nuc{48}{Ca} follow precisely opposite trends between the two analyses. However, the presence of an additional data point for \nuc{42}{Ca} --- which was absent in the TAMU analysis --- that follows the same general trend as the other three isotopes found in the present work inspires greater confidence in our results. These data, thus, conclusively exclude the possibility of a positive $K_\tau$ value for the calcium nuclei.

Also presented in Fig. \ref{RCNP KA} are the $K_A$ values derived from the \nuc{40,42,44,48}{Ca} strength distributions predicted by the FSUGarnet \cite{piekarewicz_private,piekarewicz_calcium_isotopes,fsugarnet_commissioned} relativistic interaction. The $K_\tau^\infty$ has a moderate value of $- 247.3$ MeV for this particular interaction and, accordingly, the $K_A$ values are observed to decrease over the isotopic chain, qualitatively similar to the experimental results. This trend is indeed expected, and observed, for the overwhelming majority of interactions and models \cite{sagawa_private}.

%

%
%
%
%
%
%
%
%
%
%

%
%
\chapter{Conclusions and future work} \label{conclusions}

The isoscalar giant monopole resonance is a direct means by which nuclear physicists can constrain the density dependence of the nuclear equation of state (EoS) close to saturation density. The EoS is a constitutive equation that describes the thermodynamic link between the particle density and nucleon energy in bulk systems. Figures \ref{intro:eos_symmetric_figure} and \ref{mass_radius_plot} show but a small subset of possible nonrelativistic interactions which yield unique equations of state and mass radius relations for gravitationally-bound pure neutron matter; the takeaway from this is that there is an ever-present demand for increased constraints on the density-dependence of the EoS and specifically, for the EoS of asymmetric nuclear matter. The theory which underpins modern-day efforts to extract bulk properties of infinite nuclear matter from finite nuclei is predicated on a smoothly-varying ISGMR response over the chart of nuclides as predicted by the hydrodynamical model \cite{harakeh_book}; it is thus of paramount importance to fully investigate and comprehend any and all structure effects which manifest in the nuclear chart which might influence the extrapolation from finite nuclei to nuclear matter.

In this work, we have studied two such structure effects: the first being the evasive and provocative question as to the origin of the open-shell softness in the tin \cite{Li_PRL,Li_PRC} and cadmium \cite{patel_cd} isotopes, and the second being the highly-unexpected result of a positive asymmetry component $K_\tau$ in the expansion of the finite nuclear incompressibility as reported by Ref. \cite{TAMU_44Ca}. The first question was investigated via a systematic experimental campaign to extract the ISGMR strengths of \nuc{94,96,97,98,100}{Mo} via inelastic $\alpha$-scattering at the Research Center for Nuclear Physics (RCNP); the second study was completed on \nuc{40,42,44,48}{Ca} to reproduce the measurement of Ref. \cite{TAMU_44Ca} using the same experimental methodology at RCNP.

In either case, inelastic angular distributions were measured for each reaction over an excitation energy range of $10-30$ MeV, and with an angular range of $0-10^\circ$. Multipole decomposition analyses (MDA) were completed in which the overlapping giant resonance responses were disentangled on the basis of characteristic angular distributions as predicted by Distorted Wave Born Approximation calculations using optical models which were constrained by contemporaneous measurements of the elastic scattering channels. The MDA output yielded, for each energy bin, the percentages of the energy-weighted sum rules (EWSR) which were exhausted by the giant resonances as measured by their contributions to the experimental spectra. This methodology culminated in direct measurements of the strength distributions of the isoscalar giant resonances for each of the nuclei of interest.

The present results suggest that the ISGMR response of the nuclei in the molybdenum region of the nuclear chart prefer softer interactions with $K_\infty \lesssim 230$ MeV. It should be noted that even SkM$^*$, with $K_\infty = 217$ MeV, is unable to reproduce the positioning of the ISGMR strength for the molybdenum isotopes. Even further, inspection of the ISGMR energies extracted from direct fitting and the centroid energies show that FSUGarnet --- which is an interaction carefully designed to crisply extract properties of bulk nuclear matter --- tends to put the responses of \nuc{94-96}{Mo}, \nuc{112}{Cd}, and \nuc{116}{Sn} on similar footing with regards to the value for the nuclear incompressibility which ought to reproduce the ISGMR responses of these open-shell nuclei. This culminates in the conclusion of the first portion of this dissertation: it seems that the consistency between the amounts by which the presented interactions tend to overestimate the ISGMR energies of the molybdenum nuclei --- as measured by the peak centers and the centroid energies --- conclusively indicates that the molybdenum isotopes are soft in the styles of the tin and cadmium nuclei. 

While the experimental results for \nuc{94-100}{Mo} suggest clearly that the molybdenum isotopes are soft in a manner similar to those found in the cases of the tin and cadmium nuclei, the explanation for this softness remains elusive. The data presented here offer a few clues for possible future measurements that may be of use in elucidating the origin of this softness. First, a close observation of Figs. \ref{FSUGarnet_peak_positions} and \ref{FSUGarnet_centroids} suggests that \nuc{90}{Zr} itself may be soft in relation to \nuc{208}{Pb}. Further work in this direction is clearly necessary to explain this effect; previous works have suggested that perhaps \nuc{208}{Pb} instead presents with a stiffer ISGMR response than other nuclei \cite{khan_MEM,khan_pairing}, and it is possible that the responses of \nuc{90}{Zr} and \nuc{208}{Pb} can be consistently accounted for with modifications to the symmetry energy which has has been shown to be correlated with the nuclear incompressibility \cite{piekarewicz_correlating_GMR_to_incompressibility,piek_fluffy_physRev}. It is further possible that measurements of remaining isotopes between zirconium and tin, such as the ruthenium nuclei, may shed further light onto the phenomenon as pertains to the softening which has now been conclusively documented in three isotopic chains within this region of the nuclear chart over the past two decades.

In the case of the extraction of $K_\tau$ from the ISGMR responses of the calcium nuclei, the results conclusively discount the possibility of a positive $K_\tau$ and further, are entirely consistent with pre-existing measurements for the quantity. This result does, of course, beg the question as to the origin of the egregious differences in the extracted ISGMR responses for \nuc{40-48}{Ca} relative to those obtained by the TAMU group in Refs. \cite{TAMU_40Ca,TAMU_44Ca,TAMU_48Ca}. There have been suggestions that the origin of these discrepancies lies within the methodology for subtracting the instrumental background and accounting for the physical continuum between the two experimental techniques; it would be supremely helpful if a third-party measurement were completed to serve as an independent arbiter on these results. Nonetheless, $K_\tau$ is extracted from the calcium nuclei to be $-510 \pm 115$ MeV, and is consistent with the values obtained in the tin \cite{Li_PRL,Li_PRC} and cadmium \cite{patel_cd} nuclei. In summary, the ISGMR strength distributions of \nuc{40-48}{Ca}, and the metrics that are generally used to characterize the excitation energy of the ISGMR, follow expected trends. It may be concluded, therefore, that there are no local structure effects on the ISGMR strength distribution in the calcium region of the nuclear chart and that a {\em positive} value for the asymmetry term of nuclear incompressibility, $K_{\tau}$, is ruled out.

%


\appendix

%
%
%
%
%
%
%
%
%
%

%
%

\chapter{Angular acceptance of the spectrograph, collimators, and calculation of the average spectrographic scattering angle}
\label{averaging_appendix}

\section{Angular acceptance and bin choices}

It is a feature of spectrographs that they have a finite angular acceptance along each axis in a two-dimensional angular plane. The solid angle, given a horizontal and vertical acceptance, is merely:
\begin{align}
\Delta \Omega = \Delta \phi \Delta \theta,
\end{align}
where $\Delta \phi$ and $\Delta \theta$ are the respective widths of the acceptances in the vertical and horizontal scattering directions in the calculation.

Experimentally, a series of collimators was employed to constrain the scattered beam so as to have a well-defined solid angle from which the cross-section could be extracted. At the exit of the scattering chamber of Grand Raiden, collimators with vertical acceptance $\Delta \phi = 40$ mrad and $\Delta \phi = 60$ mrad were used. The horizontal acceptance was specified via offline analysis gates. For the $0$ degree runs, the horizontal angle $\theta$ was subdivided into three equally spaced bins between $-0.6^\circ \leq \theta \leq 0.6^\circ$, each with horizontal acceptance $\Delta \theta = 0.4^\circ$. The symmetric bins from $0.2^\circ \leq |\theta| \leq 0.6^\circ$ were merged due to symmetry considerations and to optimize the statistics of the zero-degree measurement. For all other runs, horizontal collimation allowed for subdivision into four bins within the range $-0.8^\circ \leq \theta \leq 0.8^\circ$.


\newcommand{\GRANGLE}{\theta_\text{GR}}

\section{Angle averaging}
The finite acceptances of the spectrograph in both the vertical and horizontal directions require that the angle assigned to particles scattering into a solid angle $\Delta \Omega$ is averaged over all possible polar angles taken by the possible accepted scattering trajectories. The possible horizontal scattering angles for a given bin were assigned in terms of the the angular setting of Grand Raiden, denoted $\GRANGLE$, and the bounds of the horizontal acceptance for the bin, as described in the previous section. So, the center of the horizontal bin, $\theta_\text{cent}$ is given by:

\begin{align}
  \theta_\text{cent} & = \GRANGLE + \text{mid}\left(\theta_\text{low},\theta_\text{high}\right),
\end{align}
where $\text{mid}(x,y)=(x+y)/2$, and $\theta_\text{low}$, $\theta_\text{high}$ are the lower and upper bounds, respectively, of the horizontal angular acceptance of the bin in question.

Given this, the possible scattering angles $\theta_x$ in the horizontal direction for a given bin are constrained via
\begin{align}
  \theta_\text{cent} - \frac{1}{2} \Delta \theta \leq \theta_x \leq \theta_\text{cent} + \frac{1}{2} \Delta \theta.
\end{align}
In all cases, the vertical scattering angles are constrained solely by the vertical acceptance of the collimator employed during that particular run, as described in the previous section. Thus,
\begin{align}
- \frac{1}{2} \Delta \phi \leq \phi_y \leq \frac{1}{2} \Delta \phi.
\end{align}

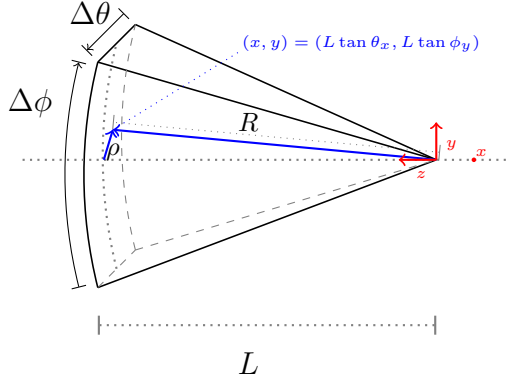
\begin{figure}
  \centering
  \begin{tikzpicture}
    \coordinate (O) at (0.5,0.5);
    \draw[semithick] (0,0) -- (4.5,1.7);
    \draw[dashed,color=gray] (O) -- ($(4,1)+(O)+(0,0.2)$);
    \draw[semithick] (0,3) -- (4.5,1.7);
    \draw[semithick] ($(0,3)+(O)$) -- ($(4,2)+(O)+(0,-0.8)$);
    \draw[semithick] (0,3) -- ($(0,3)+(O)$);
    \draw[dashed,color=gray] (0,0) -- (O);
    \draw[semithick] (0,0) arc (194.036:165.964:6.185);
    \draw[dashed,color=gray] (O) arc (194.036:165.964:6.185);

    \draw[|<-|,color=black] (-0.15,3.15)-- (0.35,3.65) ; 
    \draw[|->|,color=black] (-0.25,0) arc (194.072:165.928:6.185); 

    \draw (-0.9,2.45) node {$\Delta \phi$};
    \draw (-0.1,3.65) node {$\Delta \theta$};
    \draw[|-|,thick,color=gray,dotted] (0,-0.5) -- (4.5,-0.5); 

    \draw[->,color=blue,thick] (4.5,1.7) -- (0.20,2.1); 
    \draw [|-|,color=gray,dotted] (4.55,1.8) -- (0.20,2.2); 
    \draw[thick, dotted,color=gray] (0.25,0.25) arc (194.036:165.964:6.185);
    \draw[thick, dotted, color=gray] (-1.0,1.7) -- (5.5,1.7);

    \draw (2,-1) node {$L$};
    \draw (3.5,3.25) node [color=blue] { \tiny $(x,y) = (L \tan \theta_x, L \tan \phi_y)$};
    \draw [dotted,blue] (0.2,2.1) -- (1.9,3.15);

    \coordinate (origin) at (4.7,1.7);

    \draw [thick,color=red,->] (4.5,1.7) -- (4.0,1.7);
    \draw [thick,color=red,->] (4.5,1.7) -- (4.5,2.2);
    \node[circle,draw=red, fill = red, line width=0.25mm, inner sep=0pt,minimum size=1pt] (1) at(5.0,1.7) {};
    \draw [color = red] (5.1,1.8) node {\tiny $x$};
    \draw [color = red] (4.7,1.9) node {\tiny $y$};
    \draw [color = red] (4.3,1.5) node {\tiny $z$};

    \draw [->,color = blue,thick] (0.08,1.7) -- (0.2,2.1);
    \node [] at (0.2,1.8) {\footnotesize $\rho$};
    \draw [] (2.0,2.2) node {\footnotesize $R$};
    \node[circle,fill=none,line width = 1.0mm,minimum size=3pt] (2) at (5.0,1.7) {};
  \end{tikzpicture}
  \caption[Schematic of a possible scattering trajectory into the acceptance of the collimator.]{Schematic of scattering trajectory into acceptance of collimator. Shown are the coordinate system and origin used in the derivation in the text and lines of increasing $\phi_y$ and increasing $\theta_x$, with the maximum bounds of each angular range shown. Also shown in blue is a possible scattering trajectory, with its $x$ and $y$ positions decomposed using the scattering angles and the given geometry.}
\end{figure}

The angles $\theta_x$ and $\phi_y$ uniquely specify, respectively, the horizontal and vertical positions of the scattered particle at the entrance of the spectrograph, located a distance $L$ downstream from the scattering chamber. It is to be emphasized that one cannot average these angles directly: the angular space in either direction constitute a space wherein the angles that define the coordinates are multi-valued. To correctly average these quantities, they should be mapped to Cartesian space\footnote{For example, particles scattering at horizontal angles $\pm 0.3^\circ$ could be mistakenly taken to have an arithmetic mean of $0^\circ$, but equivalently, one of $180^\circ$.}.  In this Cartesian representation, one can then average over the bounds of the collimator acceptance and the offline gates. Following this logic, the horizontal and vertical positions $x$ and $y$ are given via:
\begin{align}
    x &= L \tan \theta_x \\
    y &= L \tan \phi_y.
\end{align}
From this, the 2-dimensional distance from the center of the acceptance can be calculated:
\begin{align}
  \rho^2 &= x^2 + y^2 \\
  &= L^2 \left(\tan^2 \theta_x + \tan^2 \phi_y\right). \label{rho2}
\end{align}
The distance from the scattering chamber to the position at the acceptance is denoted by $R$. In a spherical coordinate system centered at the scattering point, then $\rho$ is written in terms of the polar angle $\Theta$,
\begin{figure}[t]
  \centering
  \includegraphics[width=1.0\linewidth]{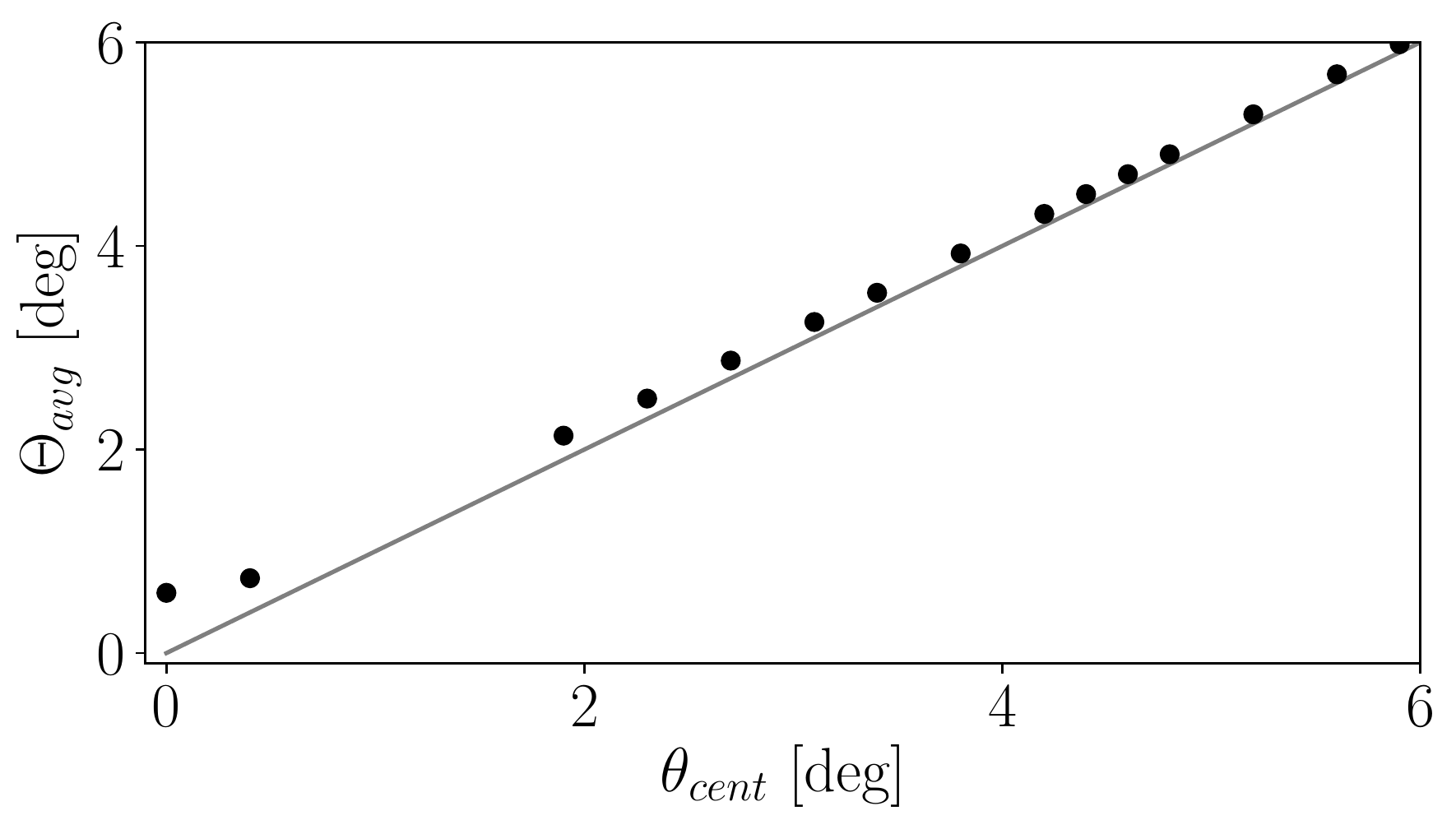}
  \caption{Overlay of potential assigned-scattering angles for both $\theta_\text{cent}$ as well as $\Theta_\text{avg}$ using $\Delta \phi$, $\Delta \theta$ described in the text.}
  \label{averaging_plot}
\end{figure}
\begin{align}
\rho = R \sin \Theta.
\end{align}
Recall that the relationship between the radial coordinate in cylindrical and spherical coordinates is $\rho = R \sin \Theta$. This, combined with the fact that $z = L$ in our cylindrical coordinate system and that in spherical polar coordinates, $z=R \cos\Theta$, we have that $L = \rho \cos\Theta / \sin\Theta$. Thus, we can write Eq. \eqref{rho2} in the form:
\begin{align}
  \rho^2 &= L^2 \left(\tan^2 \theta_x + \tan^2 \phi_y\right) \notag \\
  &= \frac{\rho^2}{\tan^2\Theta} \left(\tan^2 \theta_x + \tan^2 \phi_y \right),
\end{align}
and thus we have
\begin{align}
  \Theta(\theta_x,\phi_y) = \arctan \left(\sqrt{\tan^2 \theta_x + \tan^2 \theta_y} \right).
\end{align}

Application of the Mean-Value Theorem for integration yields the averaged scattering angle, taking into account the finite acceptance in both the vertical and horizontal directions,
\begin{align}
\Theta_\text{avg} = \frac{1}{\Delta \theta \Delta \phi} \int_{-\Delta\phi/2}^{\Delta\phi/2} \int_{\theta_\text{cent}-\Delta\theta/2}^{\theta_\text{cent}+\Delta\theta/2} \Theta(\theta_x,\phi_y) \diff \theta_x \diff \phi_y . \label{averaged_angle_result}
\end{align}

Observation of Fig. \ref{averaging_plot} indicates that the difference between an assignment of the scattering angle as the horizontal bin center, without correctly averaging over both directions, is largest at forward angles where the bounds of integration in Eq. \eqref{averaged_angle_result} are most comparable. At all but the most forward angles, $\GRANGLE$ is large in comparison to the vertical acceptance $\Delta \phi$, which constrained by the fixed geometry of the collimator.

As discussed in Chapter \ref{experimental}, in order to optimally constrain the ISGMR on the basis of angular distribution data, it is necessary for one to have sufficiently forward-angle cross sections. In order to achieve angular distribution data sufficiently close to $0^\circ$, it is thus critical to constrain the vertical and horizontal acceptances subject to Eq. \eqref{averaged_angle_result}; this fact, in combination with the constraint on available beamtime and simultaneous demand for small statistical uncertainties, informed the choice of both horizontal and vertical $\pm 20$ mrad collimators for the $0^\circ$ measurements of this thesis work.

%

%
%
%
%
%
%
%
%
%
%

%
%

\chapter{Results of multipole decomposition analyses}
\label{MDA_results}

As discussed in Chapter \ref{Data Analysis}, the experimental double-differential cross sections were decomposed into contributions from the various isoscalar modes (see Eq. \eqref{MDA_defined} and the surrounding discussion)
\begin{align}
  \frac{\diff^2\sigma^\text{exp}(\theta_\text{c.m.},E_x)}{\diff \Omega\, \diff E} & = \sum_\lambda A_\lambda (E_x) \frac{\diff^2\sigma^\text{DWBA}_\lambda(\theta_\text{c.m.},E_x)}{\diff\Omega \, \diff E}.
\end{align}

The results of the multipole decomposition analysis are presented in this Appendix, for each nucleus (\nuc{94,96,97,98,100}{Mo}, and \nuc{40,42,44,48}{Ca}) for the entirety of the excitation energy range over which angular distribution data were extracted. For the molybdenum isotopes, a typical bin width was approximately 500 keV, whereas for the calcium isotopes, 200 keV bins were utilized (1 MeV for \nuc{48}{Ca}).

\foreach \mass in {94,96,97,98,100}{

\newcommand{\numPlots}{3}
\ifnum\mass=96
\renewcommand{\numPlots}{4}
\fi

\begin{figure}
  \centering
  \includegraphics[width=1.0\linewidth]{./figures/MDA_results_figures/\mass Mo_gridPlot_vertical_first.pdf}
  \caption[Multipole decompositions for \nuc{\mass}{Mo} (1/\numPlots)]{Multipole decompositions for \nuc{\mass}{Mo} (1/\numPlots). Shown are contributions from the ISGMR (red), ISGDR (blue), ISGQR (green), and higher multipoles (cyan), alongside the response of the IVGDR (dot-dashed).}
  \label{\mass first}
\end{figure}
\begin{figure}
  \centering
  \includegraphics[width=1.0\linewidth]{./figures/MDA_results_figures/\mass Mo_gridPlot_vertical_second.pdf}
  \caption[Multipole decompositions for \nuc{\mass}{Mo} (2/\numPlots)]{Multipole decompositions for \nuc{\mass}{Mo} (2/\numPlots). Shown are contributions from the ISGMR (red), ISGDR (blue), ISGQR (green), and higher multipoles (cyan), alongside the response of the IVGDR (dot-dashed).}
\end{figure}

\begin{figure}
  \centering
  \includegraphics[width=1.0\linewidth]{./figures/MDA_results_figures/\mass Mo_gridPlot_vertical_third.pdf}
  \caption[Multipole decompositions for \nuc{\mass}{Mo} (3/\numPlots)]{Multipole decompositions for \nuc{\mass}{Mo} (3/\numPlots). Shown are contributions from the ISGMR (red), ISGDR (blue), ISGQR (green), and higher multipoles (cyan), alongside the response of the IVGDR (dot-dashed).}
\end{figure}
\ifnum\mass=96
\begin{figure}
  \centering
  \includegraphics[width=1.0\linewidth]{./figures/MDA_results_figures/\mass Mo_gridPlot_vertical_fourth.pdf}
  \caption[Multipole decompositions for \nuc{\mass}{Mo} (4/\numPlots)]{Multipole decompositions for \nuc{\mass}{Mo} (4/\numPlots). Shown are contributions from the ISGMR (red), ISGDR (blue), ISGQR (green), and higher multipoles (cyan), alongside the response of the IVGDR (dot-dashed).}
\end{figure}
\fi
}

\newcommand{\numPlots}{7}
\foreach \mass in {40,42,44}{
  \begin{figure}
    \centering
    \includegraphics[width=1.0\linewidth]{./figures/MDA_results_figures/\mass Ca_gridPlot_vertical_first.pdf}
    \caption[Multipole decompositions for \nuc{\mass}{Ca} (1/\numPlots)]{Multipole decompositions for \nuc{\mass}{Ca} (1/\numPlots). Shown are contributions from the ISGMR (red), ISGDR (blue), ISGQR (green), and higher multipoles (cyan), alongside the response of the IVGDR (dot-dashed).}
  \end{figure}
  \clearpage

  \begin{figure}
    \centering
    \includegraphics[width=1.0\linewidth]{./figures/MDA_results_figures/\mass Ca_gridPlot_vertical_second.pdf}
    \caption[Multipole decompositions for \nuc{\mass}{Ca} (2/\numPlots)]{Multipole decompositions for \nuc{\mass}{Ca} (2/\numPlots). Shown are contributions from the ISGMR (red), ISGDR (blue), ISGQR (green), and higher multipoles (cyan), alongside the response of the IVGDR (dot-dashed).}
  \end{figure}

  \begin{figure}
    \centering
    \includegraphics[width=1.0\linewidth]{./figures/MDA_results_figures/\mass Ca_gridPlot_vertical_third.pdf}
    \caption[Multipole decompositions for \nuc{\mass}{Ca} (3/\numPlots)]{Multipole decompositions for \nuc{\mass}{Ca} (3/\numPlots). Shown are contributions from the ISGMR (red), ISGDR (blue), ISGQR (green), and higher multipoles (cyan), alongside the response of the IVGDR (dot-dashed).}
  \end{figure}

  \begin{figure}
    \centering
    \includegraphics[width=1.0\linewidth]{./figures/MDA_results_figures/\mass Ca_gridPlot_vertical_fourth.pdf}
    \caption[Multipole decompositions for \nuc{\mass}{Ca} (4/\numPlots)]{Multipole decompositions for \nuc{\mass}{Ca} (4/\numPlots). Shown are contributions from the ISGMR (red), ISGDR (blue), ISGQR (green), and higher multipoles (cyan), alongside the response of the IVGDR (dot-dashed).}
  \end{figure}

  \begin{figure}
    \centering
    \includegraphics[width=1.0\linewidth]{./figures/MDA_results_figures/\mass Ca_gridPlot_vertical_fifth.pdf}
    \caption[Multipole decompositions for \nuc{\mass}{Ca} (5/\numPlots)]{Multipole decompositions for \nuc{\mass}{Ca} (5/\numPlots). Shown are contributions from the ISGMR (red), ISGDR (blue), ISGQR (green), and higher multipoles (cyan), alongside the response of the IVGDR (dot-dashed).}
  \end{figure}

  \begin{figure}
    \centering
    \includegraphics[width=1.0\linewidth]{./figures/MDA_results_figures/\mass Ca_gridPlot_vertical_sixth.pdf}
    \caption[Multipole decompositions for \nuc{\mass}{Ca} (6/\numPlots)]{Multipole decompositions for \nuc{\mass}{Ca} (6/\numPlots). Shown are contributions from the ISGMR (red), ISGDR (blue), ISGQR (green), and higher multipoles (cyan), alongside the response of the IVGDR (dot-dashed).}
  \end{figure}

  \begin{figure}
    \centering
    \includegraphics[width=1.0\linewidth]{./figures/MDA_results_figures/\mass Ca_gridPlot_vertical_seventh.pdf}
    \caption[Multipole decompositions for \nuc{\mass}{Ca} (7/\numPlots)]{Multipole decompositions for \nuc{\mass}{Ca} (7/\numPlots). Shown are contributions from the ISGMR (red), ISGDR (blue), ISGQR (green), and higher multipoles (cyan), alongside the response of the IVGDR (dot-dashed).}
  \end{figure}

}

\renewcommand{\numPlots}{2}
\begin{figure}
  \centering
  \includegraphics[width=1.0\linewidth]{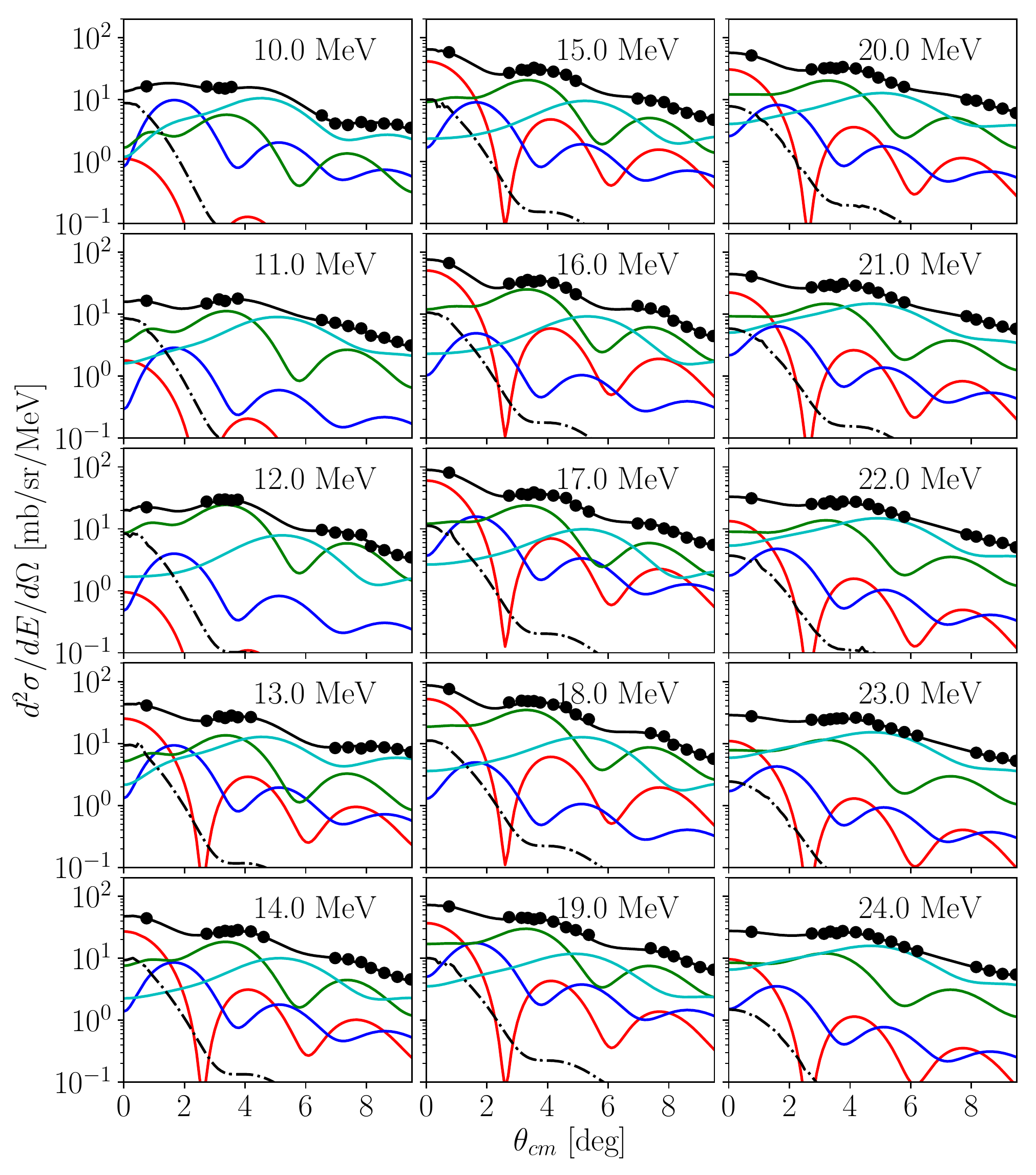}
  \caption[Multipole decompositions for \nuc{48}{Ca} (1/\numPlots)]{Multipole decompositions for \nuc{48}{Ca} (1/\numPlots). Shown are contributions from the ISGMR (red), ISGDR (blue), ISGQR (green), and higher multipoles (cyan), alongside the response of the IVGDR (dot-dashed).}
\end{figure}

\begin{figure}
  \centering
  \includegraphics[width=1.0\linewidth]{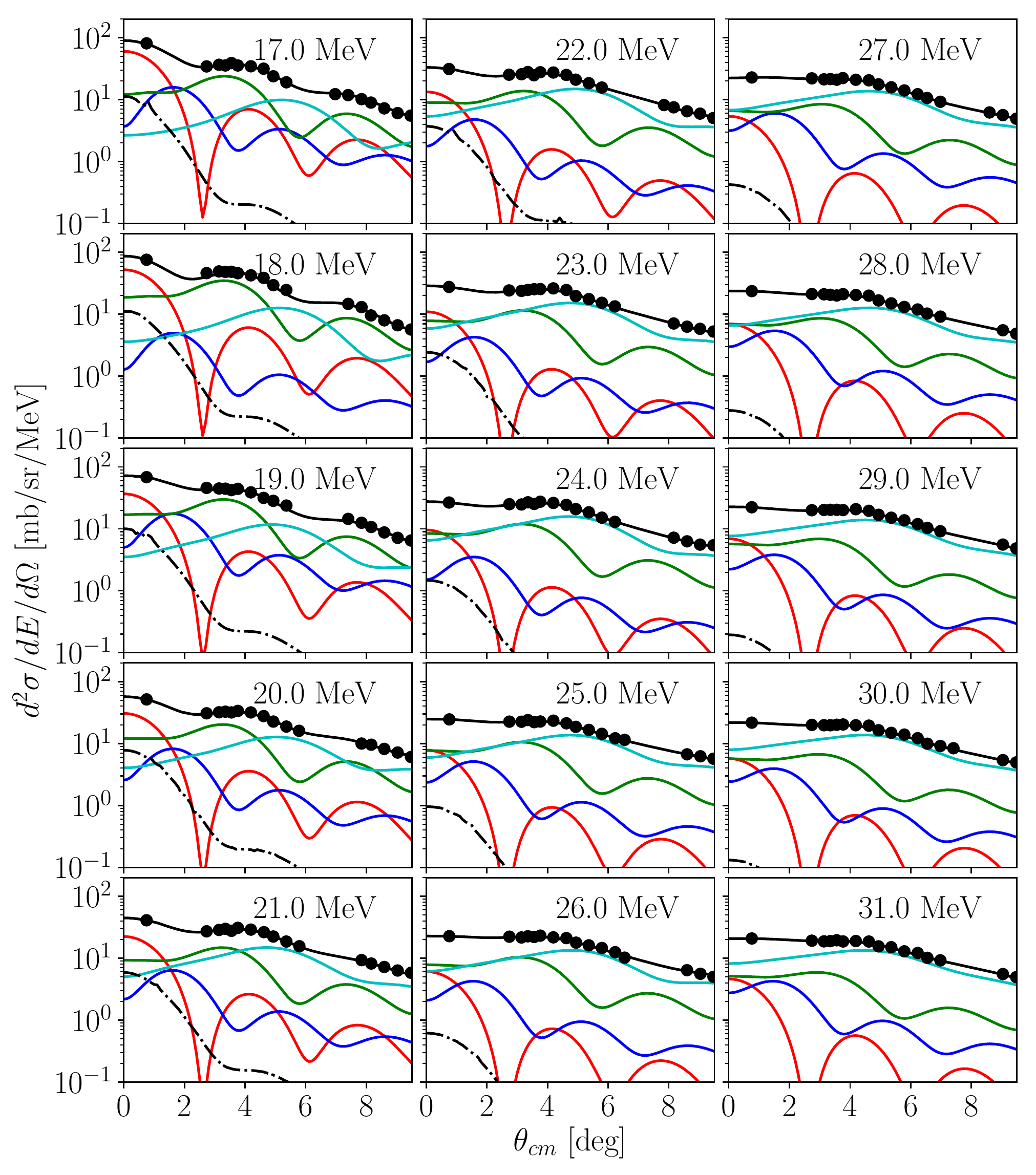}
  \caption[Multipole decompositions for \nuc{48}{Ca} (2/\numPlots)]{Multipole decompositions for \nuc{48}{Ca} (2/\numPlots). Shown are contributions from the ISGMR (red), ISGDR (blue), ISGQR (green), and higher multipoles (cyan), alongside the response of the IVGDR (dot-dashed).}
\end{figure}
%

  %
%
%
%
%
%
%
%
%
%

%
%

\chapter{Extracted isoscalar giant dipole and quadrupole strength distributions}
\label{strength_results}

The multipole decomposition procedure that is described in Chapter \ref{Data Analysis} is capable of extracting the strength distributions for the isoscalar giant dipole and quadrupole resonances in addition to those of the giant monopole resonances. The extracted $A_\lambda$ coefficients described in the definition of the MDA (cf. subsection \ref{MDA_definition_section}) are scaled by the full EWSR as described in Eq. \eqref{conversion_from_ewsr_fraction}, and the resulting strength distributions were analyzed in a manner similar to that of the analysis of the ISGMR strength extracted within each experiment.

\section{ISGDR and ISGQR in the molybdenum nuclei}
\begin{figure}[t!]
  \centering
  \includegraphics[width=0.9\linewidth]{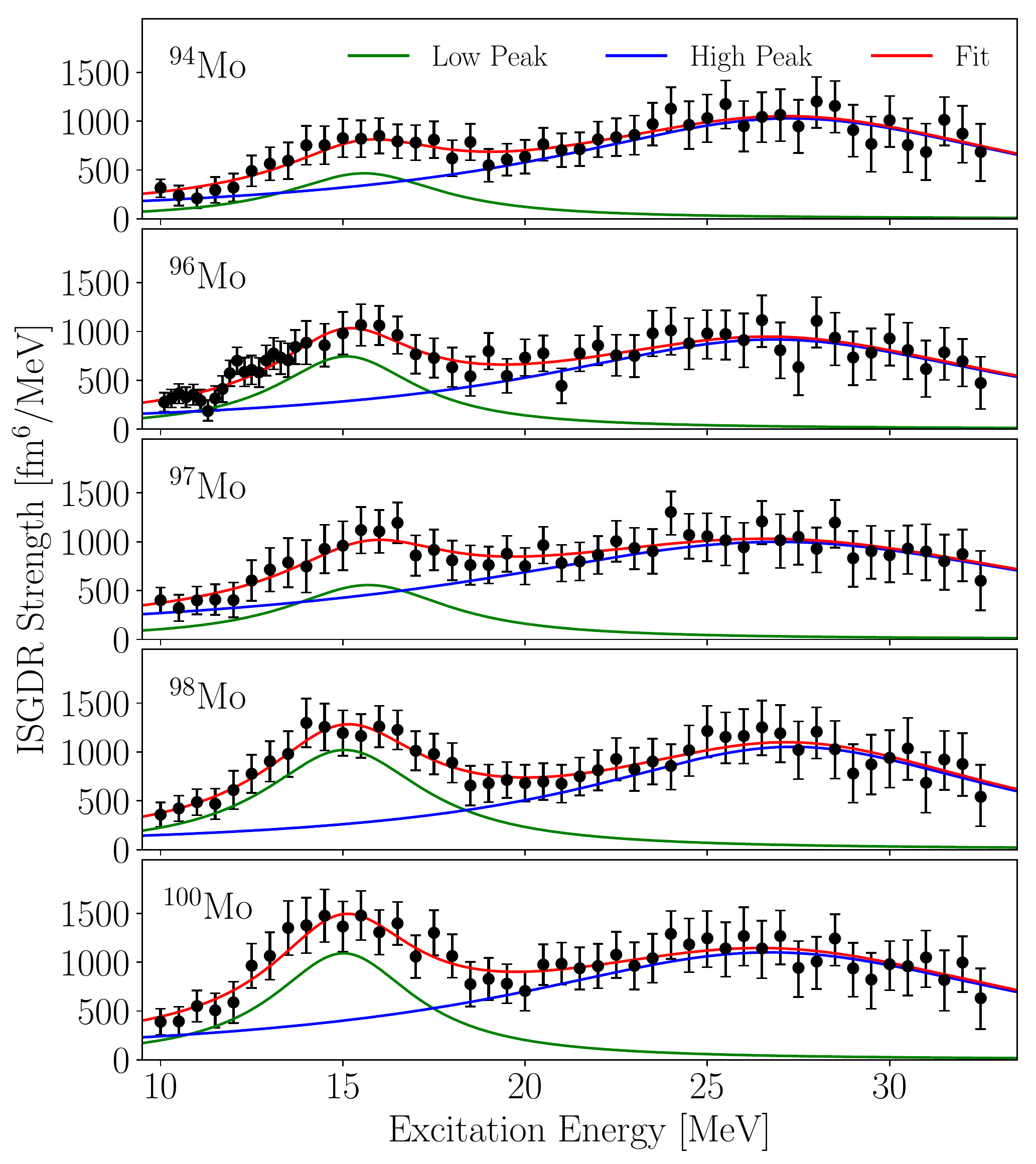}
  \caption{Extracted ISGDR strength distributions for \nuc{94-100}{Mo}, in addition to the fit distributions described in the text.}
  \label{molly_ISGDR_strength}
\end{figure}

In the case of the molybdenum nuclei, the ISGDR shows the characteristic two-peak structure which has been observed in prior experiments \cite{itoh_sm_PRC,itoh_sm_nucA,uchida_208Pb_ISGDR,hunyadi_208Pb_ISGDR}, as depicted in Fig. \ref{molly_ISGDR_strength}, with excitation energies approximately given by $1 \hbar \omega$ and $3\hbar \omega$ (cf. Table \ref{selection_rules}). The two-peak Lorentzian distributions that were fit to these distributions are given in Table \ref{molly_ISGDR_fit_parameters}. The general features of the extracted distributions and the corresponding fit parameters are in agreement, subject to the additional $\sim 20\%$ uncertainties that are necessarily quoted in the extracted strengths \cite{Li_PRC}; this fact is especially critical for interpretation of the high-energy component of the ISGDR strength distributions, as even small statistical fluctuations or systematic uncertainties in the cross sections at higher excitation energies can result in significant changes in the extracted energy-weighted sum rule.

The nuclear continuum which underlies the experimental spectra, and further, the extracted strength distributions, exacerbates the quantification of the EWSR exhausted by the high-energy peak of the ISGDR. The multipole decomposition employed in this work is insensitive to, for example, pick-up and breakup channels which open toward the high-excitation-energy end of the spectra. These channels are highly forward-peaked, and so can mimic the angular distribution characters of the ISGMR and ISGDR \cite{nayak_208Pb_ISGDR}. Thus, a significant portion of the strengths above $20$ MeV reported in this work can be spurious in nature. This conjecture was confirmed by measurements on the ISGDR \cite{hunyadi_208Pb_ISGDR,nayak_208Pb_ISGDR} on \nuc{208}{Pb}, in which both the instrumental background and nuclear continuum have been suppressed by particle-decay double-coincidence measurements of the \nuc{208}{Pb}($\alpha$,$\alpha^\prime$ $p$) \nuc{207}{Ti} reaction. In this measurement, the general positioning of the ISGDR was consistent between the singles measurement (conducted with the present methodology) and the doubles measurement, with the principle difference being the absence of the falsely-attributed ISGDR strength in the latter measurement.

It is understood that only the high-energy peak of the ISGDR corresponds to a compressional oscillation (see Ref. \cite{itoh_sm_nucA} and references cited therein). Due to this fact, while the full distribution of the ISGDR strength is well-modeled in all cases by the combination of the two peaks, the moment ratios for the ISGDR which are presented in Table \ref{molly_ISGDR_moment_ratios_total_EWSRS} are calculated in the same manner as was described for the corresponding \nuc{94-100}{Mo} ISGMR moments in Chapter \ref{results}, but for only this high-energy peak.

\begin{figure}[t!]
  \centering
  \includegraphics[width=0.9\linewidth]{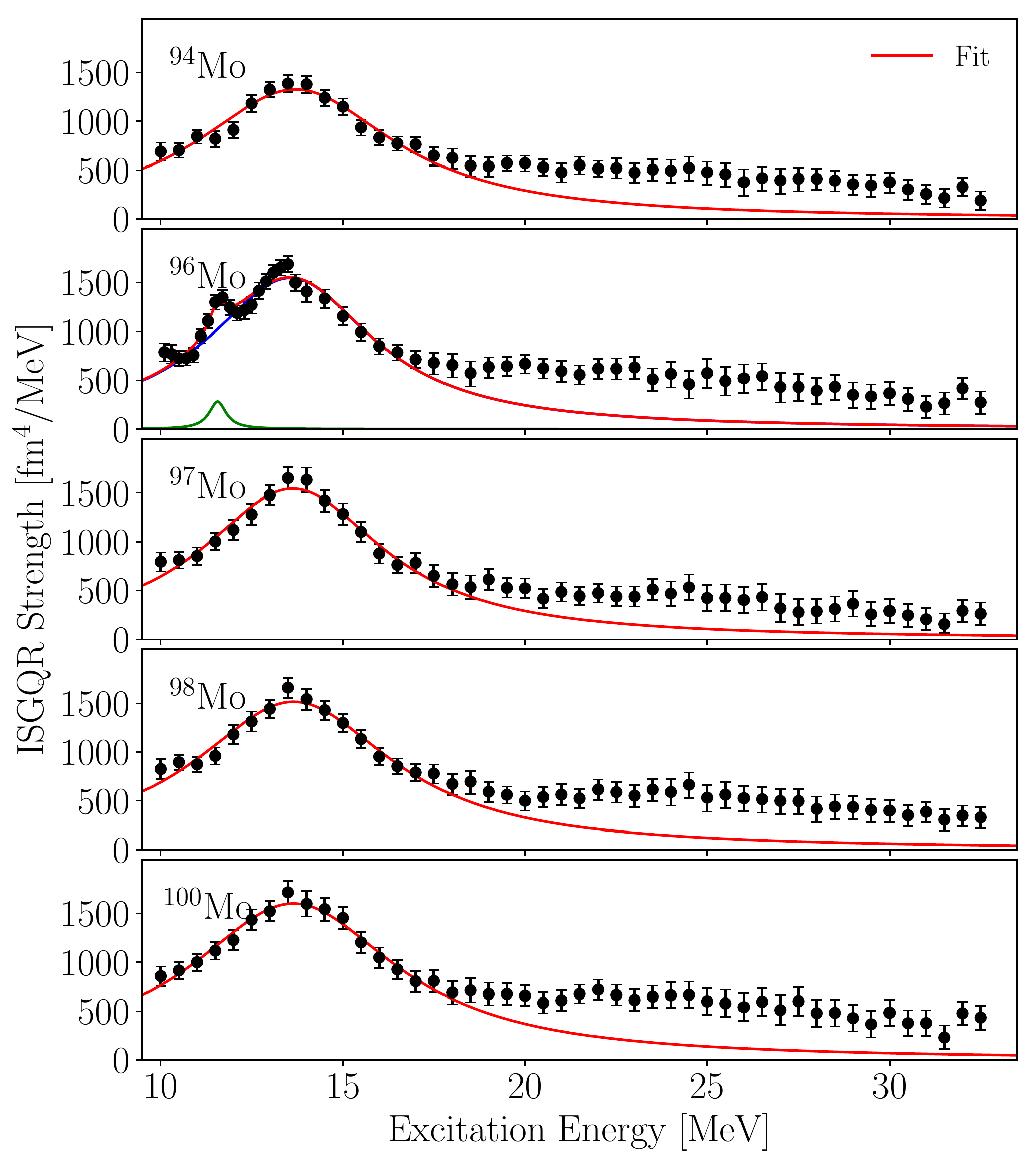}
  \caption{Extracted ISGQR strength distributions for \nuc{94-100}{Mo}, in addition to the fit distributions described in the text.}
  \label{molly_ISGQR_strength}
\end{figure}

In the case of the non-compressional ISGQR strengths, \nuc{94-100}{Mo} each have broadly-peaked distributions of strength which slightly widen with increasing mass number. This is perhaps attributable to axial deformation which causes splitting of the ISGQR into $K=0$, $1$, and $2$ components; it is possible that the coupling of the $K=0$ component of the ISGQR to the ISGMR is possible for the double-peak structure in the ISGMR strength distributions in the higher-mass molybdenum nuclei as described in Chapter \ref{results}.

In the case of \nuc{96}{Mo}, a sharp peak was manifest in the ISGQR strength distribution at $\sim 11.5$ MeV; this peak was modeled simultaneously with the broader ISGQR structure. The fitted Lorentzian distributions for the ISGQR are presented in Table \ref{molly_ISGQR_strength}, with corresponding moment ratios for the fit distributions given in Table \ref{molly_ISGQR_moment_ratios_total_EWSRS}.

\begin{table}[t!]
  \centering
  \caption[Lorentzian-fit parameters for the ISGDR in \nuc{94-100}{M\lowercase{o}}]{Lorentzian-fit parameters for the ISGDR in \nuc{94-100}{M\lowercase{o}} and integrated EWSR between 0 --- 35 M\lowercase{e}V}
  \label{molly_ISGDR_fit_parameters}
\resizebox{\linewidth}{!}{
\begin{tabular}{@{}ccccccccccc@{}}
\toprule
             &  & \multicolumn{3}{c}{Low Peak}                       &  & \multicolumn{3}{c}{High Peak} & & Total           \\ \cmidrule(lr){3-5} \cmidrule(l){7-9} \cmidrule(l){11-11}
      &  & $E_0$       & $\Gamma$   & $m_1$             &  & $E_0$       & $\Gamma$ & $m_1$   & &  $m_1$ \\

      &  & [MeV]      &  [MeV]  &  [\%]            &  &  [MeV]      & [MeV]  & [\%]  & & [\%]\\
\cmidrule(r){1-5} \cmidrule(l){7-9} \cmidrule(l){11-11}
\nuc{94}{Mo} &  & \asymmerror{15.6}{0.7}{0.6} & \asymmerror{2.7}{1.8}{0.9} & $21 \pm 1$ &  & \asymmerror{27.3}{1.0}{0.8} & \asymmerror{8.3}{2.2}{1.7}  & $153 \pm 6$ & & $174 \pm 7$ \\ 
\nuc{96}{Mo} &  & \asymmerror{15.1}{0.3}{0.3} & \asymmerror{2.4}{0.6}{0.5} & $28 \pm 1$ &  & \asymmerror{26.8}{0.9}{0.8} & \asymmerror{7.9}{2.1}{1.6}  & $127 \pm 4$ & & $154 \pm 5$ \\ 
\nuc{97}{Mo} &  & \asymmerror{15.7}{0.7}{0.5} & \asymmerror{2.7}{2.4}{1.0} & $24 \pm 1$ &  & \asymmerror{26.9}{1.2}{1.0} & \asymmerror{10.3}{3.1}{2.4}  & $153 \pm 6$ & & $177 \pm 7$ \\ 
\nuc{98}{Mo} &  & \asymmerror{15.0}{0.3}{0.3} & \asymmerror{2.7}{0.6}{0.5} & $41 \pm 2$ &  & \asymmerror{27.3}{0.9}{0.7} & \asymmerror{7.0}{2.8}{1.6}  & $132 \pm 4$ & & $173 \pm 6$ \\ 
\nuc{100}{Mo} &  & \asymmerror{15.0}{0.3}{0.3} & \asymmerror{2.4}{0.5}{0.5} & $37 \pm 2$ &  & \asymmerror{26.7}{1.0}{0.8} & \asymmerror{8.9}{2.7}{1.9}  & $147 \pm 5$ & & $184 \pm 6$ \\  \bottomrule 
\end{tabular}
}
\end{table}

\begin{table}[t!]
  \centering
  \caption[Lorentzian-fit parameters for the ISGQR in \nuc{94-100}{M\lowercase{o}}]{Lorentzian-fit parameters for the ISGQR in \nuc{94-100}{M\lowercase{o}} and integrated EWSR between 0 --- 35 M\lowercase{e}V}
  \label{molly_ISGQR_fit_parameters}
\resizebox{\linewidth}{!}{
\begin{tabular}{@{}ccccccccccc@{}}
\toprule
             &  & \multicolumn{3}{c}{Low Peak}                       &  & \multicolumn{3}{c}{High Peak} & & Total           \\ \cmidrule(lr){3-5} \cmidrule(l){7-9} \cmidrule(l){11-11}
      &  & $E_0$       & $\Gamma$   & $m_1$             &  & $E_0$       & $\Gamma$ & $m_1$   & &  $m_1$ \\

      &  & [MeV]      &  [MeV]  &  [\%]            &  &  [MeV]      & [MeV]  & [\%]  & & [\%]\\
\cmidrule(r){1-5} \cmidrule(l){7-9} \cmidrule(l){11-11}
\nuc{94}{Mo} &  & - & - & - &  & $13.6 \pm 0.1$ & $3.3 \pm 0.2$  & $118 \pm 2$ & & $118 \pm 2$ \\ 
\nuc{96}{Mo} &  & $11.5 \pm 0.1$ & \asymmerror{0.3}{0.3}{0.1} & \asymmerror{2}{2}{1} &  & $13.6 \pm 0.1$ & $2.8 \pm 0.1$  & $111\pm4$ & & $113\pm4$ \\ 
\nuc{97}{Mo} &  & - & - & - &  & $13.6 \pm 0.1$ & $3.1 \pm 0.3$  & $120 \pm 2$ & & $120 \pm 2$ \\ 
\nuc{98}{Mo} &  & - & - & - &  & $13.6 \pm 0.1$ & $3.5 \pm 0.3$  & $125 \pm 2$ & & $125 \pm 2$ \\ 
\nuc{100}{Mo} &  & - & - & - &  & $13.6 \pm 0.1$ & $3.5 \pm 0.2$  & $132 \pm 2$ & & $132 \pm 2$ \\  \bottomrule 
\end{tabular}
}
\end{table}

\begin{table}[t!]
\centering
\caption{Moment ratios for for the high-energy component of the ISGDR of \nuc{94-100}{M\lowercase{o}} calculated between $0$ --- $35$ M\lowercase{e}V from the fit distributions of Table \ref{molly_ISGDR_fit_parameters}}
\label{molly_ISGDR_moment_ratios_total_EWSRS}
\begin{tabular}{@{}cccc@{}}
\toprule
Nucleus      & $\sqrt{m_{1}/m_{-1}}$ & $m_1/m_0$   & $\sqrt{m_{3}/m_{1}}$   \\
     & [MeV] & [MeV]  &  [MeV]  \\ \midrule
\nuc{94}{Mo} & $20.5 \pm 0.8$            & $23.5 \pm 0.6$ & $26.8 \pm 0.7$           \\
\nuc{96}{Mo} & $20.5 \pm 0.7$            & $23.4 \pm 0.5$ & $26.6 \pm 0.3$            \\
\nuc{97}{Mo} & $19.2 \pm 0.8$            & $22.6 \pm 0.6$ & $26.4 \pm 0.6$            \\
\nuc{98}{Mo} & $21.3 \pm 0.8$            & $24.1 \pm 0.5$ & $27.0 \pm 0.3$            \\
\nuc{100}{Mo} & $19.9 \pm 0.7$            & $23.0 \pm 0.5$ & $26.4 \pm 0.3$            \\ \bottomrule
\end{tabular}
\end{table}

\begin{table}[t!]
\centering
\caption{Moment ratios for the ISGQR of \nuc{94-100}{M\lowercase{o}} calculated between $0$ --- $35$ M\lowercase{e}V from the fit distributions of Table \ref{molly_ISGQR_fit_parameters}}
\label{molly_ISGQR_moment_ratios_total_EWSRS}
\begin{tabular}{@{}cccc@{}}
\toprule
Nucleus      & $\sqrt{m_{1}/m_{-1}}$ & $m_1/m_0$   & $\sqrt{m_{3}/m_{1}}$   \\
     & [MeV] & [MeV]  &  [MeV]  \\ \midrule
\nuc{94}{Mo} & $12.7 \pm 0.1$            & $14.2 \pm 0.1$ & $17.4 \pm 0.6$           \\
\nuc{96}{Mo} & $12.7 \pm 0.2$            & $13.9 \pm 0.2$ & $16.8 \pm 0.6$            \\
\nuc{97}{Mo} & $12.7 \pm 0.1$            & $14.1 \pm 0.1$ & $17.1 \pm 0.7$            \\
\nuc{98}{Mo} & $12.7 \pm 0.1$            & $14.2 \pm 0.1$ & $17.3 \pm 0.7$            \\
\nuc{100}{Mo} & $12.6 \pm 0.1$            & $14.2 \pm 0.1$ & $17.5 \pm 0.6$            \\ \bottomrule
\end{tabular}
\end{table}

\section{ISGDR and ISGQR in the calcium nuclei}

\begin{figure}[t!]
  \centering
  \includegraphics[width=\linewidth]{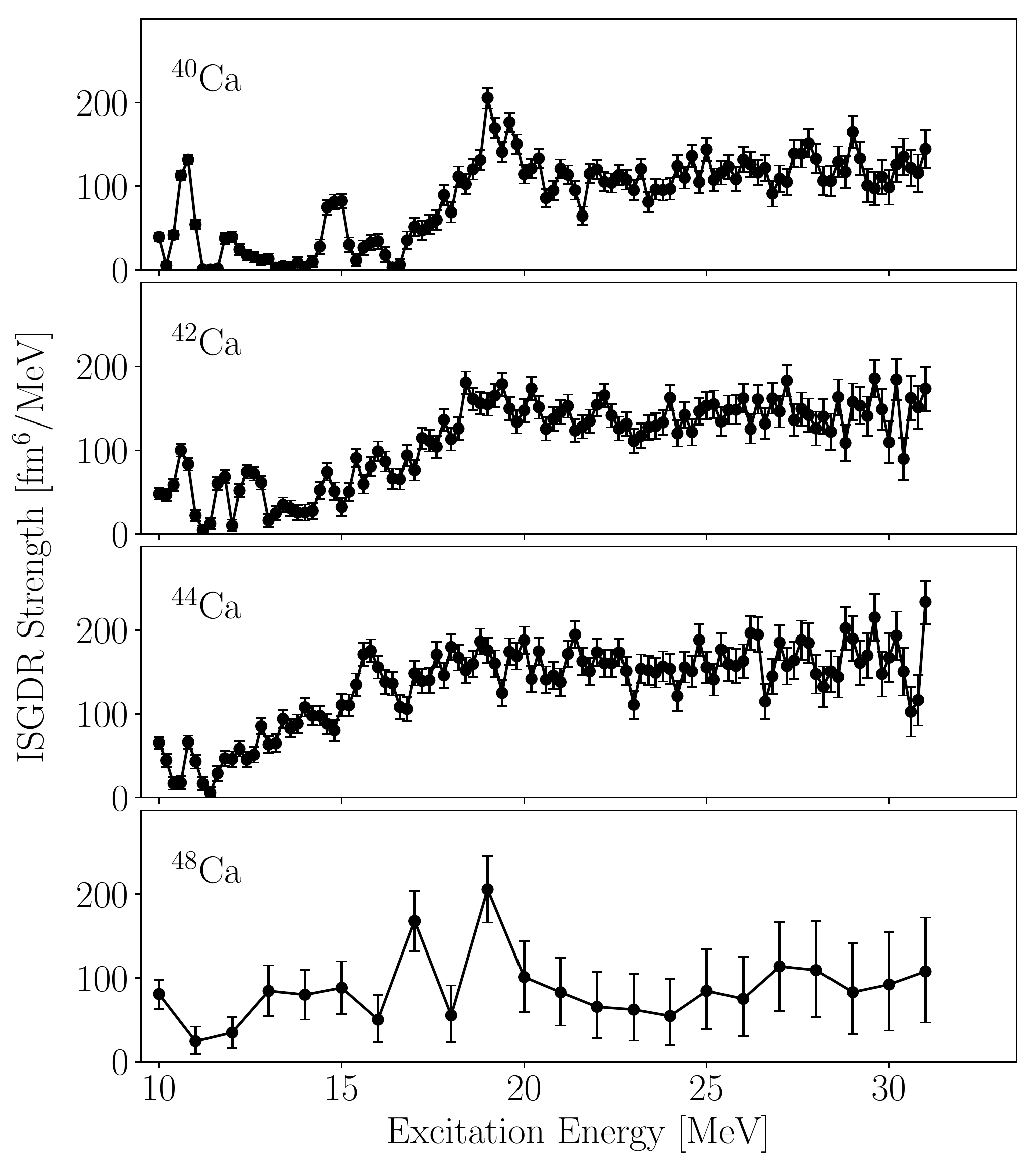}
  \caption{Extracted ISGDR strength distributions for \nuc{40-48}{Ca}.}
  \label{calcium_ISGDR_strength}
\end{figure}

\begin{figure}[t!]
  \centering
  \includegraphics[width=\linewidth]{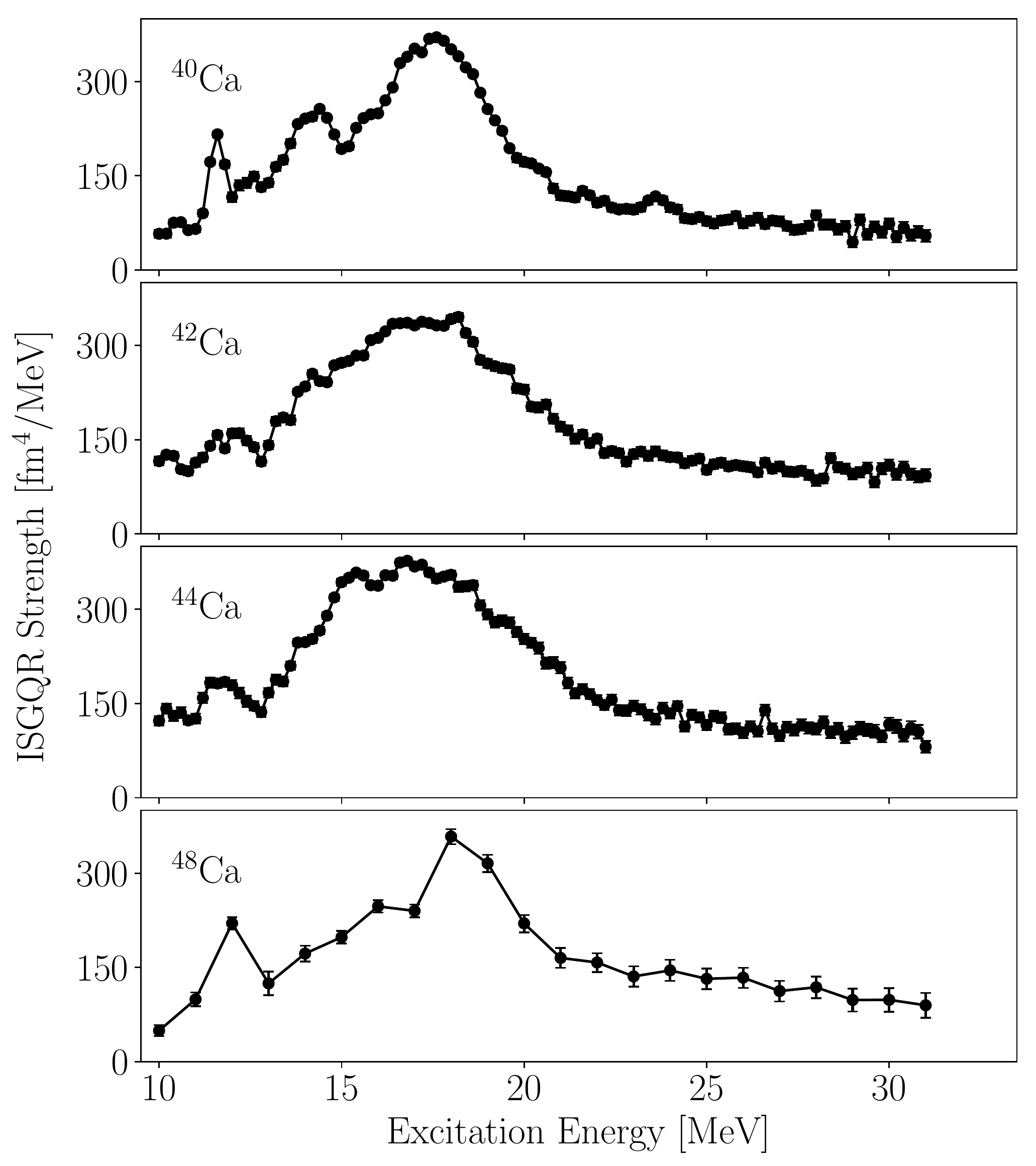}
  \caption{Extracted ISGQR strength distributions for \nuc{40-48}{Ca}.}
  \label{calcium_ISGQR_strength}
\end{figure}

For the calcium nuclei, the ISGDR and ISGQR responses were themselves similarly fragmented as was observed for the corresponding extracted monopole strength distributions. As a result, the distributions were not amenable to being described by any well-founded combination of peaks.

The extracted ISGDR strengths are depicted in Fig. \ref{calcium_ISGDR_strength}. In contrast to the molybdenum nuclei, there does not appear to be an abundance of low-energy strength until $\sim$ 20 MeV. Beyond this threshold, it is difficult to say conclusively whether the measured strength is due to the compressional response (high-energy, $3\hbar \omega$) of the ISGDR, the non-bulk (low-energy, $1\hbar\omega$) response, or some combination of the two modes. With this caveat in mind, Table \ref{calcium_ISGDR_moment_ratios_total_EWSRS} reports the moment ratios calculated from a direct Monte Carlo sampling of the probability distributions for each strength bin extracted from the multipole decomposition for excitation energies above and including $20$ MeV.

Similarly, the moment ratios for the ISGQR were calculated over the full excitation energy range in the same way, and are reported in Table \ref{calcium_ISGQR_moment_ratios_total_EWSRS}.

\begin{table}[t!]
\centering
\caption{Moment ratios for the ISGDR in \nuc{40-48}{C\lowercase{a}}, calculated between $20$ --- $31$ M\lowercase{e}V}
\label{calcium_ISGDR_moment_ratios_total_EWSRS}
\begin{tabular}{@{}cccc@{}}
\toprule
Nucleus      & $\sqrt{m_{1}/m_{-1}}$ & $m_1/m_0$   & $\sqrt{m_{3}/m_{1}}$   \\
     & [MeV] & [MeV]  &  [MeV]  \\ \midrule
\nuc{40}{Ca} & $25.6 \pm 0.1$            & $25.8 \pm 0.1$ & $26.4 \pm 0.1$           \\
\nuc{42}{Ca} & $25.5 \pm 0.1$            & $25.7 \pm 0.1$ & $26.3 \pm 0.1$            \\
\nuc{44}{Ca} & $25.5 \pm 0.1$            & $25.7 \pm 0.1$ & $26.3 \pm 0.1$            \\
\nuc{48}{Ca} & $26.1 \pm 0.5$            & $26.3 \pm 0.5$ & $26.8 \pm 0.5$            \\ \bottomrule
\end{tabular}
\end{table}

\begin{table}[t!]
\centering
\caption{Moment ratios for the ISGQR in \nuc{40-48}{C\lowercase{a}}, calculated between $10$ --- $31$ M\lowercase{e}V}
\label{calcium_ISGQR_moment_ratios_total_EWSRS}
\begin{tabular}{@{}cccc@{}}
\toprule
Nucleus      & $\sqrt{m_{1}/m_{-1}}$ & $m_1/m_0$   & $\sqrt{m_{3}/m_{1}}$   \\
     & [MeV] & [MeV]  &  [MeV]  \\ \midrule
\nuc{40}{Ca} & $18.0 \pm 0.1$            & $18.6 \pm 0.1$ & $20.6 \pm 0.1$           \\
\nuc{42}{Ca} & $18.5 \pm 0.1$            & $19.1 \pm 0.1$ & $21.2 \pm 0.1$            \\
\nuc{44}{Ca} & $18.4 \pm 0.1$            & $19.1 \pm 0.1$ & $21.2 \pm 0.1$            \\
\nuc{48}{Ca} & $19.0 \pm 0.2$            & $19.8 \pm 0.2$ & $21.9 \pm 0.3$            \\ \bottomrule
\end{tabular}
\end{table}

%

\backmatter              

\bibliographystyle{nddiss2e}
\bibliography{thesis_source}           

\end{document}